\documentclass[12pt]{article}

\usepackage[margin=1in]{geometry}
\usepackage{amsmath,amsthm,amsfonts,amssymb}
\usepackage{bm,rotating,enumitem}
\usepackage{setspace}
\usepackage[colorlinks,citecolor=blue,urlcolor=blue]{hyperref}
\usepackage{nameref}
\usepackage[authoryear]{natbib}
\usepackage{lmodern}
\usepackage{graphicx}
\usepackage{epstopdf}
\usepackage{tabularx}
\usepackage{makecell}
\usepackage{caption}
\usepackage{subcaption}
\usepackage{tikz}
\usetikzlibrary{shapes,decorations,arrows,calc,arrows.meta,fit,positioning,shadows}
\usepackage{booktabs}
\usepackage{pifont}
\usepackage{grffile}
\usepackage{xcolor}
\usepackage{comment}
\usepackage{authblk}

\theoremstyle{plain}

\theoremstyle{remark}

\newcommand{\cmark}{\ding{51}}
\newcommand{\xmark}{\ding{55}}

\newcommand{\blambda}{\bm \lambda}

\newcommand{\btheta}{\bm \theta}

\newcommand{\bpi}{\bm \pi}

\newcommand{\bnu}{\bm \nu}

\def\bSig\mathbf{\Sigma}

\newcommand{\rev}[1]{#1}

\tikzset{
    every picture/.style=thick,
    -Latex,auto,node distance =1 cm and 1 cm,semithick,
    state/.style ={ellipse, draw, minimum width = 0.7 cm},
    point/.style = {circle, draw, inner sep=0.04cm,fill,node contents={}},
    bidirected/.style={Latex-Latex,dashed},
    el/.style = {inner sep=2pt, align=left, sloped}
}

\title{Bayesian Federated Cause-of-Death Classification and Quantification Under Distribution Shift}
\author[1]{Yu Zhu}
\author[1]{Jason Teng}
\author[1]{Zehang Richard Li}
\affil[1]{Department of Statistics, University of California, Santa Cruz}
\date{}

\begin{document}
\maketitle

\begin{abstract}
In regions lacking medically certified causes of death, verbal autopsy (VA) is a widely used tool to ascertain the cause of death through interviews with caregivers. Data collected by VAs are often analyzed using probabilistic algorithms. The performance of these algorithms often degrades due to distribution shift across populations. Most existing VA algorithms rely on centralized training, requiring full access to training data for joint modeling. This can be infeasible due to privacy and logistical constraints. In this paper, we propose a novel Bayesian Federated Learning (BFL) framework that avoids data sharing across multiple training sources. Our method supports individual-level cause-of-death classification and population-level quantification of cause-specific mortality fractions in a target domain with limited or no local labeled data. The proposed framework is modular, computationally efficient, and compatible with a wide range of existing VA algorithms as base models, facilitating flexible deployment in real-world mortality surveillance systems. We validate the performance of BFL through extensive experiments on two real-world VA datasets under varying levels of distribution shift scenarios. Our results show that BFL significantly outperforms single-domain base models and performs comparably to or better than joint modeling.
\end{abstract}


\section{Introduction}
\label{sec:intro}
Understanding the distribution of causes of death is essential for public health planning, disease surveillance, and evaluating the impact of health interventions. Many low- and middle-income countries (LMICs) lack complete civil registration and vital statistics systems that can routinely produce reliable cause-of-death statistics. The World Health Organization (WHO) estimates that only $10\%$ of deaths in Africa are registered, and just $8\%$ of those have a documented cause of death \citep{world2021civil, onyango2024using}. In regions where medical certification of causes of death is unavailable, verbal autopsy (VA) is widely used to infer causes of death and to estimate trends and disparities in mortality burdens \citep{Maher2010health, Sankoh2021epidemiology, Nkengasong2020birth, chu2024temporal}. 

VA involves structured interviews with family members or caregivers of deceased individuals to collect information on symptoms, circumstances, and events preceding death. Compared to alternative cause-of-death ascertainment methods such as minimally invasive tissue sampling (MITS) \citep{bassat2017validity}, VA is the only feasible procedure in many low-resource settings, as it does not require specialized clinical infrastructure or equipment. Over the last decade, statistical algorithms have increasingly replaced expert review of VA survey responses. Various algorithmic or probabilistic VA models have been routinely used by national statistics offices and health surveillance sites around the world \citep[e.g.][]{byass2019integrated, mccormick2016probabilistic, serina2015improving}. A recent review of the the adoption and implementation of VA models can be found in \citet{chandramohan2021estimating}. 

Probabilistic VA models typically utilize supervised or semi-supervised classifiers trained on reference deaths with known causes. These models learn relationships between reported symptoms and underlying causes to classify unlabeled deaths and estimate cause-specific mortality fractions (CSMFs) in populations. The latter task, estimating the proportion of individuals with an unknown outcome, is known as quantification learning \citep{gonzalez2017review}. For both tasks, a critical limitation of existing models is the need for access to adequate reference death data. In reality, labeled VA data are scarce and are often collected by independent organizations and research projects in different countries. Consolidating data in a central location can be challenging due to privacy regulations and logistical constraints. As a result, the choice of training datasets is often dictated by accessibility rather than relevance to the target population. Analysts without access to reference deaths cannot apply many state-of-the-art VA models that require training data, and must instead rely on simpler, more rigid methods with fixed parameters. Notably, all three VA methods currently recommended by the WHO, InterVA \citep{byass2019integrated}, Tariff \citep{serina2015improving}, and InSilicoVA \citep{mccormick2016probabilistic}, fall into this category: with key parameters describing the marginal association between symptoms and causes fixed to summary statistics from a specific training dataset or to information provided by expert physicians \citep{Nichols2018who}. Several recently developed approaches in the literature \citep[e.g.,][]{tsuyoshi2017, li2021bayesian, kunihama2024bayesian} have been shown to consistently outperform the three recommended models. However, the adoption of these more advanced and flexible models is limited in real-world settings, primarily due to the need for access to training data.

A second major challenge that compounds this problem is distribution shift. VAs collected from different populations usually produce systematically different distributions in reported symptoms, even for deaths with the same cause. Such shift can be due to differences in epidemiology, healthcare access, and cultural norms. Distribution shifts can also exist within the same population when reference deaths are not representative, e.g., when subpopulations are selectively included for labeling \citep{zhu2025hierarchicallatentclassmodels}. Both cross- and within-population shifts can introduce substantial bias in CSMF quantification and significant limit in the generalizability of VA algorithms \citep{li2021bayesian}. Existing methods to address distribution shift require either a diverse collection of reference deaths from multiple populations \citep{li2021bayesian, wu2021tree} or sufficient local reference deaths within the target population \citep{datta2020regularized, fiksel2020generalized, pramanik2023modeling}. Therefore, both solutions effectively demand access to even more labeled data.

Lastly, even in scenarios where data access is unrestricted, computational constraints create another barrier to adopting advanced VA models. The recent VA algorithms usually rely on complex Bayesian hierarchical models to jointly model both labeled and unlabeled data, which are computationally expensive to refit as new VAs and reference deaths are collected, especially in continuous mortality surveillance systems. Currently, there is no established guidance on how to select training datasets or determine how frequently to retrain models, leading to ad hoc and inconsistent practices.

In this paper, we address these challenges by introducing a Bayesian Federated Learning (BFL) framework that eliminates the need for centralized labeled data in VA analysis. Our approach allows analysts to leverage reference death datasets from multiple domains, e.g., regions, countries, or time periods, without the need for direct data sharing. The framework addresses a common but previously unexplored scenario: reference deaths exist across multiple domains but can only be analyzed locally, while target datasets may have few or no local labeled deaths. 
Our BFL approach trains separate base models independently within each domain and shares only summarized model parameters with the analyst in the target population. These model summaries are combined via a Bayesian latent-class aggregation model that explicitly models shifts in cause prevalences and the conditional distributions of symptoms given causes. 
The framework offers three key advantages over existing analysis approaches: it is fully modular and supports various VA algorithms as base models, adds negligible computational cost once the base models have been trained, and performs robustly across different levels of distribution shift and label availability. We evaluate the framework and through extensive experiments on two VA datasets under different degrees of distribution shift and label availability scenarios.

The remainder of the paper is organized as follows. Section \ref{sec:background} reviews related work in both VA methods and federated learning. Section \ref{sec:method} introduces the proposed modeling framework for verbal autopsy data. Section \ref{sec:phmrc} evaluates the proposed approach through three sets of experiments under varying levels of distribution shift, based on the Population Health Metrics Research Consortium (PHMRC) gold-standard VA dataset \citep{murray2011population}. Section \ref{sec:champs} presents a case study of the Child Health and Mortality Prevention Surveillance (CHAMPS) neonatal dataset \citep{blau2019overview} using the proposed method. Section \ref{sec:discuss} concludes the paper with a discussion of key findings and directions for future research.

\section{Background and related work}\label{sec:background}
\rev{Cause-of-death assignment based on VA data was traditionally performed by physicians, who review each questionnaire and certify a likely cause, a process known as physician-coded verbal autopsy \citep{chandramohan2021estimating}. Physician coding is expensive and slow, diverts scarce clinical capacity in the settings where VA is most needed, and physicians reviewing the same record frequently disagree \citep{lozano2011performance}. These limitations have motivated a large literature on automated cause-of-death classification over the past three decades.}

\rev{VA instruments contain several hundred questions, producing high-dimensional and sparse binary symptom vectors, while cause lists typically contain dozens of categories. Many cause-symptom combinations are therefore observed only a handful of times in a typical VA dataset.} As a result, early automated VA methods typically operated with highly simplified parametric models \citep{byass2019integrated, mccormick2016probabilistic, Serina2015VA, miasnikof2015naive} or decision rules \citep{kalter1990validation}, where symptom-cause relationships are derived from domain knowledge or single reference datasets. For example, the widely used InterVA \citep{byass2019integrated} and InSilicoVA algorithms \citep{mccormick2016probabilistic} assume that the probability of observing each symptom given any cause of death is provided by experts or estimated from training data. This information is shared as fixed model parameters alongside the software and cannot be easily updated for local contexts.
Thus, these methods are highly susceptible to shifts in conditional symptom distributions across datasets. However, they remain widely used in practice largely because they  require only predetermined parameters rather than training datasets.

\rev{In parallel, generic machine learning classifiers have also been applied to VA data, including early neural networks \citep{boulle2001case} and random forests \citep{flaxman2011random}, and \citet{KingandLu2008} proposed estimating CSMFs directly without performing individual-level classification. More recently, natural language processing methods have been explored to extract causes of death from free-text VA narratives \citep{jeblee2019automatically, blanco2020extracting, chen2025lava}.}

\rev{A more recent line of Bayesian generative models introduces additional flexibility in modeling symptom dependence and population heterogeneity. Bayesian factor models \citep{tsuyoshi2017} and FARVA \citep{moran2021bayesian} represent dependence through low-dimensional latent factors. LCVA and DoubleTree use latent-class structures to model heterogeneous symptom distributions within causes and across domains \citep{li2021bayesian, wu2021tree}. Dimension-grouped tensor-decomposition approaches have also been proposed to characterize symptom dependence more flexibly \citep{zhu2025flexiblebayesiantensordecomposition}. These models can improve individual classification and CSMF estimation, but fitting them jointly across populations typically requires centralized access to individual-level reference data and may be computationally expensive.}

A different line of work circumvents the need for individual-level training data by calibrating pretrained algorithms with labeled data \citep{datta2020regularized, fiksel2020generalized}. While these calibration methods can be applied in decentralized settings, their modeling approach differs fundamentally as they focus exclusively on estimating misclassification rates and the population-level CSMF, without modeling the data generating process of observed VA data or performing individual-level classification. We provide detailed comparisons with our BFL approach in Section \ref{sec:calibratedVA}.

\rev{
Beyond predictive accuracy, deploying VA methods in practice raises a distinct set of challenges. VA data are usually distributed across institutions with different governance and data-sharing requirements, and may contain sensitive information about the deceased and their families. Questionnaire versions, translations, skip patterns, symptom definitions, and cause lists may differ across settings, requiring careful harmonization before models can be transferred \citep{Nichols2018who, world2022verbal}. Locally labeled deaths may also be selectively collected, as deaths with reference causes are usually sampled from health facilities or surveillance catchment areas and are not representative of community deaths. This creates a form of verification bias long recognized in the diagnostic testing literature \citep{begg1983assessment}, and recently documented in VA settings \citep{zhu2025hierarchicallatentclassmodels}. At the same time, VA is increasingly embedded in routine and rapid mortality surveillance systems \citep{gomes_nationwide_2017, world2020revealing, duarte2021rapid}, which require timely, reproducible, and computationally affordable analysis, along with outputs that practitioners can interpret and audit. Methods that require refitting a large joint model as new data arrive are difficult to sustain in these settings.
}

Outside of VA, similar challenges with sensitive data and data-model trade-offs exist in many clinical prediction tasks. Federated learning (FL) has emerged as a promising solution in such settings \citep[e.g.,][]{dayan2021federated, rieke2020future, han2025federated}. FL is a distributed machine learning paradigm that enables multiple decentralized parties to collaboratively train a model without sharing their local data. Formally, given $M$ clients, each with data drawn from $p_m(X, Y)$, the goal is usually to find a global model that minimizes the average expected loss across all clients using locally trained predictive models \citep{mcmahan2023FedAvg, achituve2021personalizedfederatedlearninggaussian, zhang2022personalizedfederatedlearningvariational, kotelevskii2023Fedpop}. 
When client-level data distributions are heterogeneous, client models are often reweighted adaptively \citep[e.g.,][]{ghari2023POFL, dinh2022PFL,Chen2021HFL}.
In the Bayesian framework, several approaches have been developed to approximate the global posterior distribution from multiple local posterior distributions given partial data. For example, \citet{chen2021fedbemakingbayesianmodel} combined the local predictions by assuming $p(Y \mid X) \approx \int_\phi p(Y \mid X; \phi)dq(\phi)$, where $q(\phi)$ is a parametric approximation to the distributions of the locally estimated parameters $(\hat\phi_1, ..., \hat\phi_M)$ across the $M$ clients. 
This approach was further developed in \citet{bhatt2023FLU} to account for the uncertainty of locally estimated parameters. A recent review of the Bayesian FL literature can be found in \citet{cao2023bayesianfederatedlearningsurvey}. 

 Table \ref{tab:method_comparison} summarizes existing VA methods in terms of their capabilities for handling distribution shift and data access requirements. Current approaches face a fundamental trade-off: simpler methods avoid data sharing but cannot adapt to distribution shift, while sophisticated methods handle data heterogeneity at the cost of requiring centralized data access. So far, no VA method has addressed modeling under decentralized
data settings.  \rev{The BFL framework introduced in the next section aims to relax this trade-off. Each domain shares fitted model components rather than individual records, and the target-domain model learns how much each domain contributes to each cause.}

\begin{table}[htb]
\centering
\begingroup
\footnotesize
\setlength{\tabcolsep}{4pt}
\renewcommand{\arraystretch}{1.25}
\renewcommand{\tabularxcolumn}[1]{m{#1}}
\begin{tabularx}{\textwidth}{
    >{\raggedright\arraybackslash}X
    >{\raggedright\arraybackslash}m{0.26\textwidth}
    >{\centering\arraybackslash}m{0.17\textwidth}
    >{\centering\arraybackslash}m{0.15\textwidth}}
\toprule
\textbf{Method(s)} &
\textbf{Training-data access} &
\textbf{Handles training-domain heterogeneity} &
\textbf{Individual classification} \\
\midrule
\makecell[l]{
Tariff/SmartVA \citep{Serina2015VA} \\
InterVA \citep{byass2019integrated}  \\
InSilicoVA \citep{mccormick2016probabilistic}}  
        & No training data required; estimated parameters only. & \xmark & \cmark \\ \midrule
\makecell[l]{BF \citep{tsuyoshi2017} \\
    FARVA \citep{moran2021bayesian}} & Full training data required. & \xmark & \cmark \\ \midrule
\makecell[l]{LCVA \citep{li2021bayesian} \\
DoubleTree \citep{wu2021tree}} & Full training data required. & \cmark & \cmark \\ \midrule
\makecell[l]{BTL \citep{datta2020regularized} \\
GBQL \citep{fiksel2020generalized}} & Local labeled data required. & \cmark & \xmark \\ \midrule
\textbf{BFL} & No training data required; local labeled data optional. & \cmark & \cmark \\
\bottomrule
\end{tabularx}
\endgroup
\caption{Comparison of major existing VA algorithms and the proposed BFL model.}
\label{tab:method_comparison}
\end{table}


\section{Bayesian federated learning for VA data} \label{sec:method}

Consider $M$ training datasets from different studies, locations, or time periods. We refer to these as $M$ training domains throughout the paper. For the $m$-th domain with $n_m$ deaths, let $X_i^{(m)} \in \{0, 1\}^p$ denote the $p$-dimensional vector of binary signs/symptoms collected by VA and $Y_i^{(m)} \in \{1, ..., C\}$ denote the reference cause of death for the $i$-th death. We consider the situation where a pre-defined list of $C$ mutually exclusive causes is available. Individual domains may not contain deaths from all $C$ causes. However, we assume that pooling across all domains yields deaths from each cause. For the target domain, let $X_i^{(0)} \in \{0, 1\}^p$ and $Y_i^{(0)} \in \{1, ..., C\}$ denote the signs/symptoms and the cause of death for the $i$-th death for $i = 1, ..., n_0$. The goal of the analysis is to produce both individual-level cause-of-death assignments $\hat Y_i^{(0)}$ and population-level CSMF estimates $\hat\bpi^{(0)} = (\hat\pi^{(0)}_1, ..., \hat\pi^{(0)}_C)$, where $\hat\pi^{(0)}_c$ is the prevalence of the $c$-th cause in the target domain.

\subsection{Joint modeling of VAs from multiple domains}

We first review joint modeling approaches for pooled data, as these methods directly motivate our BFL framework. Joint modeling of VA data across multiple sources was first considered in the LCVA algorithm developed in \citet{li2021bayesian}. The key insight is that deaths from the same cause may manifest different symptom patterns across domains, but these patterns can be represented as domain-specific mixtures of shared latent symptom profiles. LCVA assumes, for both the target and training domains, i.e., $m = 0, .., M$,
\begin{align*}
Y_i^{(m)}  &\sim \mbox{Cat}(\bpi^{(m)}), \\
Z_i\mid Y_i^{(m)} = c &\sim \mbox{Cat}(\bnu_c^{(m)}), \\
X_{ij}^{(m)} \mid  Y_i^{(m)} = c, Z_i = k &\sim \mbox{Bern}(\theta_{ckj}),
\end{align*}
where $Z_i \in \{1, ..., K\}$ is a latent class indicator for each death and $\bnu_c^{(m)}$ are $K$-dimensional domain-specific mixing weights of the latent classes with a stick-breaking prior. \rev{The  introduction of $Z_i$ allows deaths due to the same cause to exhibit different symptom patterns and enables flexible characterization of the dependence structures among symptoms given each cause \citep{li2021bayesian}.} The only component shared across domains is $\btheta = \{\theta_{ckj}\}$, which represents $K$ latent symptom profiles for each cause. LCVA uses a sparse latent class model \citep{zhou2015bayesian} to characterize the heterogeneous distributions of symptoms and causes across domains. A spike-and-slab prior was placed on $\btheta$ to encourage these latent profiles to be similar for most symptoms. 
Different modeling choices can be made for $\bnu^{(0)}$ to borrow information from the training domains. For example, when $M > 1$, the domain-level mixture model in \citet{li2021bayesian} assumes $\bnu_c^{(0)} = \sum_{m=1}^M \eta_{m} \bnu_c^{(m)}$. When $M = 1$, the model reduces to a single training domain and we assume the training and target mixing weights are the same, i.e., $\bnu_c^{(0)} = \bnu_c^{(1)}$.
The shared collection of symptom profiles $\btheta$ is central to the formulation of LCVA. It provides a concise summary of data patterns that are shared across domains. More sophisticated models for $\btheta$ have also been explored in \citet{zhu2025flexiblebayesiantensordecomposition}, though they usually perform similarly to LCVA in practice. Regardless of the modeling choices for the latent parameters, the shared symptom profiles require data from all training domains to be modeled jointly.

\subsection{The BFL model for VA data}
We now describe our proposed BFL approach for modeling VA data.
We start with $M$ pretrained base models using data from each domain. We treat these pretrained models as black-box characterizations of the joint distribution of symptoms and causes within each domain. 
That is, they provide a model-based approximation of $p_m(X, Y)$ for $m = 1, ..., M$. We first consider the fully unsupervised analysis with all $Y_i^{(0)}$ unobserved, and
adopt a latent class model framework similar to LCVA for the target domain, assuming the conditional distribution of symptoms given each cause can be represented as a weighted average of those from the training domains. For some nonnegative weights $\blambda = \{\lambda_{cm}\}_{c = 1, ..., C; \; m = 1, ..., M}$ with $\sum_m \lambda_{cm} = 1$ for all $c$, we let
\[
p_0(X \mid Y = c) = \sum_{m = 1}^M \lambda_{cm} p_m(X \mid Y = c).
\]
This formulation assumes that the $M$ pretrained models each capture their respective local conditional symptom patterns, $p_m(X \mid Y)$, and the probability of observing any set of symptoms in the target domain is within the convex hull of the conditional probabilities estimated across all training domains. This is usually a reasonable assumption if we have access to a sufficiently diverse collection of pretrained models. In the special case where $\lambda_{cm} = 1$ for some $m$, the model reduces to assuming no shift in the conditional distribution of symptoms between the target domain and the $m$-th training domain. 

When the pretrained base model is a latent class model, e.g., single-domain variant of LCVA with $K$ latent classes, the BFL model induces another latent class model for the conditional distribution of symptoms, i.e.,   
\[
\hat p_0(X_i^{(0)} \mid Y_i^{(0)} = c) = \sum_{m=1}^M\sum_{k=1}^K \lambda_{cm}\nu_{ck}^{(m)} \prod_{j = 1}^p (\theta_{ckj}^{(m)})^{X_{ij}} (1 - \theta_{ckj}^{(m)})^{1 - X_{ij}}, 
\]
where $\btheta^{(m)}$ are the symptom profiles estimated from the $m$-th domain. This latent class representation is similar to what is assumed by LCVA, except that the shared symptom profiles $\btheta$ are replaced by a collection of $M$ domain-specific symptom profiles $\btheta^{(m)}$. Therefore, \rev{under the assumption of LCVA and with sufficient} sample size and diversity of deaths in each domain, the domain-specific $\btheta^{(m)}$ will approach the global symptom profile $\btheta$, and the proposed BFL approach will behave similarly to the pooled analysis using LCVA.

We assume the CSMF of the target domain,  $p_0(Y)$, is independent of all training domain CSMFs. This ensures the predictions are robust to arbitrary label shift across domains. This is especially useful when CSMFs in the training domains are imbalanced or when some causes are not observed in certain domains. 
We note that if known structural relationships exist among domains, the CSMFs from training domains may also be leveraged to improve the inference for the target domain \citep{zhu2025hierarchicallatentclassmodels}. 
We leave this aspect for future work.

Putting everything together, our BFL model can be expressed as a nested latent class model on the target domain, given the pretrained conditional likelihood functions $\hat p_{m}(X \mid Y)$, 
\begin{align*}
Y_i^{(0)}  &\sim \mbox{Cat}(\bpi) \\
H_i\mid Y_i^{(0)} = c &\sim \mbox{Cat}(\blambda_c) \\
p(X_i^{(0)} \mid  Y_i^{(0)} = c, H_i = m) &= \hat p_{m}(X_i^{(0)} \mid Y_i^{(0)} = c),
\end{align*}
where $H_i \in \{1, 2, ..., M\}$ is a latent indicator of which source model contributed to the $i$-th death in the target domain. The cause-specific weights $\blambda_c$  measure the overall contribution from each source domain among deaths due to the $c$-th cause. We refer to this stage as the global model, as it ensembles information from multiple base models. The global model stage may also be viewed as fine-tuning the $M$ pretrained models with local data while ``freezing'' their model parameters describing $p_m(X \mid Y)$. A schematic diagram of information sharing under the BFL framework is shown in Figure \ref{fig:BFL_workflow}.

\rev{\begin{figure}[!tb]
    \centering
    \includegraphics[width = \textwidth]{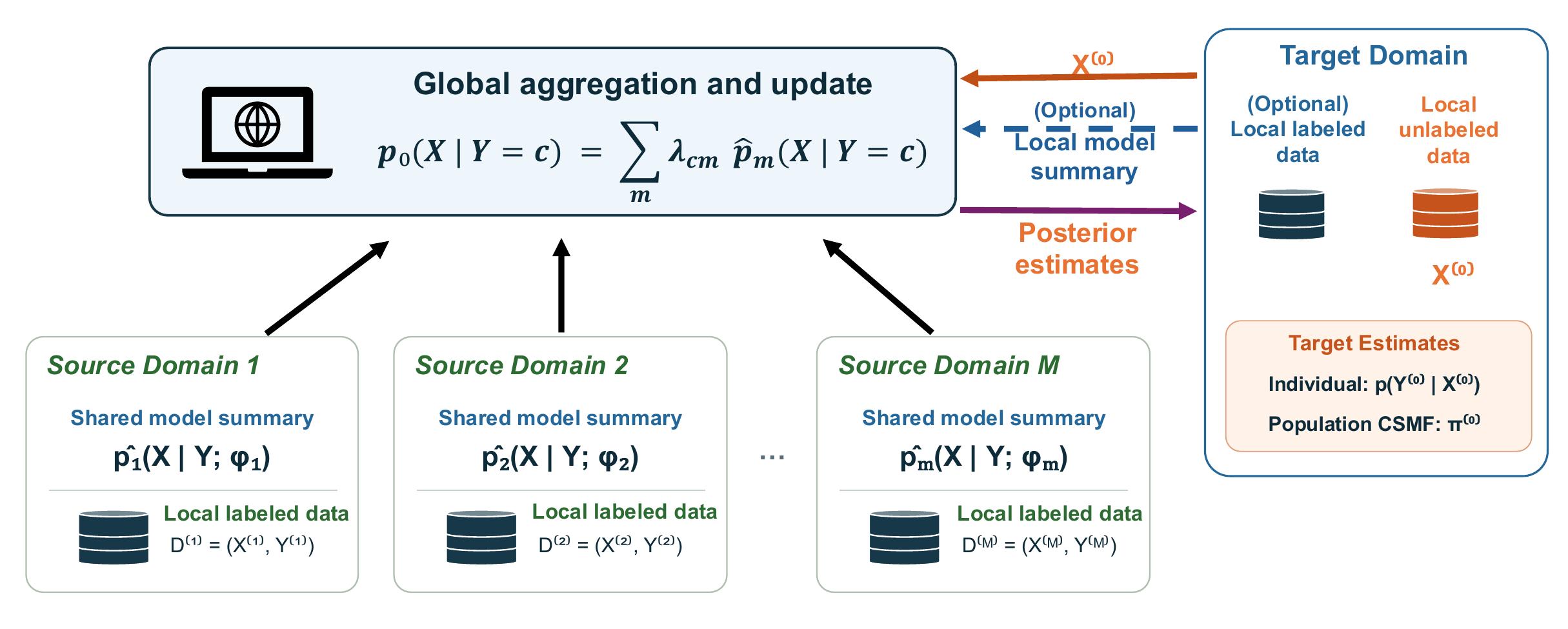}
    \caption{The BFL workflow for VA analysis. The solid black arrows indicate the information shared from training domains to the global model. The dashed arrow indicates the information sharing from the optional local base model trained on the target domain (in the \emph{BFL-domain} and \emph{BFL-mix} models).}
    \label{fig:BFL_workflow}
\end{figure}
}
 
It is worth noting that the only assumption we make about the pretrained models is that we can evaluate the conditional likelihood, $p_m(X \mid Y)$, for any cause that exists in the $m$-th domain.  Unlike most FL approaches, which focus on ensembling the prediction function $p(Y \mid X)$, our model focuses on the generative distribution $p(X \mid Y)$. This formulation allows the decomposition of the two sources of distribution shift across domains: label shift in cause-of-death distributions $p_m(Y)$, and the heterogeneity in conditional symptom distributions $p_m(X \mid Y)$. Learning the latter can be viewed as a federated representation learning task \citep{wang2024desiderata, jiang2022invariant} and directly extends the concept of the symptom-cause-information (SCI) in the VA literature \citep{Clark2018}, from individual symptom to multivariate symptom vectors.

Generative classification models, also called anti-causal models \citep{scholkopf2012causal}, have been empirically shown to be more robust to distribution shift than discriminantive models estimating $p(Y \mid X)$ \citep{ jaini2023intriguing, li2024generative}. Specifically for VA data, the generative factorization corresponds more closely to the natural data generating process where most VA symptoms are consequences of the underlying disease. Therefore, most existing VA algorithms are generative classification models. A variety of different parametric models have been developed to characterize $p(X \mid Y)$, including conditional independence model \citep{mccormick2016probabilistic}, latent Gaussian factor model \citep{tsuyoshi2017}, PARAFAC decomposition \citep{zhu2025hierarchicallatentclassmodels}, sparse PARAFAC decomposition \citep{li2021bayesian}, and collapsed Tucker decomposition  \citep{zhu2025flexiblebayesiantensordecomposition}, etc. Therefore, the class-conditional likelihood $p(X \mid Y )$ can usually be easily computed for most choice of base models. 

\subsection{Prior specification and posterior inference}
We now complete the model with prior specifications for $\bpi$ and $\blambda$. For the target domain CSMF, we let $\bpi \sim \mbox{Dir}(1, ..., 1)$. \rev{Since some causes may not exist in all source domains, we denote the $m$-th model as active for the $c$-th cause if the cause exists in the $m$-th domain. We denote $\boldsymbol\tilde\lambda_c$ as the subvector of $\blambda_c$ on the dimensions where the corresponding model is active. Then we let
\[
  \boldsymbol\tilde\lambda_c \sim \mathrm{Dirichlet}(\mathbf{1}_{M_c}),
  \quad c = 1,\dots,C,
\]
where $M_c \le M$ is the number of models active for cause $c$. Inactive entries of $\blambda_c$ are set to $0$. 
Posterior inference can be conducted efficiently given the pretrained model. It is straightforward to construct a Gibbs sampler with the latent class indicator $H$, which we describe in the supplementary materials.
In the supplementary materials, we also consider a logistic-normal prior on $\blambda_c$ to allow more flexible specification of across-domain dependence. We also consider a sampler for the marginal posterior of $\bpi$ and $\blambda$, using the probabilistic programming language Stan \citep{carpenter2017stan}. The different priors and implementations yield identical results in the experiments in this paper, but the Gibbs sampler is over $10$ times faster than Stan. All these options have been implemented in the {\tt R} package {\tt BFL} \citep{bfl2026}.}

\subsection{Supervised fine-tuning with local labeled data} \label{sec:partial-label}
When some labeled data exist in the target domain, they can be used to improve parameter estimation. Local labeled data may provide information on both $p_0(X \mid Y)$ and $p_0(Y)$. To fix notation, let $Y_i^{(0)}$ be known for $i = 1, ..., n_L$ and unknown for $i = n_L + 1, ..., n_0$. We consider the following variations of the BFL models to utilize the labeled data in different ways. 

One simple option for incorporating local labeled data is to train one more base model with the local data. We then proceed with $M + 1$ models in the BFL procedure from before. This essentially treats the local labeled data as a new domain, thus ignoring any information in the observed cause-of-death distribution among the labeled data. The advantage of this approach is that for the additional model based on local data, there is no distribution shift in terms of $p(X \mid Y)$ compared to the distribution in the target population, so the convex hull assumption is automatically satisfied. When the size of the local labeled data, $n_L$, is small, this additional local model may not be very well estimated. But as $n_L$ increases, the global model is expected to put more weight on this local base model, thus downweighting the influence from out-of-domain observations. We refer to this approach as \emph{BFL-domain} model.

Another natural approach for handling partially known labels is to jointly model the labeled and unlabeled data in the target domain. That is, we let  
$Y_i^{(0)}  \sim \mbox{Cat}(\tilde\bpi)$ for $i = 1, ..., n_L$ and $Y_i^{(0)}  \sim \mbox{Cat}(\bpi)$ for $ i = n_{L} + 1, ..., n_0$.  The posterior joint distribution of parameters is
\begin{align*}
p(\bpi, \blambda \mid X^{(0)}, Y_1^{(0)} = y_1, ..., 
Y_{n_L}^{(0)} = y_{n_L}) 
=&  p(\bpi)p(\tilde\bpi) p(\blambda)
\prod_{i=1}^{n_L}\sum_m \hat\phi_{iy_im} \tilde\pi_{y_i}\lambda_{y_im}
\\
&\times \prod_{i=n_L + 1}^{n_0} \sum_c\sum_m \hat\phi_{icm} \pi_c\lambda_{cm},
\end{align*}
We refer to this approach as \emph{BFL-partial} model to highlight the model is estimated on a target dataset with labels that are partially known. If the labeled data are known to be a random sample of the deaths in the target domain, we let $\tilde\bpi = \bpi$, and otherwise we put independent Dirichlet priors on $\bpi$ and $\tilde\bpi$. For this approach, the labeled data contribute directly to the estimation of weights $\blambda$ and possibly $\bpi$, but it cannot expand the conditional symptom profiles $p_m(X \mid Y)$ estimated by the base models. Therefore, if distribution shift is severe and the convex hull assumption is violated, this approach may not be effective even with a large number of labeled deaths. 

In practice, the two approaches can also be combined. We also consider splitting the labeled data into two subsets; one subset is used to train a local model and the other is used as partial labels in the local joint model. In this work, we consider the simple case where we split the local labeled data into two equal-sized subsets, which we denote as \emph{BFL-mix}.


\subsection{Comparing BFL with VA calibration} \label{sec:calibratedVA}
With supervised fine-tuning, the proposed BFL model has a setup similar to the calibration methods for CSMF quantification, including the BTL algorithm from \citet{datta2020regularized} and GBQL algorithm from \citet{fiksel2020generalized}. We focus on GBQL algorithm in our comparison since it includes BTL as a special case.  Both calibration algorithms were originally designed to ensemble multiple different pretrained black-box models. In our FL setting, one may extend their formulation to ensemble any VA models trained on different datasets. Consider the case with $M$ base models and let $a_{ic}^{(m)} = \hat p_m(Y_i^{(0)} = c \mid X_i^{(0)} = x_i)$ denote the predicted probability of the $c$-th cause for the $i$-th death, estimated by the $m$-th base model. 
The calibration method estimates CSMF in the target domain by modeling the confusion matrices  $\mathcal{M}^{(m)}$ for $m = 1, ..., M$, where $\mathcal{M}_{cc'}^{(m)} = \mathbb{E}(a_{ic'}^{(m)} \mid Y_i^{(0)} = c)$. The approach was also extended to estimate confusion matrices for multiple target populations \citep{pramanik2023modeling}.

The proposed BFL framework differs from the calibration methods in several key aspects.
First, local labeled data are essential for calibration methods; without them, the misclassification matrix cannot be estimated. 
\citet{fiksel2020generalized} showed that under their proposed prior, $\mathcal{M}^{(m)}$ reduces to the identity matrix when there are no labeled data and the calibrated CSMF estimates reduce to the average individual-level predicted probabilities. In such cases, the proposed BFL approach can still learn the weights of baseline models based on individual-level likelihood. Intuitively, $p_0(Y)$ can still be estimated because we observe the empirical $p_0(X)$ from the unlabeled data and multiple candidates of $\hat p_m(X \mid Y)$ from training domains. The trade-off is that our BFL framework makes more parametric assumptions on $p_0(X \mid Y)$, whereas the calibration approaches only parameterize the confusion matrix, which is of lower dimensions. Modeling assumptions on $p_0(X \mid Y)$, however, are inevitable if individual-level classification is of interest. Unlike BFL, the calibration methods cannot produce individual-level cause-of-death assignments.

Second, BFL makes different assumptions on the local labeled data compared to the calibration methods. The key assumption behind the calibration approach is that $p(X \mid Y)$ is the same across the labeled and unlabeled subsets of the target domain, i.e.,  $p(X_i^{(0)} \mid Y_i^{(0)}, L_i = 0) = p(X_i^{(0)} \mid Y_i^{(0)}, L_i = 1)$, where $L_i$ is a binary indicator of whether the $i$-th death in the target domain is labeled. This can be violated if $L_i$ depends directly on $X_i$. For example, this occurs if the local labeled data are sampled conditional on the presence of certain symptoms. In the BFL framework, when treating local labeled data as partial labels, valid inference requires selection into the labeled data to be conditionally ignorable given observed quantities. Thus, BFL can still be used when labeled data selection depends directly on symptoms, as long as it does not depend on the unknown cause of death $Y$.

Finally, as do most ensemble learning approaches, the calibration methods in \citet{datta2020regularized} and \citet{fiksel2020generalized} extract $p(Y \mid X)$ from each model as input. This predicted quantity can be heavily biased if the CSMFs of the source domain differ significantly from the target domain, unless label shift is accounted for in the base model. When the confusion matrix is far from a diagonal matrix, calibration is difficult with a limited number of observations. On the other hand, BFL extracts the conditional likelihood, i.e., $p(X \mid Y)$, from each base model, explicitly removing the information about source domain CSMFs.

Beyond these conceptual differences, our empirical analysis also revealed an additional issue with the shrinkage priors used in previously published calibration methods. \citet{fiksel2020generalized} and \citet{datta2020regularized} assume that for $c = 1, ..., C$,
\begin{align*}
(\mathcal{M}^{(m)}_{c1}, ..., \mathcal{M}^{(m)}_{cC}) &\sim \mbox{Dir}(\gamma_c^{(m)}\textbf{}(\bm I_{c*} + \epsilon \bm{1})), 
\end{align*}
where $\bm I_{c*}$ is the $c$-th row of the identity matrix, and $\gamma_c^{(m)}$ is $\gamma_c^{(m)} \sim \mbox{Gamma}(\alpha, \beta)$,
with default choice $\alpha = 5$ and $\beta = 0.5$. With a prior mean of $10$ for $\gamma_c^{(m)}$, the confusion matrices are heavily regularized to be close to the identity matrix. The rationale for the shrinkage prior is to reduce the variance from the confusion matrix estimation. However, when the number of causes, $C$, is large and the base models are inaccurate, this assumption can be overly conservative and may fail to remove the bias of base classifiers. In our analysis of the PHMRC data in Section \ref{sec:phmrc}, we found that the default prior (\emph{GBQL-0.5}) performs poorly due to shrinkage, whereas less shrinkage generally improve the estimation of CSMFs in such cases. We showed that when $\beta = 50$ (\emph{GBQL-50}) the calibration method in \citet{fiksel2020generalized} performs significantly better. Results for other choices of $\beta$ between $0.5$ and $50$ are omitted since they mostly lead to performance metrics between these two cases. However, the default shrinkage prior does lead to better performance in the analysis in Section \ref{sec:champs} where the number of causes is small, which is consistent with the findings in \citet{fiksel2020generalized}. It is unclear how to properly choose the priors for the calibration methods to achieve optimal bias-variance tradeoff. Since optimizing the calibration methods is not the focus of this paper, we leave this investigation for future work.

\section{Simulation analysis using PHMRC gold-standard dataset} \label{sec:phmrc}
In this section, we comprehensively evaluate the proposed BFL model using the PHMRC gold-standard adult VA dataset \citep{murray2011population}. The PHMRC dataset is widely used to validate and compare VA algorithms. It includes $7,841$ adult deaths from six study sites: Andhra Pradesh, India ($n = 1,554$), Uttar Pradesh, India ($n = 1,419$), Bohol, Philippines ($n = 1,259$), Mexico City, Mexico ($n = 1,586$), Dar es Salaam, Tanzania ($n = 1,726$), and Pemba Island, Tanzania ($n = 297$). The deaths are coded into $34$ non-overlapping causes of death. This dataset has been processed into $168$ binary symptoms following \citet{li2023openva}. 

To investigate realistic scenarios involving distribution shifts across domains, we preserve the original six-site structure and conduct leave-one-domain-out experiments. i.e., each of the six domains is treated as the target domain in turn, with the remaining five domains serving as the training domains. The conditional distribution of symptoms given causes is highly heterogeneous across the six sites \citep{li2021bayesian, wu2021tree}, making the leave-one-domain-out prediction a difficult task.

\subsection{Distribution shift scenarios} Within each leave-one-domain-out prediction task, we consider case with and without local labeled data. When local labeled data exist, we vary the degrees of within-domain label shift to investigate the influence of non-representative local labeled data. Specifically, we consider the following three simulation scenarios:  
\begin{enumerate}
    \item \emph{No within-target label shift}: We randomly sample $20\%$ of the target dataset to be labeled. 
    \item  \emph{Mild within-target label shift}: We first generate $\tilde\bpi$ and $\bpi$ independently from $\mbox{Dir}(1, ..., 1)$ for the labeled and unlabeled data respectively. We then sample deaths with replacement so that $20\%$ of the resampled data are labeled and and cause-of-death distributions in the labeled and unlabeled portions match the generated prevalences. 
    \item  \emph{Severe within-target label shift}: For each cause, we generate $q_c \sim \mbox{Beta}(0.2, 0.2)$ and randomly sample $q_c n_c$ deaths to be labeled and keep the remaining unlabeled. The sample size of labeled data vary across datasets. The labeled and unlabeled prevalence are negatively correlated by design. 
\end{enumerate}

In the first scenario, we estimate the CSMF of the entire target dataset. Deaths that are used to train the local model in \emph{BFL-domain} and \emph{BFL-mix} are held out in the global model, and thus the prevalence estimates in these two models are based on fewer deaths. For a fair comparison, we included a finite-sample adjustment to account for the known causes of death used for local training. The adjusted estimate is
$\frac{n_0 - n_{h}}{n_0}\hat\pi_c+ \frac{n_{hc}}{n_0}$, 
where $n_{h}$ is the size of the held-out data used for training the local model, and $n_{hc}$ is the number of deaths from the $c$-th cause in the held-out data.  In the second and third scenarios, target of inference is the CSMF of the unlabeled subset of the target dataset, i.e., $\bpi$, so the estimates do not need to be adjusted.

In all scenarios, the individual cause-of-death assignment is evaluated on all the unlabeled deaths. We generate $30$ resampled datasets for each target site. In comparison, we also evaluate the BFL methods when no local labeled data exist. 

\subsection{Methods under comparison} 
We use single-domain LCVA with constant mixing weights as the base classifier for all BFL model variants. To comprehensively evaluate BFL performance, we compare against five baseline methods that represent different data access scenarios. We use the term `local' to denote methods that train models on individual domains without cross-domain information sharing.

\begin{enumerate}

    \item \emph{local-self}: Single-domain LCVA trained using the labeled data in the target domain only. For this approach, no information from the training domains is used. This method represents the scenario where only local data are accessible.

    \item \emph{local-avg}: Single-domain LCVA trained on one training dataset, then applied to the target domain where some labels are known. Single-domain LCVA produces estimates for the CSMF of the entire target domain. When the target of inference is the CSMF of the unlabeled subset of target domain, they are estimated by the posterior predictive distribution of $\hat \pi_c = \frac{1}{n_0 - n_L} \sum_{i=n_L+1}^{n_0} \hat p(Y_i^{(0)} = c)$. The average accuracy measures of the five training models are reported. This represents the scenario where training data from only one non-target domain are accessible. 

    \item LCVA: Multi-domain LCVA model with domain-level mixing weights. 
    This is the model where all available data in all training domains and target domain are pooled and jointly modeled. This represents the scenario where data from all domains are accessible.

    \item \emph{GBQL-0.5}: The GBQL method \citep{fiksel2020generalized} using the same five base models as inputs, similar to BFL. We use the default shrinkage prior of $\mbox{Gamma}(5, 0.5)$ on $\gamma_c^{(m)}$.  

    \item \emph{GBQL-50}: The same GBQL implementation as above, but with prior of $\mbox{Gamma}(5, 50)$ on $\gamma_c^{(m)}$, which introduces little shrinkage a priori. Both GBQL models operate under the same level of data access as the BFL models.
    
\end{enumerate}

For multi-domain LCVA with five training sites, we set the number of latent classes to $K = 10$ following the suggestion in \citet{li2021bayesian}. For single-domain LCVA models, we reduce it to $K = 5$ due to the smaller sample sizes. For all LCVA models, we run MCMC for $4,000$ iterations. For the GBQL models, we run $3$ parallel MCMC chains with $4,000$ iterations each, as suggested by the original paper. For the prediction stage in the BFL models, we run $4$ parallel MCMC chains with $4,000$ iterations each. Convergence of MCMC chains is satisfactory under these settings, and the results are insensitive to the choice of MCMC parameters in all fitted models.

When evaluating the population-level quantification, we use the CSMF accuracy,  a widely used metric in VA analysis that measures how well the estimated CSMFs match the true population fractions. It is defined as
\begin{align*}
\text{CSMF}_{\text{acc}}(\hat{\bpi}) &= 1 - \frac{\sum_{c=1}^C |\hat{\pi}_c - \pi_c^{\mbox{true}}|}{2(1 - \min_c \pi_c^{\mbox{true}})},
\end{align*}
where $\pi_c^{\mbox{true}}$ denotes the true CSMF from the target dataset. This metric quantifies the difference between the estimated and true CSMF, scaled between 0 and 1, with higher values indicating more accurate prevalence estimation. $1 - \text{CSMF}_{\text{acc}}$ is also known as the normalized absolute error in the quantification learning literature \citep{gonzalez2017review}.

For individual-level classification, the main metric we consider is the top cause accuracy, i.e., the proportion of the predicted cause of death being correct,
\begin{align*}
    \text{Top Cause Accuracy} &= \frac{1}{n} \sum_{i=1}^{n} \bm 1_{\hat Y_i = Y_i} 
\end{align*}
In addition, we evaluate the classification results using the balanced accuracy, which is computed as the average per-class recall  across all causes. \rev{The results are presented in the supplementary materials.}
\subsection{Results without within-target label shift}\label{sec:no-local-shift}

\begin{figure}[!htb]
    \centering
    \includegraphics[width = \textwidth]{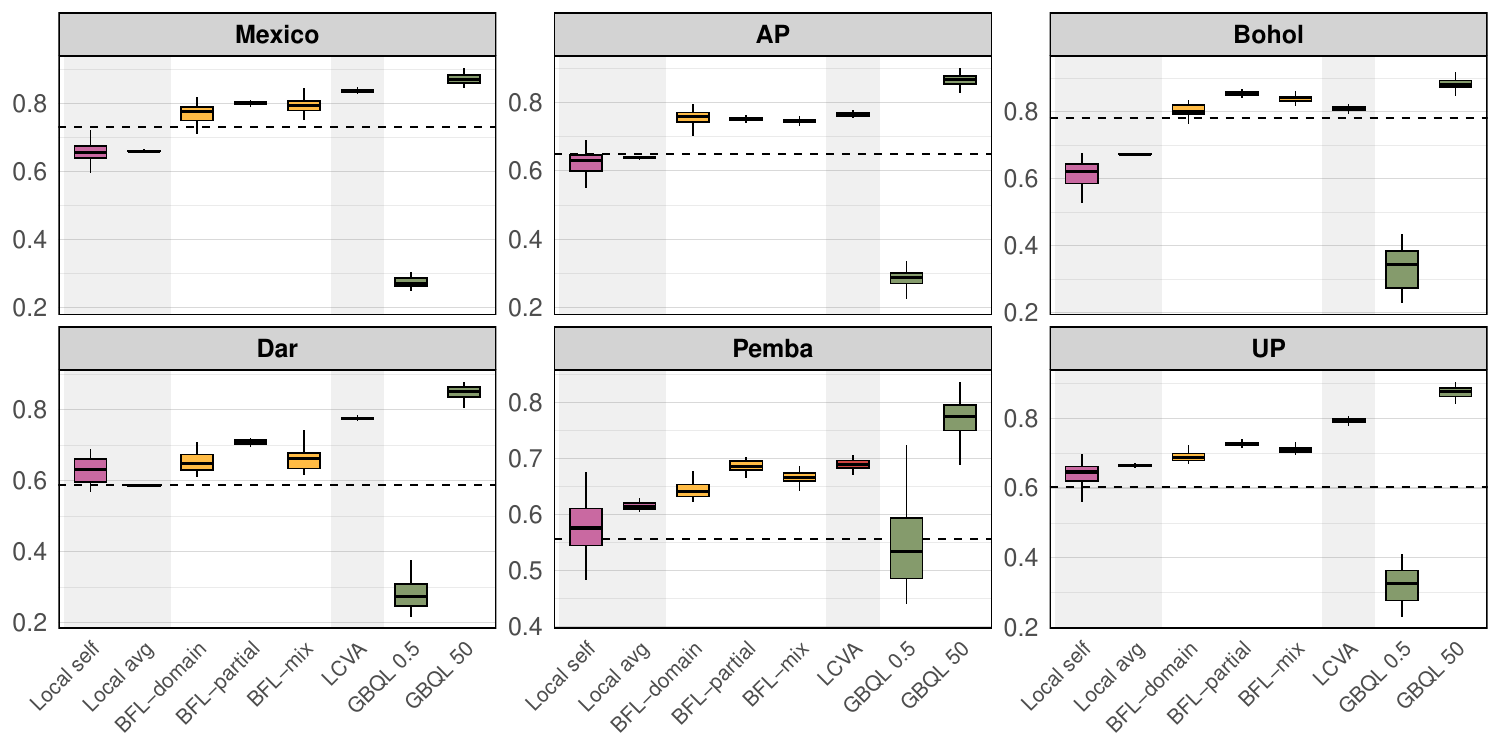}
    \caption{\textbf{No within-target label shift}: comparison of CSMF accuracy. The dashed line shows the CSMF accuracy from BFL without local labeled data. The modified \emph{GBQL-50} and LCVA are the two highest-perofrmaning methods in all sites.}
    \label{fig:CSMF_balanced_0.8}
\end{figure}

Figure \ref{fig:CSMF_balanced_0.8} shows CSMF accuracy under the first scenario without local label shift. The dashed reference lines indicate BFL performance without local labeled data. First, as detailed in Section \ref{sec:calibratedVA}, GBQL results are highly sensitive to prior choice: the default strong shrinkage prior (\emph{GBQL-0.5}) performs worst across all six experiments, while weak shrinkage (\emph{GBQL-50}) achieves the best performance. This suggests that calibration with minimal shrinkage is optimal for CSMF quantification when local deaths are randomly sampled for labeling. However, this superior performance does not generalize to later settings with non-representative local labeled data in Section \ref{sec:local-shift} or when the number of causes is small in Section \ref{sec:champs}. Among the rest of the models, single-domain models (\emph{local-avg} and \emph{local-self}) perform worst overall. Full data pooling (LCVA) yields the highest CSMF accuracy across most sites, with BFL variants performing between local methods and LCVA. This is consistent with their intermediate level of information sharing. All BFL models consistently outperform their constituent base models, highlighting the substantial benefits of ensemble prediction over individual models.  

\rev{
\begin{figure}[!htb]
    \centering
    \includegraphics[width = \textwidth]{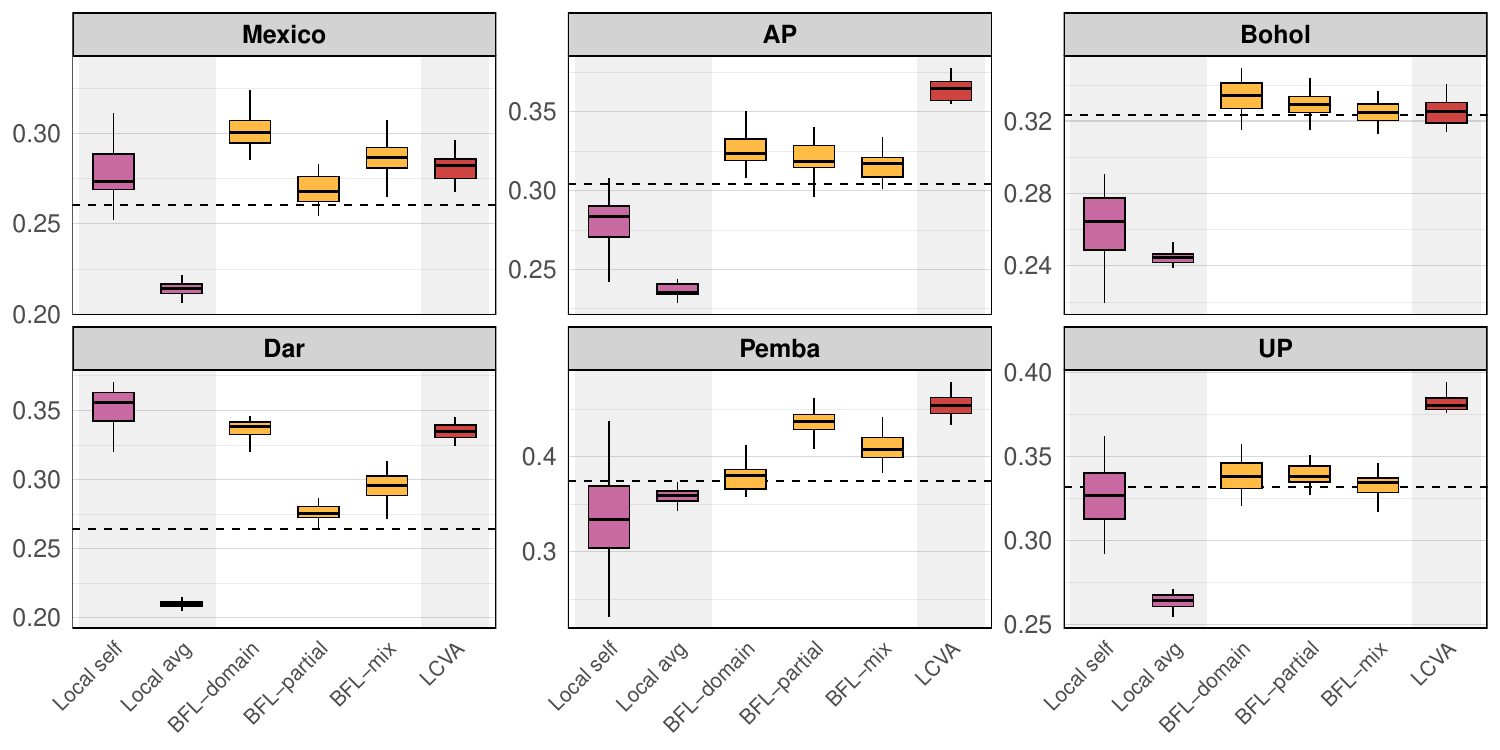}
    \caption{\textbf{No within-target label shift}: comparison of top cause accuracy. The dashed line shows the top cause accuracy from BFL without local labeled data. LCVA with full data pooling achieves the highest top cause accuracy in three sites. In Mexico City and Bohol, BFL-domain achieves the highest top cause accuracy.}
    \label{fig:ACC_balanced_0.8}
\end{figure}
}

Figure \ref{fig:ACC_balanced_0.8} shows the same comparison for top-cause accuracy, and balanced accuracy results are included in the supplementary materials. The relative performance is largely similar across models, except for local-only models (\emph{local-self}), where the top cause accuracy can be misleadingly high. This discrepancy likely reflects a tendency to favor prevalent causes as the local-only model is biased toward predicting the more prevalent causes.
Among the target sites, Mexico City and Dar es Salaam exhibit stronger distribution shift in symptom patterns $p(X \mid Y)$, making non-local models less effective. Including a locally-trained model (\emph{BFL-domain}) appears to better satisfy the convex hull assumption (compared to \emph{BFL-partial}) by expanding the model space to include local patterns \rev{in all sites except Pemba}.
The \emph{BFL-mix} model, which combines both approaches, mostly performs between the other two BFL variants, as expected.

\subsection{Results with within-target label shift}\label{sec:local-shift}

\rev{
\begin{figure}[!htb]
    \centering
    \includegraphics[width = \textwidth]{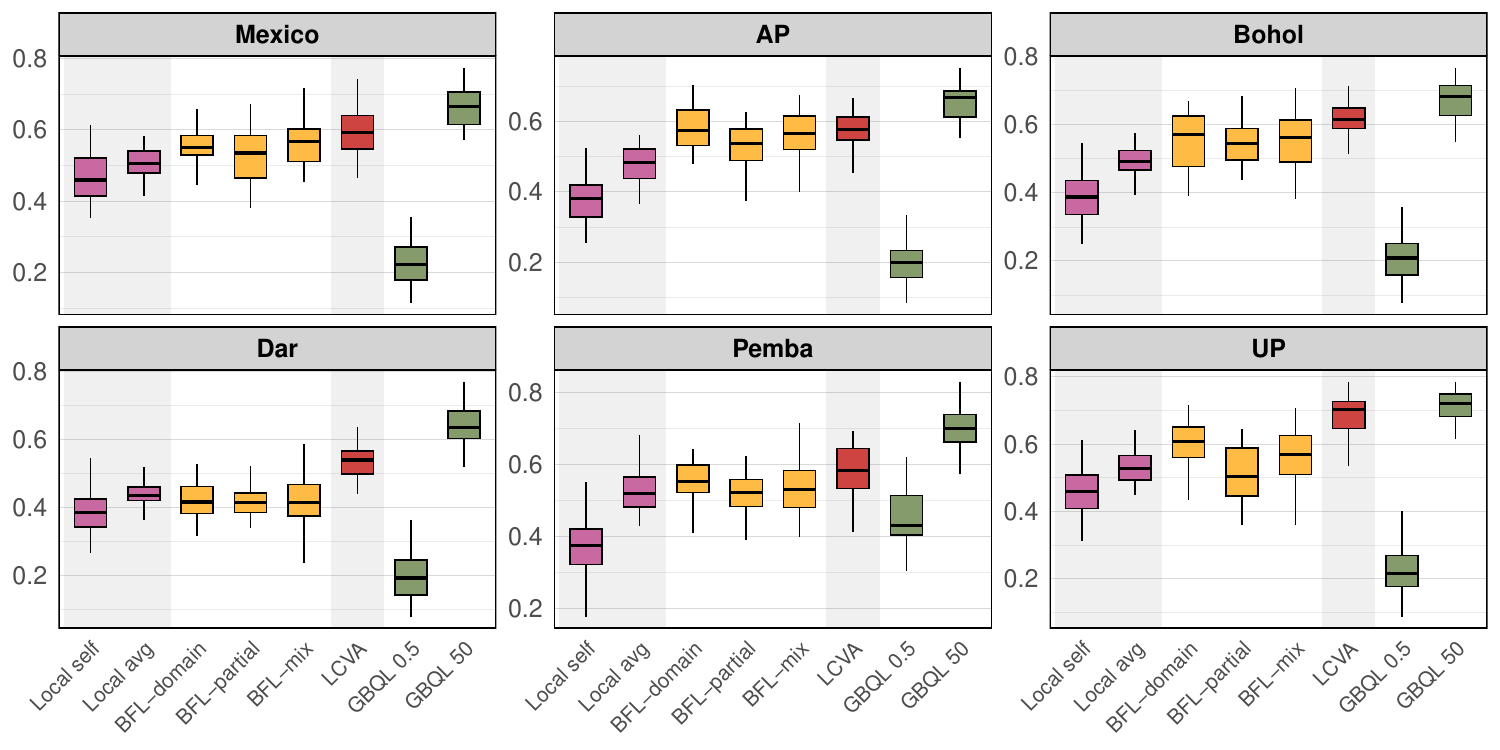}
    \caption{\textbf{Mild within-target label shift}: comparison of CSMF accuracy. The modified \emph{GBQL-50}, LCVA, and \emph{BFL-domain} achieve the highest CSMF accuracy on different sites.}
    \label{fig:CSMF_unbalanced_0.8}
\end{figure}
}
\rev{
\begin{figure}[!htb]
    \centering
    \includegraphics[width = \textwidth]{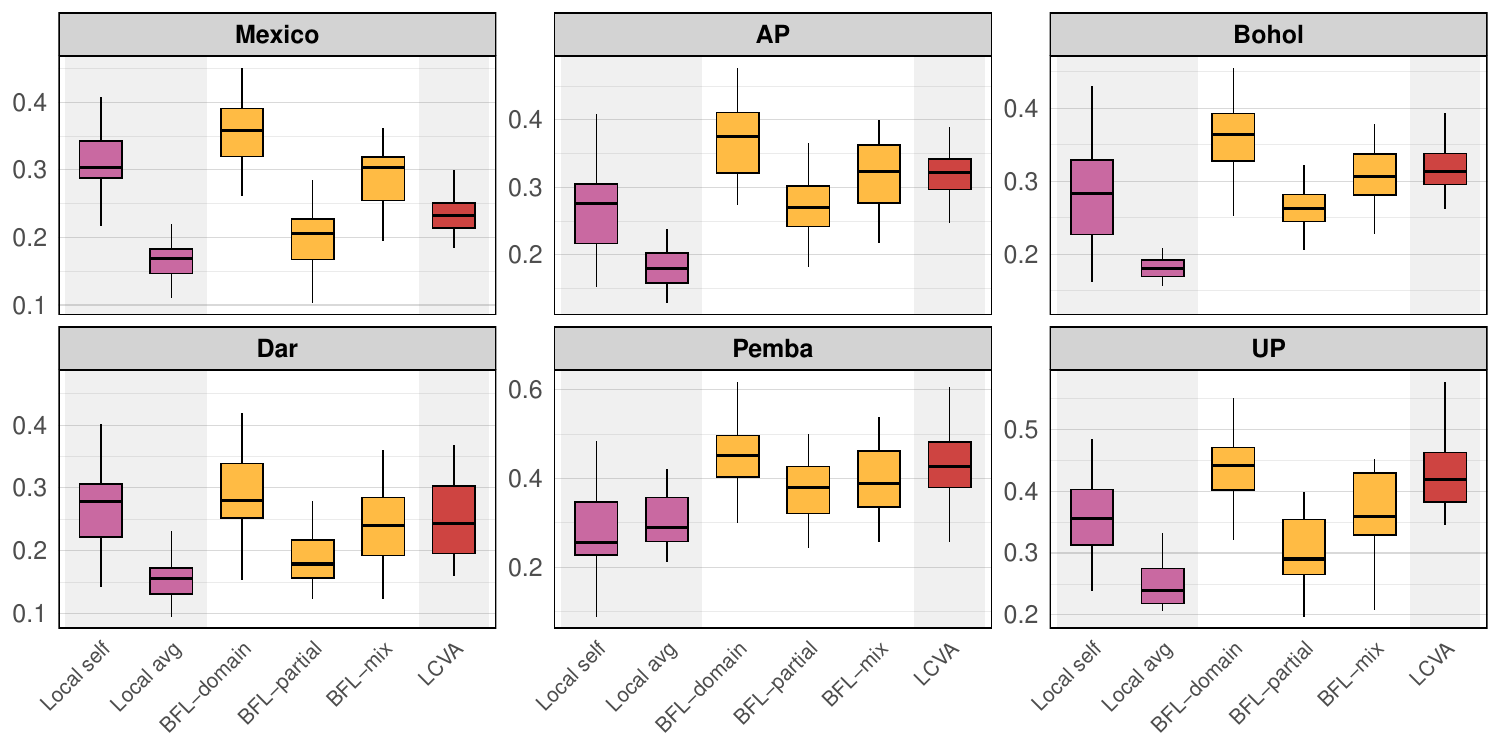}
    \caption{\textbf{Mild within-target label shift}: comparison of top cause accuracy. \emph{BFL-domain} achieves highest average accuracy in all sites.}
    \label{fig:ACC_unbalanced_0.8}
\end{figure}
}
Figures \ref{fig:CSMF_unbalanced_0.8} and \ref{fig:ACC_unbalanced_0.8} show results under mild within-target label shift, where we estimate CSMF for only the unlabeled subset since labeled data are non-representative of the full population. Overall performance deteriorates compared to the random labeling scenario, with increased variability due to greater uncertainty in target prevalences. The relative ranking among methods remains largely similar, but performance gaps narrow substantially. BFL variants become more competitive. Notably, \rev{\emph{BFL-domain} achieves the highest top cause accuracy in all domains}. This demonstrates that that the local fine-tuning in BFL can compensate for the loss of information in federated learning settings.

\rev{
\begin{figure}[!htb]
    \centering
    \includegraphics[width = \textwidth]{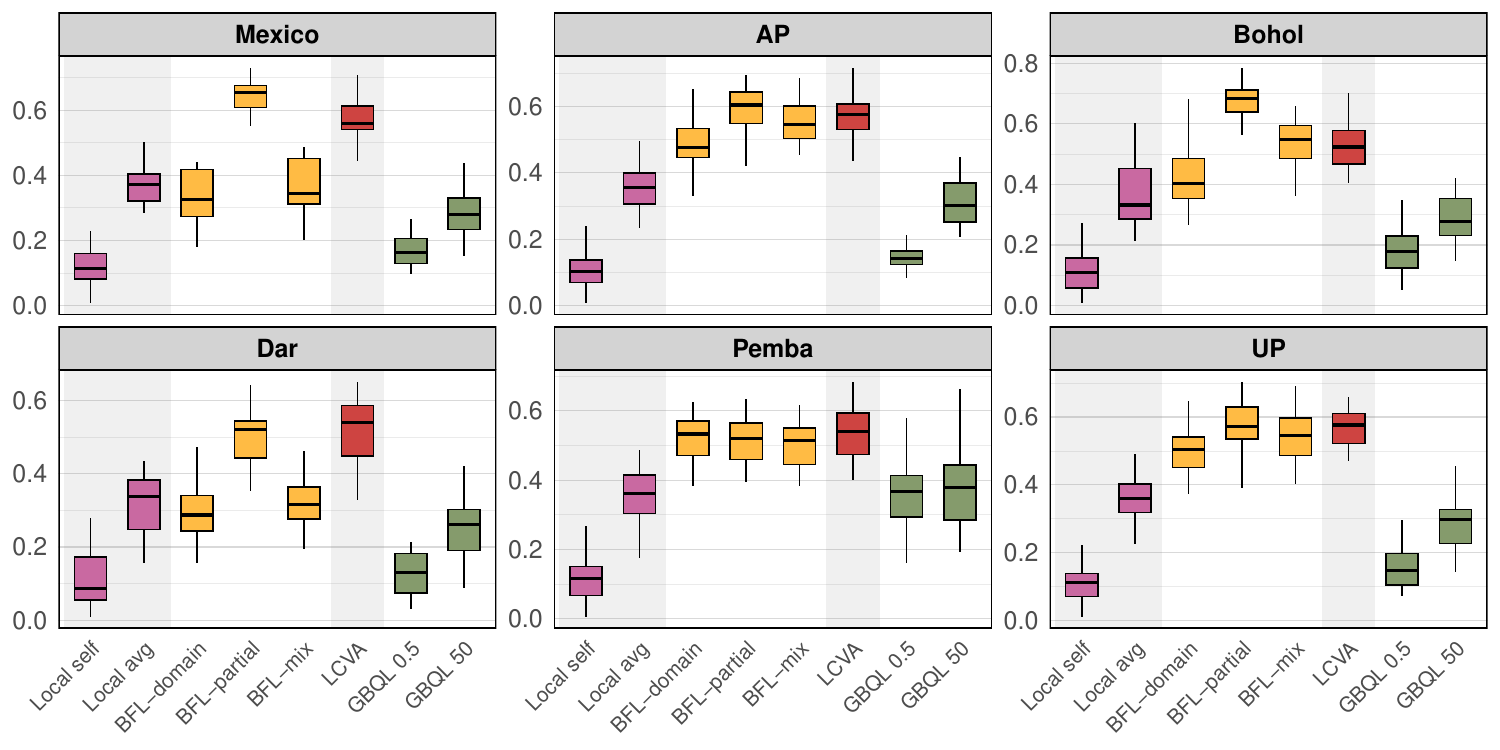}
    \caption{\textbf{Severe within-target label shift}: comparison of CSMF accuracy. \emph{BFL-partial} and LCVA achieve the highest CSMF accuracy on different sites.}
    \label{fig:CSMF_extreme_unbalanced}
\end{figure}
}
\rev{
\begin{figure}[!htb]
    \centering
    \includegraphics[width = \textwidth]{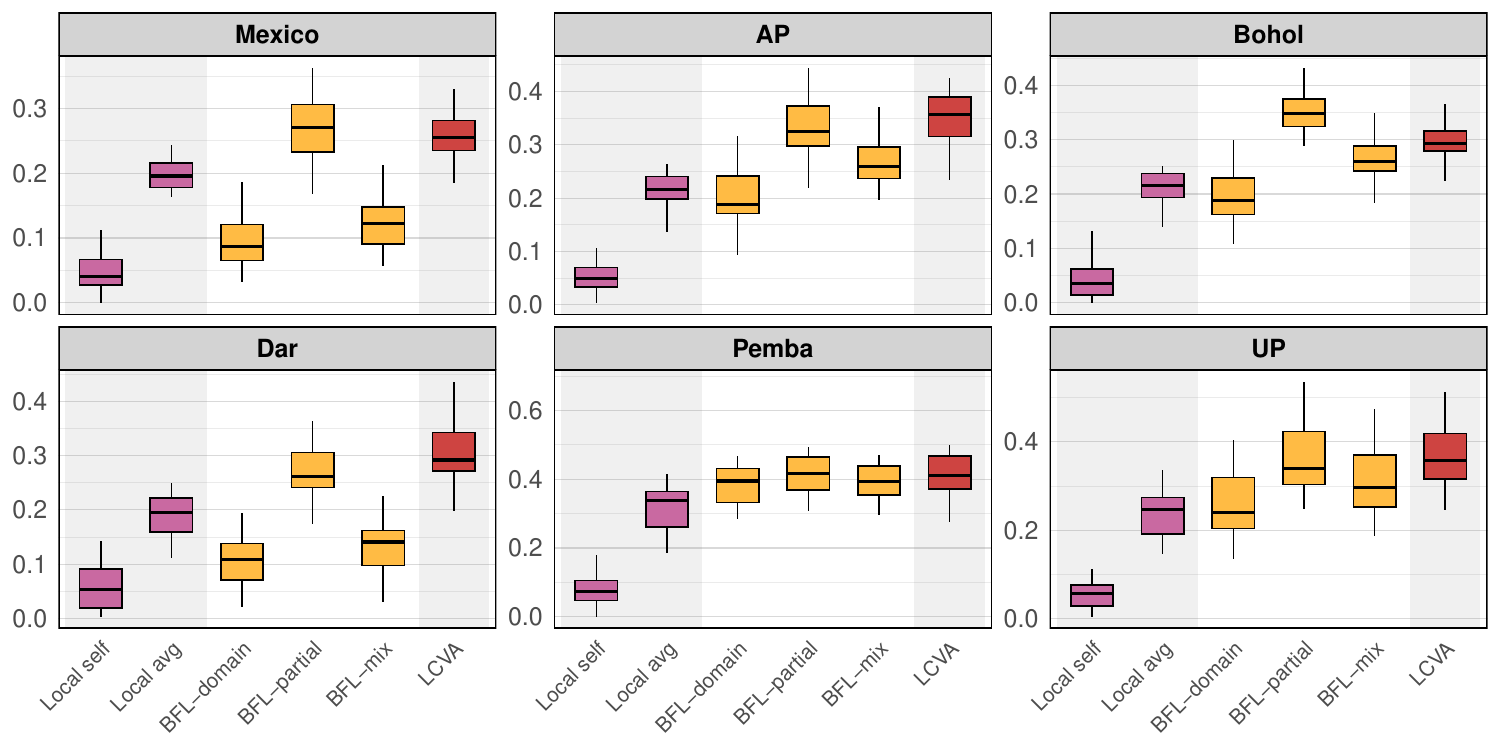}
    \caption{\textbf{Severe within-target label shift}: comparison of top cause accuracy. \emph{BFL-partial} achieves the highest accuracy in Mexico City and Bohol, and show similar accuracy to LCVA in the other four sites.}
    \label{fig:ACC_extreme_unbalanced}
\end{figure}
}

Figures \ref{fig:CSMF_extreme_unbalanced} and \ref{fig:ACC_extreme_unbalanced} show results under severe within-target label shift, where performance degrades further. For CSMF accuracy, BFL models remain competitive with LCVA, with \emph{BFL-partial} slightly outperforming LCVA in \rev{four of the six domains, while showing comparable CSMF accuracy in Dar es Salaam and Pemba.} 
The previous advantage of \emph{GBQL-50} disappears under this extreme scenario. While GBQL theoretically does not require the labeled and unlabeled subsets to have the same CSMF, the severe label distribution makes the estimation of the misclassification matrix challenging in finite samples. Dealing with this type of difficult scenario for calibration-based approaches is an importnat direction for further future research. \rev{For top cause accuracy, \emph{BFL-partial} and LCVA achieve the highest accuracy consistently and their differences are mostly small, except in Bohol, where \emph{BFL-partial} shows a clear advantage.} \rev{Unlike the case with mild shift, the performance of \emph{BFL-partial} is consistently better than \emph{BFL-domain} in this simulation. This} reflects the extreme sampling design: causes common in the local labeled subset are rare in unlabeled data. This limits the contribution of the local model in \emph{BFL-domain} since it is trained primarily on causes rare in the target dataset. 


To summarize, across all three scenarios, LCVA with full data pooling generally performs the best or \rev{close to the best}, though its advantage diminishes as label shift intensifies. BFL models provide robust federated alternatives that generally outperform single-domain approaches, demonstrating the value of ensembling multiple non-local models. Modified GBQL with weak shrinkage performs well under no label shift. \rev{However, its performance decreases as the labeled data becomes more unbalanced, especially under severe label shift}. The lack of a clear winner for all cases is not entirely surprising, as all methods involve trade-offs between different distributional assumptions. The preferred method to use should depend on the specific degree of distribution shift in both $p(Y)$ and $p(X \mid Y)$ for the given problem.

Among the BFL variants, performance depends on how the local labeled data are used and on the severity of within-target label shift. \rev{\emph{BFL-domain} uses the labeled data to estimate an additional target-domain base model, $\hat p_{\mathrm{local}}(X\mid Y)$, thereby expanding the collection of candidate conditional symptom distributions. In contrast, \emph{BFL-partial} keeps the base models fixed and incorporates the labeled observations directly into the global likelihood to improve estimation of $\blambda$ and, when the labeled observations are representative of the target population, $\bpi$. When the labeled observations are sampled randomly, both approaches can be effective. Their relative performance depends on whether the labeled data are more informative for estimating a target-specific base model or for directly estimating the global mixing weights and CSMFs. Under mild label shift with sufficient overlap between the causes represented in the labeled and unlabeled subsets remains, and our experiments favor \emph{BFL-domain}. The locally trained model captures target-specific symptom patterns and provides a useful additional component. Under severe label shift, however, causes prevalent in the unlabeled population may be absent or poorly represented in the labeled subset. The additional local base model then contributes little to prediction for the causes that dominate the unlabeled population. In this setting, \emph{BFL-partial} is more effective because it uses the labeled observations directly to update the relevant mixing weights without relying on a separately estimated local base model, while retaining the source-domain models to represent causes that are underrepresented locally.}
In all scenarios, \emph{BFL-mix} offers a pragmatic compromise when the degree of label shift is uncertain or when both classification and quantification are important. Overall, variant selection should be guided by the analysis objective and the anticipated relationship between the labeled and unlabeled cause distributions in the target domain.


\section{Analysis of CHAMPS neonatal Dataset} \label{sec:champs}
We now apply our methods on the Child Health and Mortality Prevention Surveillance (CHAMPS) neonatal dataset. 
CHAMPS is a global health initiative dedicated to understanding and preventing child mortality, particularly in regions with high under-five death rates. Launched in 2015, CHAMPS operates across $19$ catchment areas in eight countries within Africa and South Asia. 
The CHAMPS neonatal dataset includes $1,573$ VAs. We processed the data into $353$ binary symptom indicators following the WHO 2016 standard format \citep{Nichols2018who}. The dataset consists of deaths from surveillance sites in seven countries: Bangladesh (BD), Kenya (KE), South Africa (ZA), Mozambique (MZ), Mali (ML), Ethiopia (ET), and Sierra Leone (SL). Each VA is paired with cause-of-death diagnoses assigned by a multidisciplinary panel
of experts based on VA, clinical records, and lab results.  The cause-of-death distribution in this dataset is highly imbalanced, with the majority of deaths attributed to intrapartum-related events (IPRE). Due to the limited sample size, we grouped the assigned underlying causes into $8$ broad categories shown in Figure \ref{fig:CSMF_no_partial_champs_neonatal}. For our experiment, we consider the \rev{three} countries with larger sample sizes, Kenya (KE), Mozambique (MZ), and South Africa (ZA) as source domains, and aggregate the remaining four countries into a single target domain. A summary of sample sizes by cause is included in the supplementary materials. 

\rev{
\begin{figure}[!htb]
    \centering
    \includegraphics[width = \textwidth]{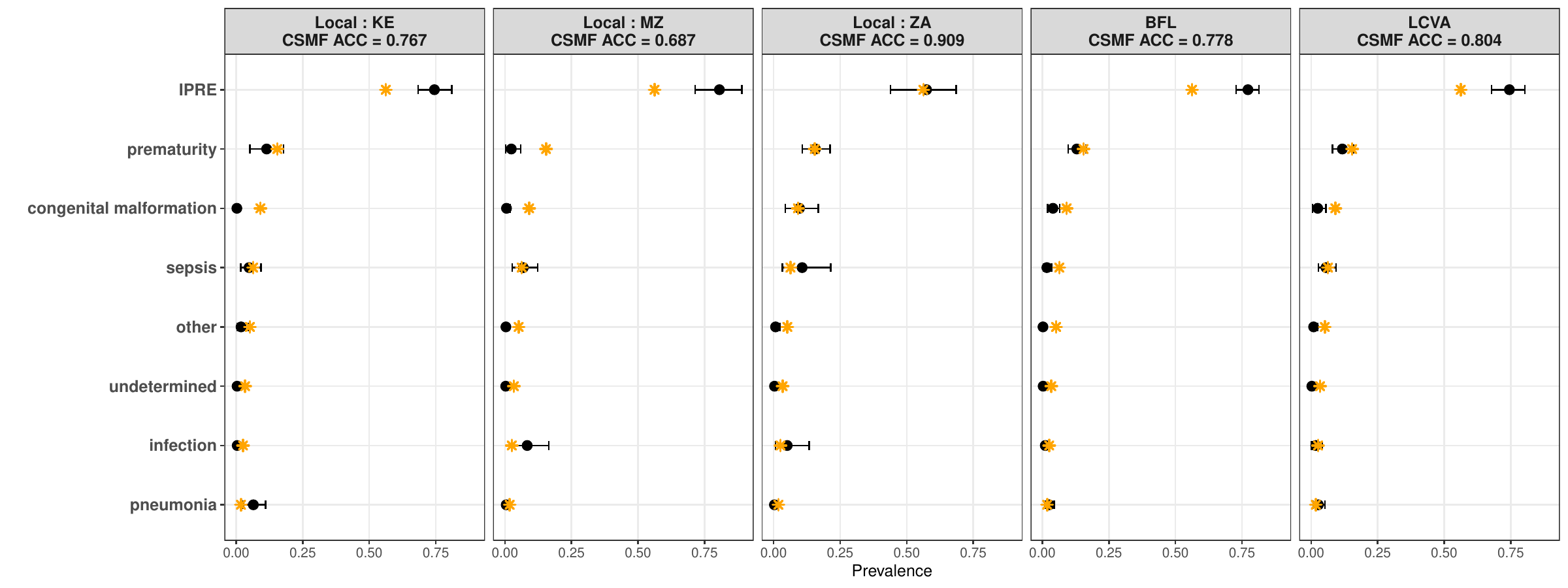}
    \caption{CHAMPS neonatal prevalence estimation with CSMF accuracy comparison. The yellow asterisks are the actual CSMF, and the black dots and lines are the posterior estimated mean and 95\% credible intervals of CSMF.}
    \label{fig:CSMF_no_partial_champs_neonatal}
\end{figure}
}

We first examine the case where the target domain is fully unlabeled. Figure \ref{fig:CSMF_no_partial_champs_neonatal} shows estimated CSMFs and 95\% credible intervals for LCVA trained on each of the three training domains, the BFL model with these three base models, and the full multi-domain LCVA. Among individual models, single-domain LCVA trained on South Africa data achieves the highest CSMF accuracy, while BFL outperforms the other two base models and performs nearly as well as multi-domain LCVA. BFL also yields narrower credible intervals than individual models, reflecting reduced uncertainty from leveraging multiple source domains.

Following a similar setup as in Section \ref{sec:phmrc}, we examine scenarios with partial labeling in the target domain by randomly sampling $20\%$ and $40\%$ of target data for labeling. We focus on random labeling due to the small sample size and severe class imbalance, which make it challenging to design realistic label shift simulations.
\rev{Figure \ref{fig:champs_neonatal} summarizes the CSMF accuracy and top cause accuracy across $500$ resampled datasets. Among the BFL variants, the highest CSMF accuracy is achieved by \emph{BFL-partial} and the highest top cause accuracy is achieved by \emph{BFL-domain}, both achieving similar or better performance compared to LCVA. It is worth noting that with a small number of causes and an imbalanced cause distribution, the \emph{local-self} model achieves high top cause accuracy when labeled data are abundant. However, it remains to be ineffective in estimating the CSMF, likely due to overfitting to the more prevalence causes.} 

The calibration methods were less accurate in estimating CSMFs compared to joint modeling approaches in this analysis. The default \emph{GBQL-0.5} approach yields better performance than the modified version with weak shrinkage. This contrasts with findings on the PHMRC dataset, but aligns with the original GBQL design and reported findings. With only $8$ causes, the assumption that confusion matrices resemble identity matrices becomes more reasonable and the stronger shrinkage prior indeed helps. 

\rev{
\begin{figure}[!htb]
    \centering
    \includegraphics[width = 0.85\textwidth]{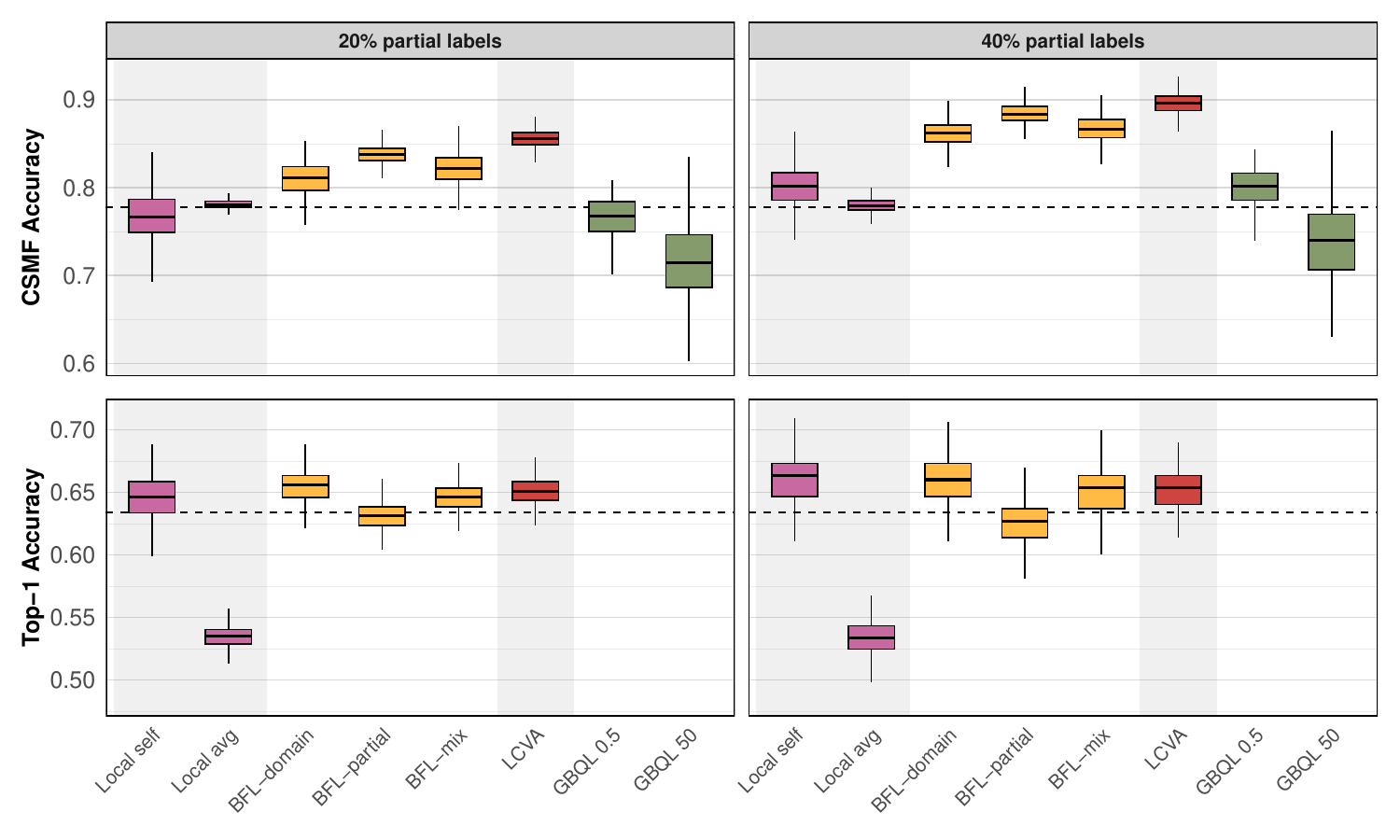}
    \caption{CHAMPS neonatal data: comparison of CSMF accuracy (top row) and top cause accuracy (bottom row). The dashed lines correspond to BFL without local labeled data.}
    \label{fig:champs_neonatal}
\end{figure}
}


Finally, Figure \ref{fig:heatmap_lambda_neonatal_balanced_full_causes_miss_prop0.8_site_Target} shows the posterior mean of the weights $\blambda$ for each domain under different BFL variants. The models tend to assign higher cause-specific weights to domains where that cause is more prevalent, despite having no direct access to source domain CSMFs. This suggests BFL effectively identifies which domains provide the most informative symptom patterns for each cause. Given the relatively small sample size in this dataset, the ability to identify and leverage symptom profiles estimated from larger samples is an appealing feature.

\rev{
\begin{figure}[!htb]
    \centering
    \includegraphics[width = 0.95\textwidth]{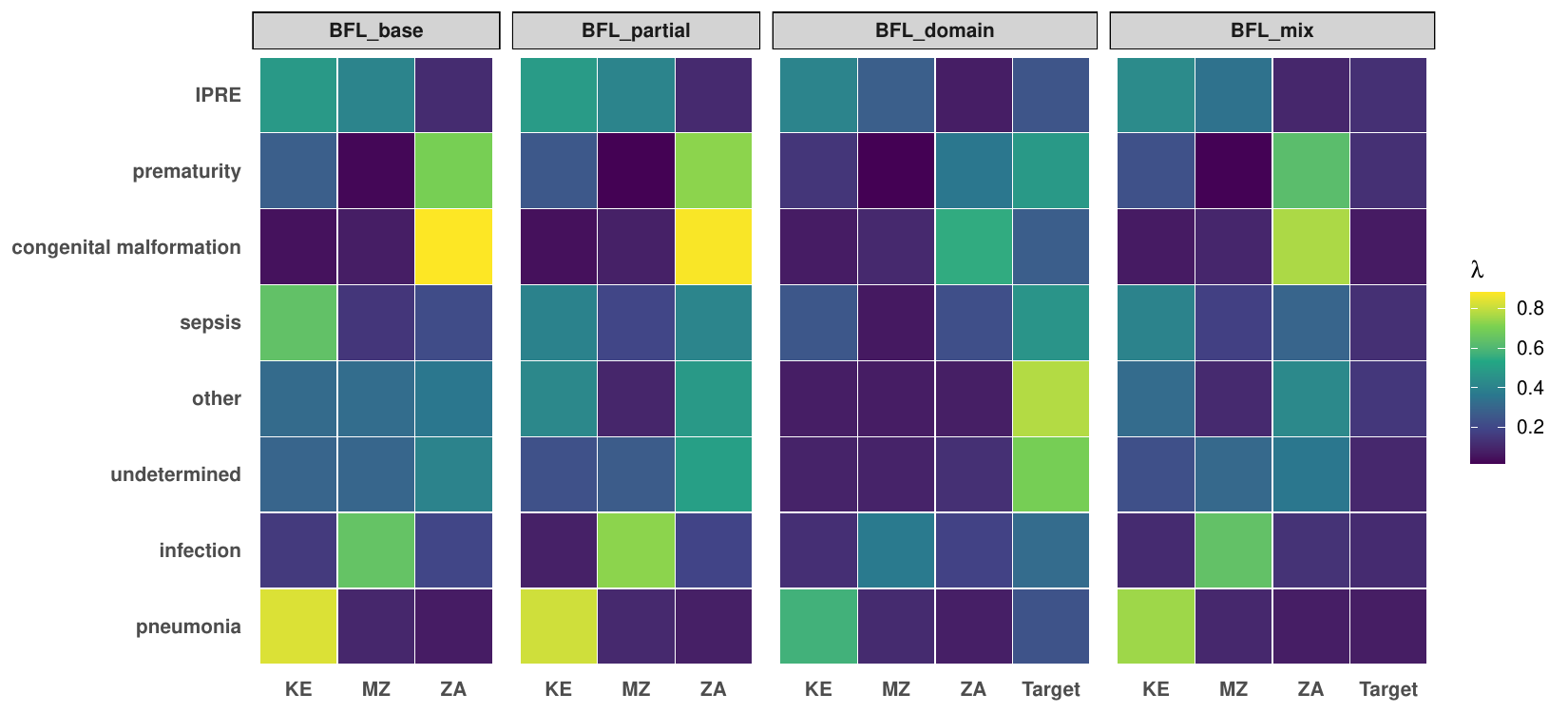}
    \caption{Posterior mean of the estimated domain weights $\blambda$ for predicting the target dataset, including four countries. The causes are arranged in decreasing order of target prevalence from top to bottom. For \emph{BFL-domain} and \emph{BFL-mix}, the additional local model is included in the candidate models (the Target column).}
    \label{fig:heatmap_lambda_neonatal_balanced_full_causes_miss_prop0.8_site_Target}
\end{figure}
}

\section{Discussion} \label{sec:discuss}
In this paper, we introduced a novel Bayesian federated learning framework for cause-of-death assignment using verbal autopsy data, without data sharing across studies. Our model accounts for both across-domain heterogeneity and within-domain label shift. Our approach uses a generative model to ensemble conditional symptom distributions $p(X \mid Y)$ rather than prediction functions $p(Y \mid X)$, enabling knowledge transfer that is robust to differences in cause-of-death profiles across datasets. We provide three strategies to fine-tune the model with local labeled data in the target domain. \rev{With pretrained models, BFL requires only a few seconds to fit on a laptop for datasets in this paper, and can scale easily to large datasets.} 
Our experiments on PHMRC and CHAMPS datasets demonstrate BFL's robustness across diverse settings. BFL consistently improves upon single-domain models and often matches or exceeds performance with full data pooling. 

While we focus on modeling the structured questionnaire data in this paper, similar challenges also exist when analyzing VA narratives \citep{cejudo2023cause, fan2024narratives}, and the FL framework can be adapted in such cases too. More broadly, our study also highlights the potential of FL approaches in global health research, where data are often fragmented, sparse, and difficult to share. The core formulation behind our model is generally applicable for generative classification and quantification models.  

Finally, we conclude with future work and open challenges. First, our model currently does not assume any structure among the candidate models. Spatial or temporal dependence among training domains can be easily incorporated through structured priors on $\blambda$.
Second, the framework extends naturally to sub-population CSMF estimation \citep{zhu2025hierarchicallatentclassmodels} via hierarchical priors on $\bpi$, offering improved scalability over joint modeling approaches. 
Third, we only considered the case where a single model is fitted in each domain. Future applications may involve fitting multiple different models in each domain and using the BFL framework to perform both domain and model selection. 
\rev{Fourth, we note that like any Bayesian models, the posterior credible intervals from BFL may undercover when parametric assumptions are violated. Valid frequentist coverage may be achieved with additional debiasing based on local labeled data \citep{angelopoulos2023prediction}.}

Lastly, our analysis reveals that no method dominates universally and model performance depends crucially on the type and extent of distribution shifts among the available data. It is an important future area of research to better understand how to select which model to use, or design more efficient data collection schemes, when ground truth is not available. In the meantime, this variability can create confusion among practitioners evaluating complex models, particularly models adapted from machine learning and artificial intelligence. Systematic and robust evaluation procedures are needed to guide model selection in practice. Ultimately, model assessment remains challenging without sufficient reference deaths, and claims of having ``solved'' cause-of-death assignment based on selective experiments should be viewed with caution. Advancing the field requires continued collection of diverse reference datasets to better understand fundamental cause-symptom relationships.


\section*{Acknowledgments}
The authors would like to thank Yue Chu for helping in preparing the CHAMPS dataset, and Abhirup Datta for helpful discussions about the GBQL algorithm.

\section*{Funding}
YZ, JT, and ZRL were supported by grant R03HD110962 from the Eunice Kennedy Shriver National Institute of Child Health and Human Development (NICHD), and in part by the Bill \& Melinda Gates Foundation. The findings and conclusions contained within are those of the authors and do not necessarily reflect positions or policies of the Bill \& Melinda Gates Foundation.

\clearpage
\appendix
\section*{Supplementary Materials}
\addcontentsline{toc}{section}{Supplementary Materials}

\section{Posterior inference}
In this section, we detail the posterior inference steps for the BFL model. For completeness, we restate the model here.
\subsection{Model}
Let $C$ denote the number of causes of death and $M$ the number of candidate models. For each individual $i = 1,\dots,N$ and each
cause $c$, the conditional likelihood for individual $i$ given
cause $c$ under model $m$ is denoted $\hat\phi_{icm}$, which are precomputed and fixed. 
The posterior joint distribution is
\begin{equation*}
  p(\bpi, \blambda \mid X^{(0)})
  = p(\bpi)p(\blambda)\prod_{i=1}^N\sum_{c=1}^{C} \pi_c \sum_{m=1}^{M} \hat\phi_{i c m}\,\lambda_{cm},
  \label{eq:likelihood}
\end{equation*}
where $\boldsymbol\pi = (\pi_1,\dots,\pi_C)\in\Delta^{C-1}$ is the
vector of cause-specific mortality fractions (CSMF) and
$\boldsymbol\lambda_c = (\lambda_{c1},\dots,\lambda_{cM})\in\Delta^{M-1}$
is the vector of model weights for cause $c$.
For a data augmentation scheme that enables Gibbs sampling, let $Y_i\in\{1,\dots,C\}$ denote the cause assignment and
$H_i\in\{1,\dots,M\}$ the latent model assignment for individual $i$. We place a symmetric Dirichlet prior on the CSMF vector,
\[
  \boldsymbol\pi \sim \mathrm{Dirichlet}(\mathbf{1}_C).
\]
Two prior options are available for the model-weight vectors
$\boldsymbol\lambda$.
\paragraph{Dirichlet prior.}
For each cause $c$, we denote $\boldsymbol\tilde\lambda_c$ as the subvector of $\blambda_c$ on the dimensions where the corresponding model is active, where model $m$ is active for cause $c$ if the cause $c$ exists in the $m$-th model. Then we let
\[
  \boldsymbol\tilde\lambda_c \sim \mathrm{Dirichlet}(\mathbf{1}_{M_c}),
  \quad c = 1,\dots,C,
\]
where $M_c \le M$ is the number of models active for cause $c$. Inactive entries of $\blambda_c$ is set to $0$. This prior yields conjugate posterior full conditionals.
\paragraph{Logistic-normal prior.}
We also consider independent normal priors on unconstrained parameters
and map to the simplex via softmax. Specifically, let
$\boldsymbol\beta_c = (\beta_{c1},\dots,\beta_{cM_c})$ collect the
active logit weights for cause $c$, with
\[
  \beta_{cm} \stackrel{\text{iid}}{\sim} \mathcal{N}(0,1),
  \quad m \in \mathcal{A}_c,
\]
where $\mathcal{A}_c$ is the active-model index set for cause $c$. Then,
\[
  \lambda_{cm} = \frac{\exp(\tilde\beta_{cm}) \mathbb{1}_{m \in \mathcal{A}_c}}
                      {\sum_{m'=1}^{M}\exp(\tilde\beta_{cm'})\mathbb{1}_{m' \in \mathcal{A}_c}}.
\]
Inactive entries in $\blambda_c$ are fixed to $0$. Under this model, $\boldsymbol\beta_c$ is updated via a Metropolis--Hastings (MH) step with a symmetric
Gaussian random-walk proposal of scale $\sigma_{\mathrm{MH}}$.
\subsection{Posterior sampling}
We use a Gibbs sampler that cycles through the following steps.
\paragraph{Step 1: Cause of death $Y_i$.}
For individuals whose cause is unobserved, draw
\[
  p(Y_i = c \mid \boldsymbol\pi, \boldsymbol\Lambda) 
  \;\propto\;  
    \pi_c \sum_{m\in\mathcal{A}_c} \hat\phi_{i c m}\,\lambda_{cm}.
\]
\paragraph{Step 2: Latent model $H_i$.}
For each individual, given $Y_i = c$, draw
\[
  p(H_i = m \mid Y_i = c,\, \boldsymbol\lambda_c)
  \;\propto\; \hat\phi_{i c m}\,\lambda_{cm},\;\;\;\; m\in\mathcal{A}_c.
\]
\paragraph{Step 3: Cause fractions $\boldsymbol\pi$.}
Let $n_c = \#\{i : Y_i = c\}$,  where the counting applies to all individuals under the assumption of no within-target label shift, and only unlabeled individuals if otherwise. Then
\[
  \boldsymbol\pi \mid \mathbf{Y}
  \;\sim\; \mathrm{Dirichlet}(1 + n_1,\dots,1 + n_C).
\]
\paragraph{Step 4a: Model weights $\boldsymbol\lambda_k$ (Dirichlet prior).}
Let $v_{cm} = \#\{i : Y_i = c,\, H_i = m\}$. Then
\[
  \boldsymbol\lambda_c \mid \mathbf{Y}, \mathbf{W}
  \;\sim\; \mathrm{Dirichlet}(1 + v_{c,m_1},\dots,1 + v_{c,m_{M_k}}),
\]
where $(m_1,\dots,m_{M_c}) = \mathcal{A}_c$. Inactive entries are set
to zero.
\paragraph{Step 4b: Model weights $\boldsymbol\lambda_k$ (logistic-normal prior).}
The log full conditional for the active logit parameters is
\begin{equation}
  \log p(\boldsymbol\beta_k \mid \mathbf{Y}, \mathbf{W})
  \;\propto\;
  -\frac{1}{2}\|\boldsymbol\beta_c\|^2
  + \sum_{m\in\mathcal{A}_c} v_{cm}
    \!\left(\beta_{cm} - \log\!\sum_{m'=1}^{M} e^{\tilde\beta_{cm'}}\right).
\end{equation}
A Metropolis--Hastings random-walk step is used: propose
$\boldsymbol\beta_c^* = \boldsymbol\beta_c + \sigma_{\mathrm{MH}}\,\boldsymbol\varepsilon$
with $\boldsymbol\varepsilon\sim\mathcal{N}(\mathbf{0},I_{M_c})$, and
accept with probability
$\min\!\left(1,\,\exp\!\left[\log p(\boldsymbol\beta_c^*\mid\cdot) -
\log p(\boldsymbol\beta_c\mid\cdot)\right]\right)$.
The model weights are then set to
$\boldsymbol\lambda_c = \mathrm{softmax}(\tilde{\boldsymbol\beta}_c)$.

\subsection{Posterior inference using Stan}
Posterior inference can also be carried out using HMC. We marginalize out $H$ and perform inference on the posterior of $\bpi$ and $\blambda$ instead. Let $\hat\phi_{icm} =  \hat p_{m}(X_i^{(0)} \mid Y_i^{(0)} = c)$. The posterior joint distribution of the parameters is
\begin{align*}
p(\bpi, \blambda \mid X^{(0)}) 
=&  p(\bpi) p(\blambda)\prod_{i} \sum_c\sum_m \hat\phi_{icm} \pi_c\lambda_{cm},
\end{align*}
We can conduct posterior inference of $\bpi$ and $\blambda$  using the probabilistic programming language Stan \citep{carpenter2017stan}. Cause-of-death assignment at the individual level is then performed by evaluating the posterior predictive distribution, 
\[
p_0(Y_i^{(0)} = c \mid X_i^{(0)} = x_i) = \int\int \frac{\sum_m  \hat \phi_{icm} \pi_c\lambda_{cm}}{\sum_{c'}\sum_m  \hat \phi_{ic'm} \pi_c\lambda_{c'm}} p(\bpi, \blambda \mid X^{(0)})d\bpi d\blambda .
\]

In the experiments considered in this paper, at 4{,}000 iterations, Gibbs sampler takes about 6.7 seconds versus about 81 seconds for Stan. The two different priors leads to indistinguishable posterior and runtime. Therefore, for the main analysis in the paper, we adopt the Gibbs sampler with the Dirichlet prior for simplicity.

\section{Additional analysis of PHMRC Gold-standard Dataset}
\subsection{Balanced accuracy comparisons under three versions of label shift}
In this subsection, we show additional results evaluating balanced accuracy, defined as
\begin{align*}
    \text{Balanced Accuracy} &= \frac{1}{C} \sum_{c=1}^{C} \frac{1}{n_c}\sum_{i: Y_i = c} \bm 1_{\hat Y_i = c}.
\end{align*}
where $C$ is the number of causes and $n_c$ is the number of deaths due to cause $c$, 
This metric is equivalent to the macro-average recall and ensures that all causes contribute equally to the overall accuracy, regardless of their prevalence. Balanced accuracy can be a more interpretable metric when the true cause distribution is highly imbalanced. 

\begin{figure}[!htb]
    \centering
    \includegraphics[width = \textwidth]{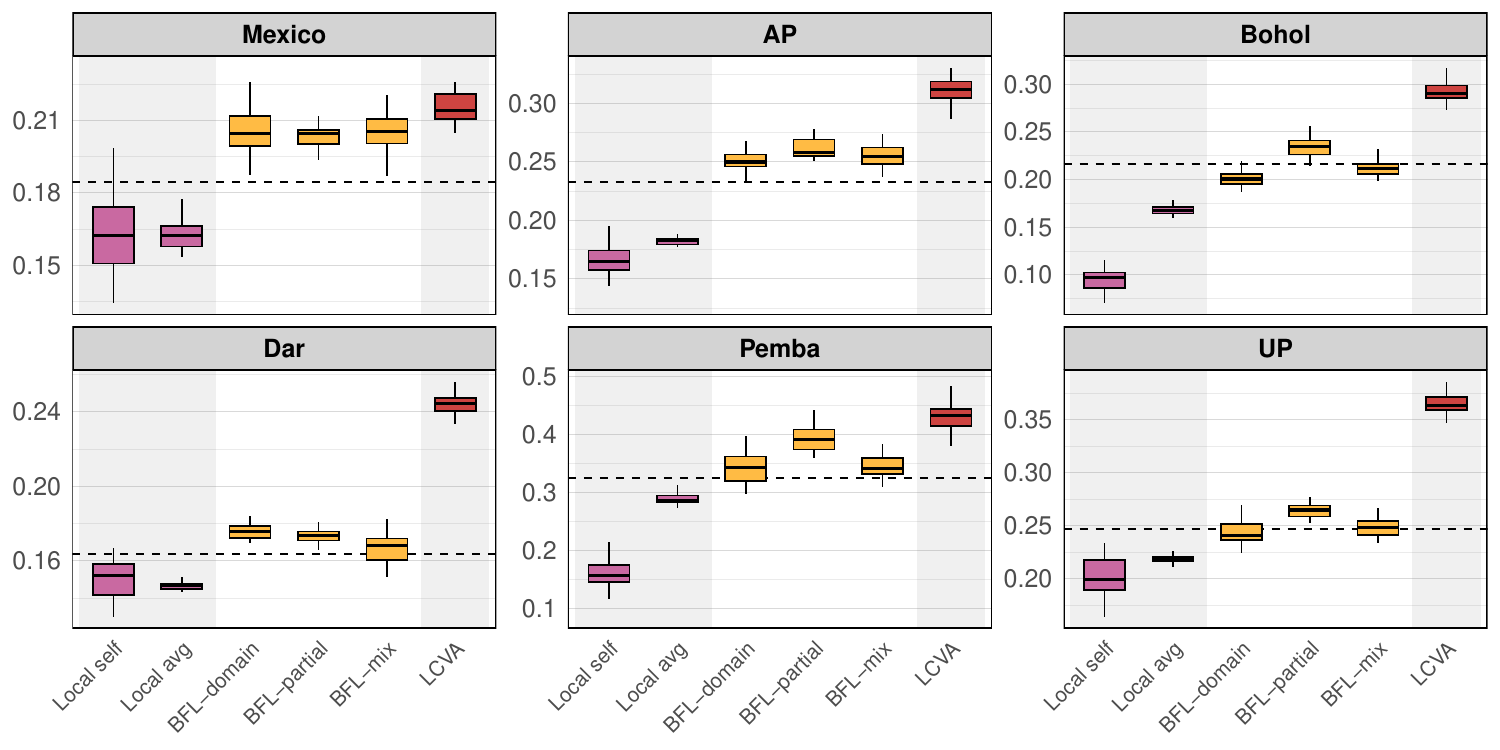}
    \caption{\textbf{No within-target label shift}: comparison of balanced accuracy across different methods. BFL models are consistently better than the base models trained on a single domain (\emph{local-self} and \emph{local-avg}). Among BFL models, \emph{BFL-partial} achieves the highest accuracy in four sites. In comparison with the top cause accuracy plot in the main paper, the improvements of \emph{BFL-domain} are primarily for the more prevalent causes, which is less pronounced in the evaluation of balanced accuracy. The dashed line shows the balanced accuracy from BFL without local labeled data.}
    \label{fig:Balanced_ACC_balanced_0.8}
\end{figure}

\begin{figure}[!htb]
    \centering
    \includegraphics[width = \textwidth]{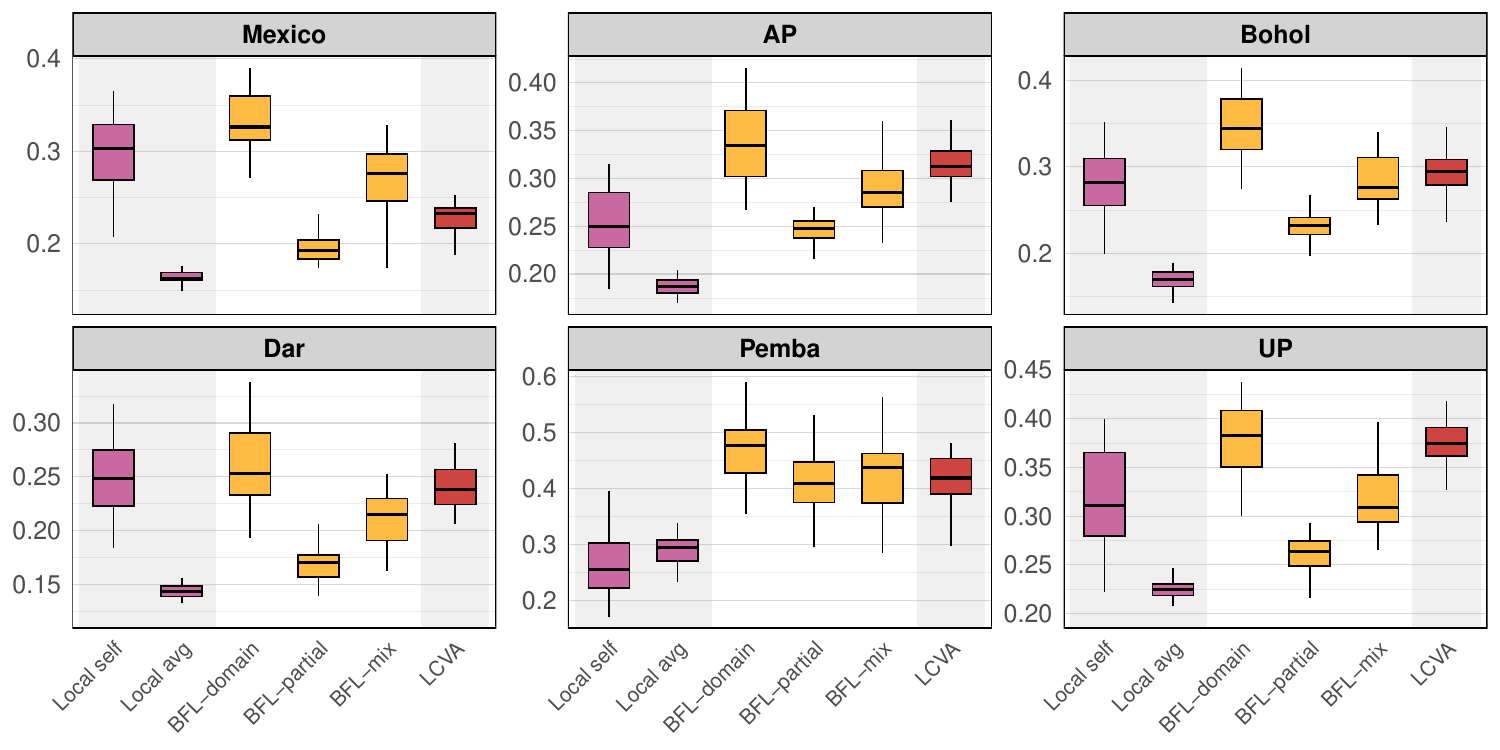}
    \caption{\textbf{Mild within-target label shift}: comparison of balanced accuracy across different methods. \emph{BFL-domain} is better than LCVA and achieves highest average accuracy in all six sites. The results are in general similar to the top cause accuracy plot in the main paper.}
    \label{fig:Balanced_ACC_unbalanced_0.8}
\end{figure}

\begin{figure}[!htb]
    \centering
    \includegraphics[width = \textwidth]{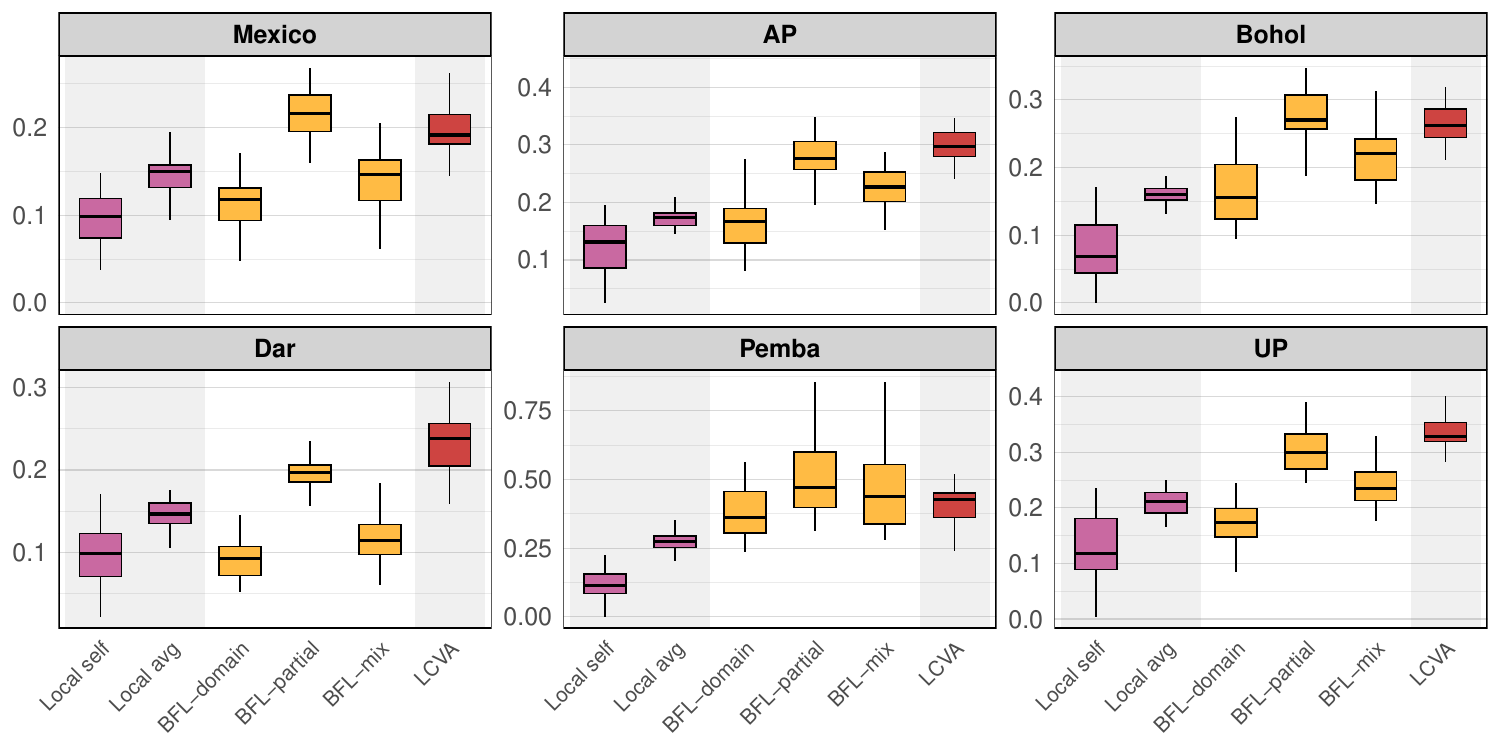}
    \caption{\textbf{Severe within-target label shift}: comparison of balanced accuracy across different methods. \emph{BFL-partial} is better than LCVA and achieves highest average accuracy in all six sites. The results are in general similar to the top cause accuracy plot in the main paper}
    \label{fig:Balanced_ACC_extreme_unbalanced_0.8}
\end{figure}

Figure \ref{fig:Balanced_ACC_balanced_0.8}, \ref{fig:Balanced_ACC_unbalanced_0.8}, and \ref{fig:Balanced_ACC_extreme_unbalanced_0.8} show the balanced accuracy comparisons under three simulation scenarios in the main paper.
When there is no label shift within the target domain, \emph{BFL-domain} model achieves slightly higher top cause accuracy as shown in the main paper, whereas \emph{BFL-partial} model achieves higher balanced accuracy in some sites, by better capturing minority causes. The differences between the two approaches in this scenario is moderate.
Under mild label shift within the target domain, \emph{BFL-domain} model achieves the best top cause accuracy and balanced accuracy in all domains. When within-target label shift increases from mild to severe, \emph{BFL-partial} model achieves the best top cause accuracy and balanced accuracy in all domains.
\clearpage
\subsection{Results ensembling likelihood conditional latent classes}
\label{sec:yz}
In this subsection, we replace the standard likelihood $P(X \mid Y)$ in the main paper with the conditional likelihood $P(X \mid Y, Z)$, given the latent sub-classes $Z$ under the LCVA framework. When LCVA is the base model, we can let 
    \[
        p_0(X\mid Y=c)
        =\sum_{m=1}^M\sum_{k=1}^K
        \tilde{\lambda}_{cmk}\,\hat p_{mk}(X\mid Y=c,Z=k).
    \]
    be the likelihood evaluated at the ensemble stage, where $Z$ represents the latent class learned by LCVA. In this way, we treat the latent symptom likelihood from each sub-class essentially as one candidate model. Since $\hat p_{mk}(X\mid Y=c,Z=k)$ is considered fixed, this model can be easily fit by treating the sub-class-specific conditional likelihood functions as if they arise from $MK$ pretrained models. 
    
\begin{figure}[!htb]
    \centering
    \includegraphics[width=\textwidth]{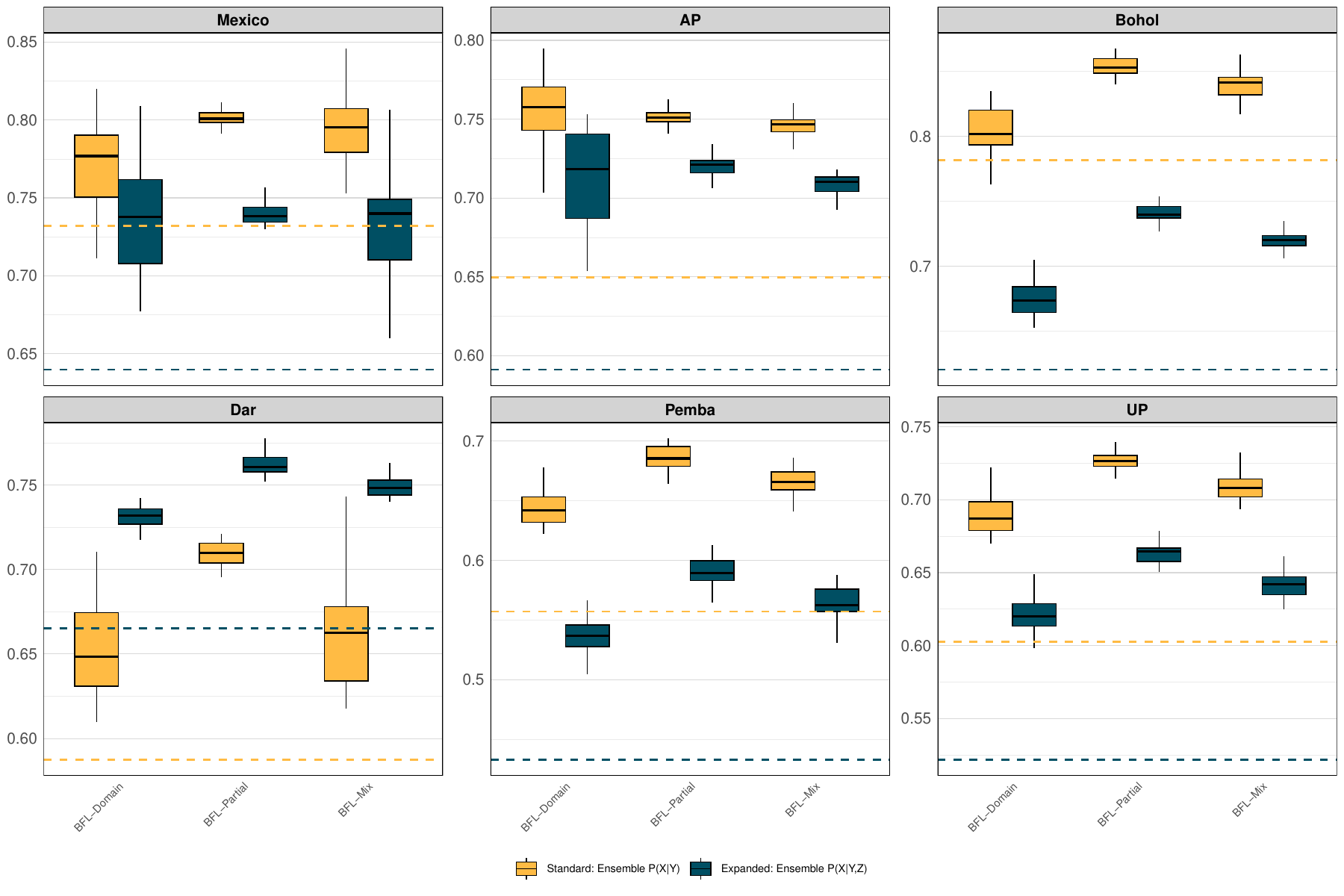}
    \caption{\textbf{No within-target label shift. CSMF accuracy}: the original BFL ($P(X\mid Y)$) outperforms the sub-class-conditional variant at all sites except Dar.}
    \label{fig:yz_no_csmf}
\end{figure}

\begin{figure}[!htb]
    \centering
    \includegraphics[width=\textwidth]{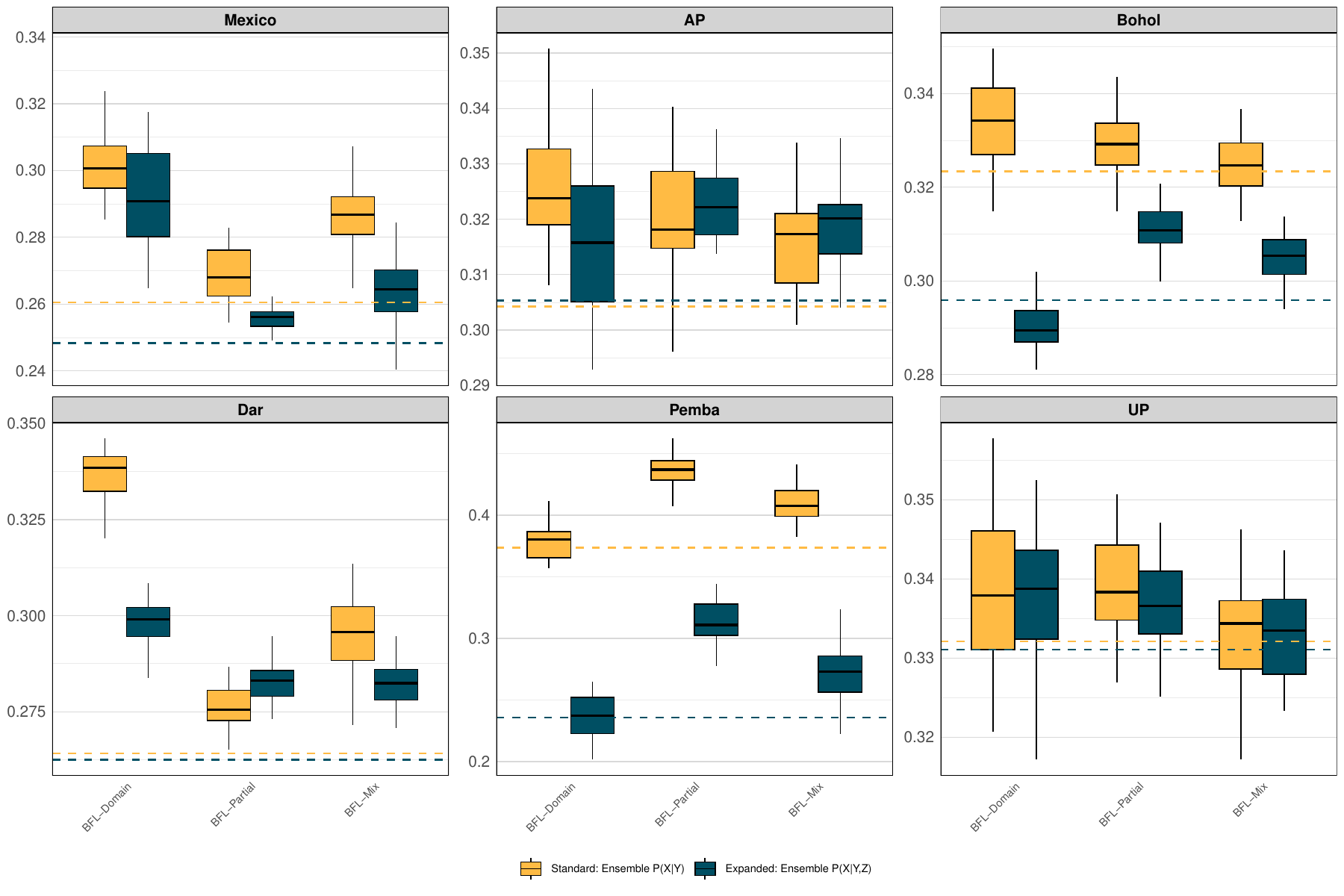}
    \caption{\textbf{No within-target label shift. Top cause accuracy}: results are competitive across sites, except at Bohol and Pemba where the original BFL is evidently better.}
    \label{fig:yz_no_acc}
\end{figure}

\begin{figure}[!htb]
    \centering
    \includegraphics[width=\textwidth]{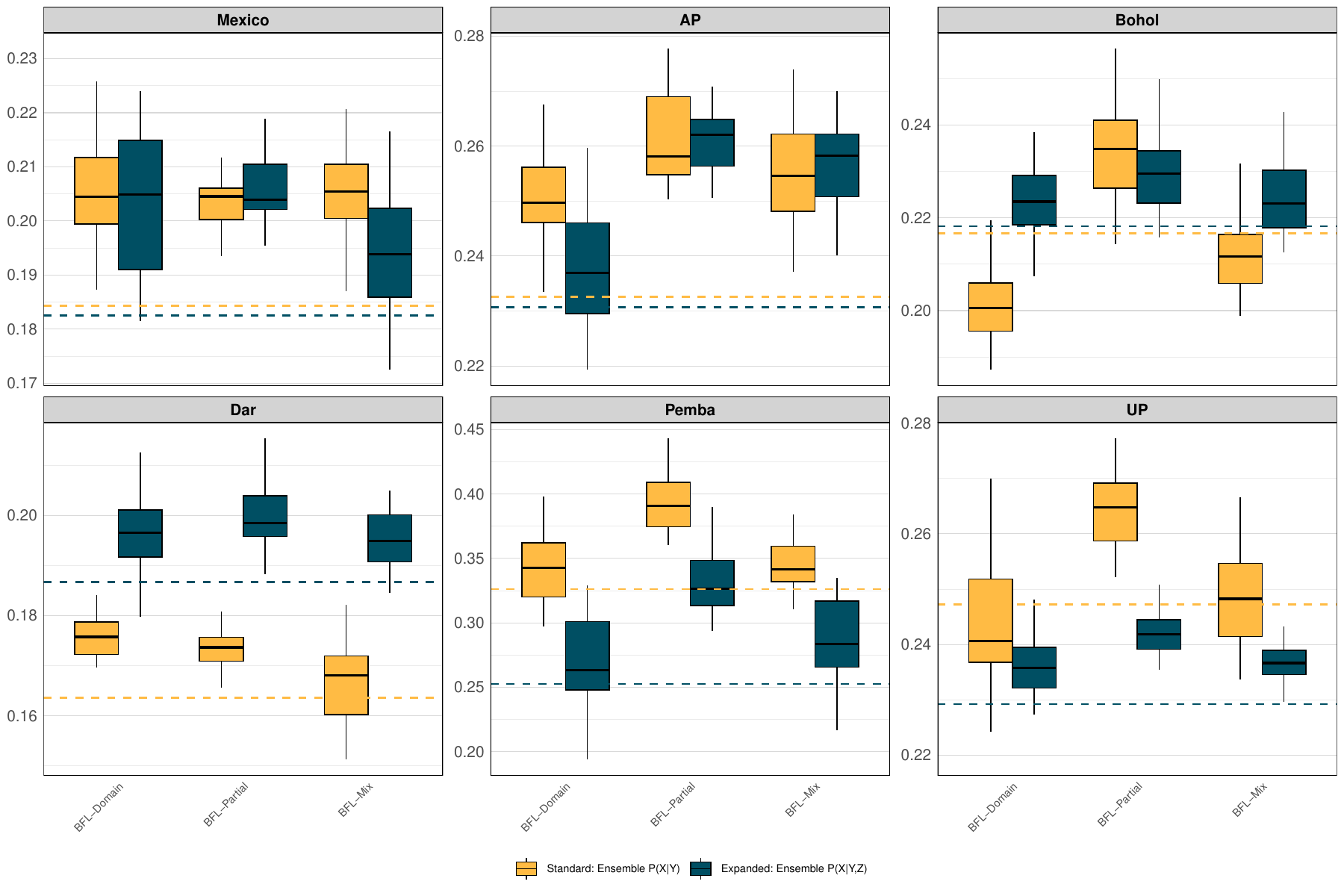}
    \caption{\textbf{No within-target label shift. Balanced accuracy}: the sub-class-conditional variant is better at Dar, the original BFL is better at Pemba and UP, and the remaining sites are competitive.}
    \label{fig:yz_no_bal}
\end{figure}

\begin{figure}[!htb]
    \centering
    \includegraphics[width=\textwidth]{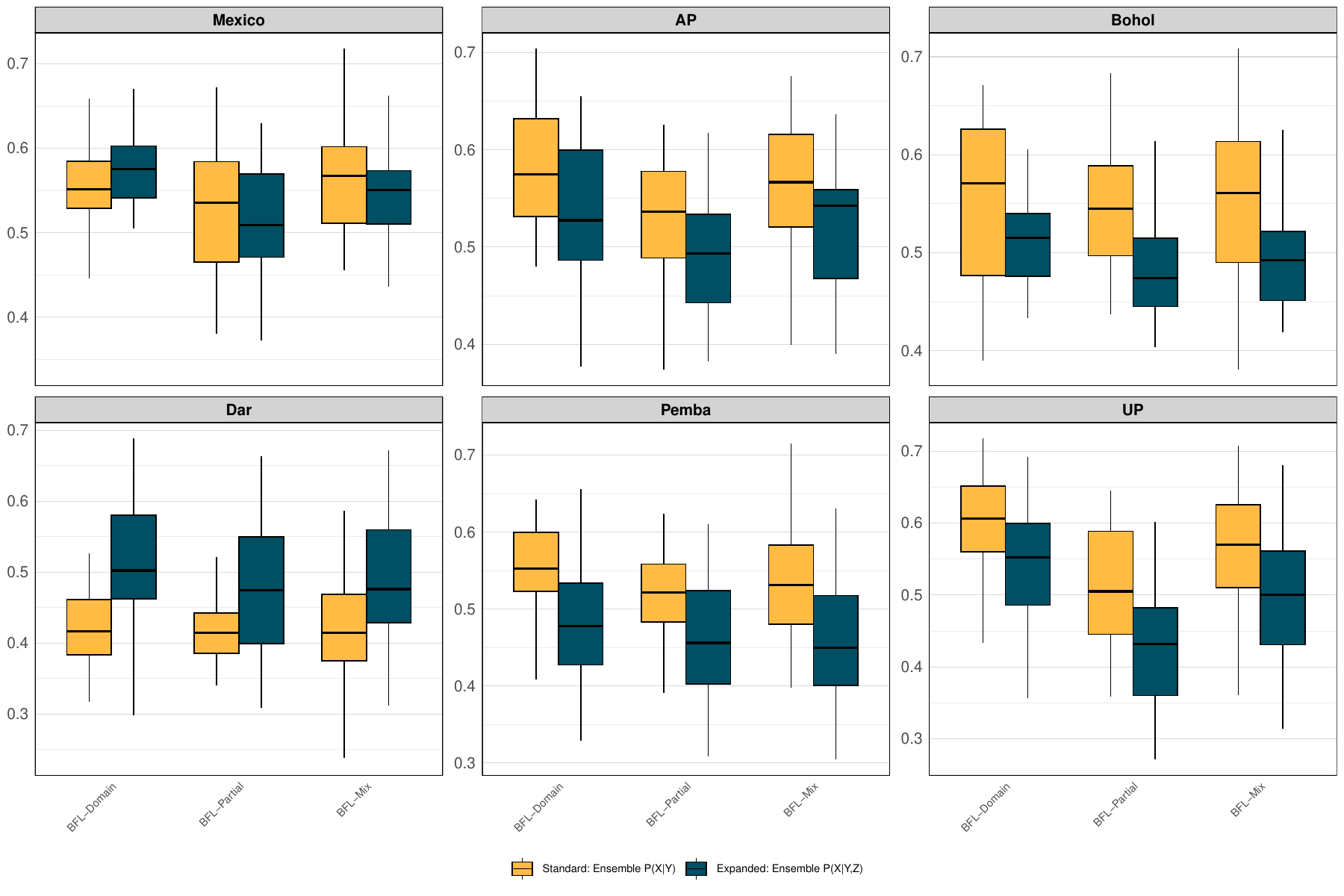}
    \caption{\textbf{Mild within-target label shift. CSMF accuracy}: the two variants are competitive across all sites.}
    \label{fig:yz_mild_csmf}
\end{figure}

\begin{figure}[!htb]
    \centering
    \includegraphics[width=\textwidth]{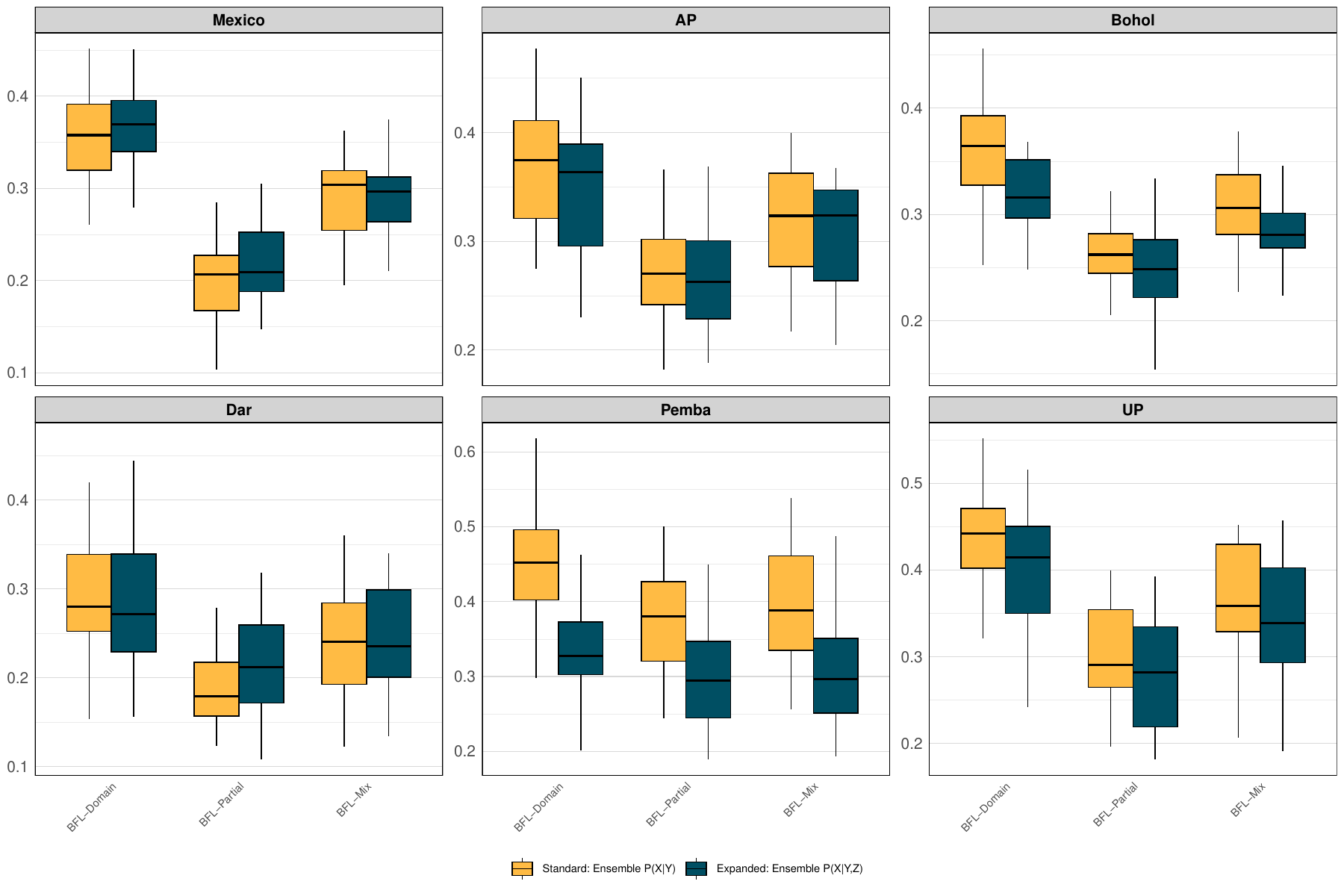}
    \caption{\textbf{Mild within-target label shift. Top cause accuracy}: the two variants are competitive across all sites.}
    \label{fig:yz_mild_acc}
\end{figure}

\begin{figure}[!htb]
    \centering
    \includegraphics[width=\textwidth]{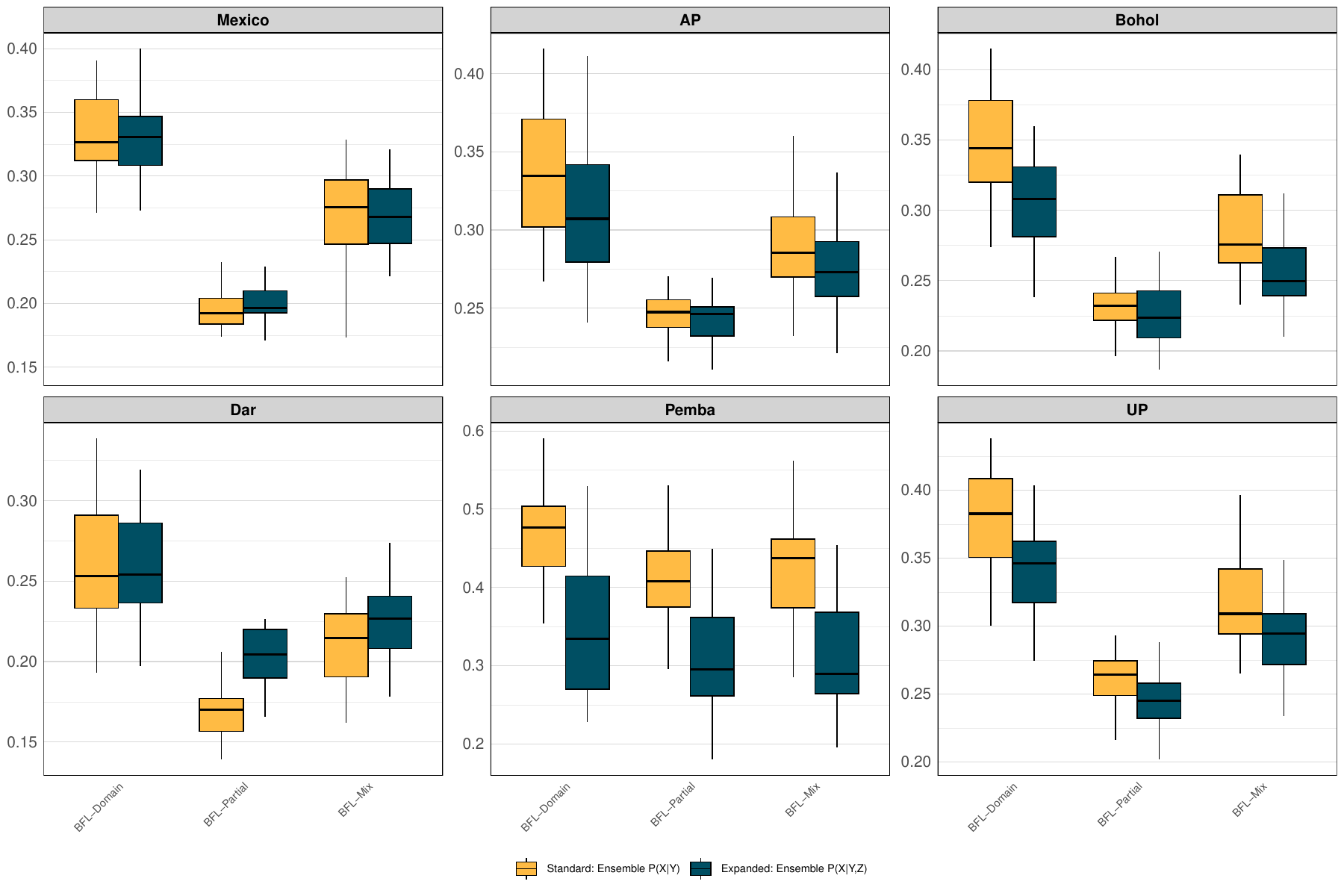}
    \caption{\textbf{Mild within-target label shift. Balanced accuracy}: the original BFL slightly outperforms the sub-class-conditional variant at four sites and ties at the remaining sites.}
    \label{fig:yz_mild_bal}
\end{figure}

\begin{figure}[!htb]
    \centering
    \includegraphics[width=\textwidth]{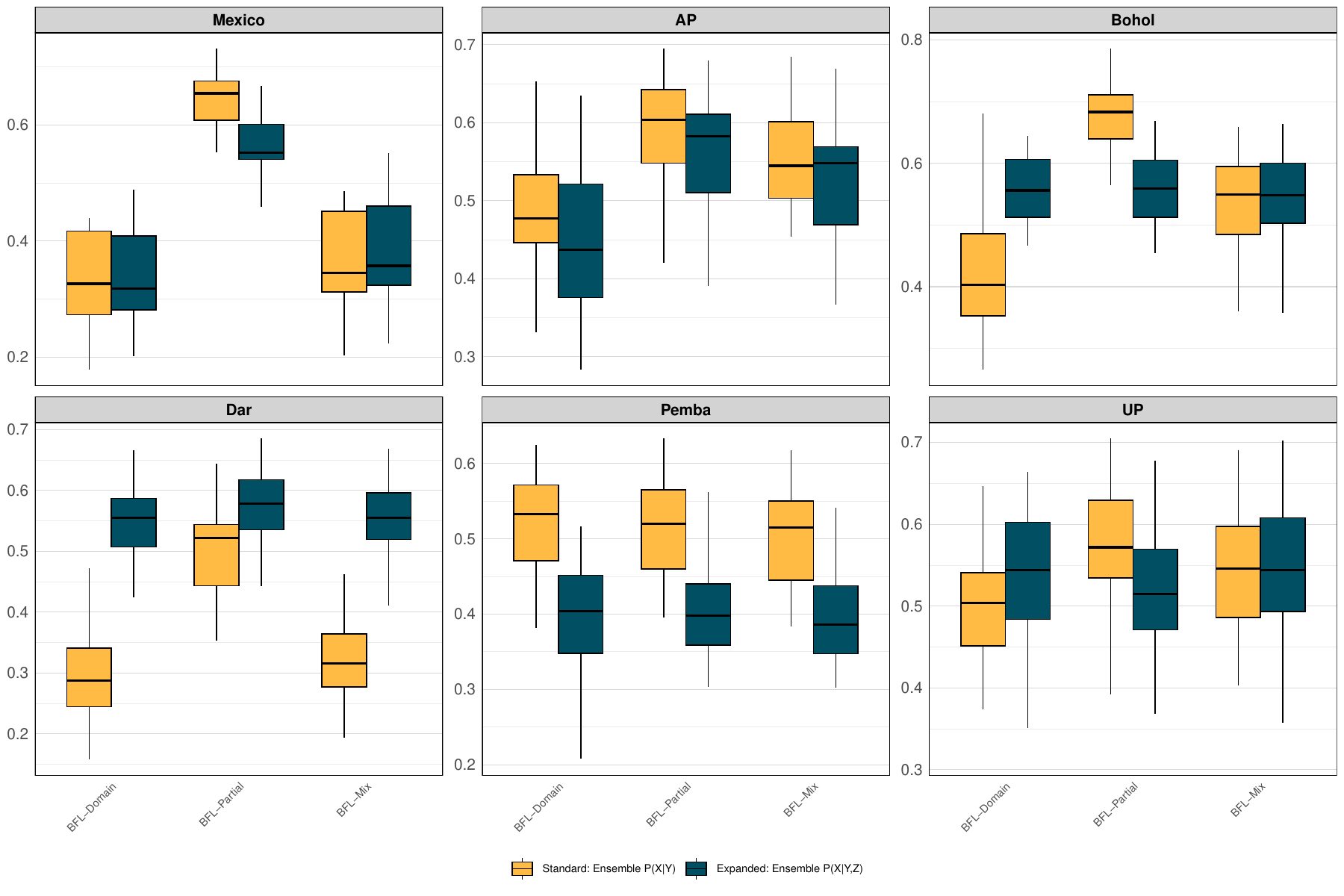}
    \caption{\textbf{Severe within-target label shift. CSMF accuracy}: the sub-class-conditional variant is better at Dar, the original BFL is better at Pemba, and the remaining sites are comparable.}
    \label{fig:yz_sev_csmf}
\end{figure}

\begin{figure}[!htb]
    \centering
    \includegraphics[width=\textwidth]{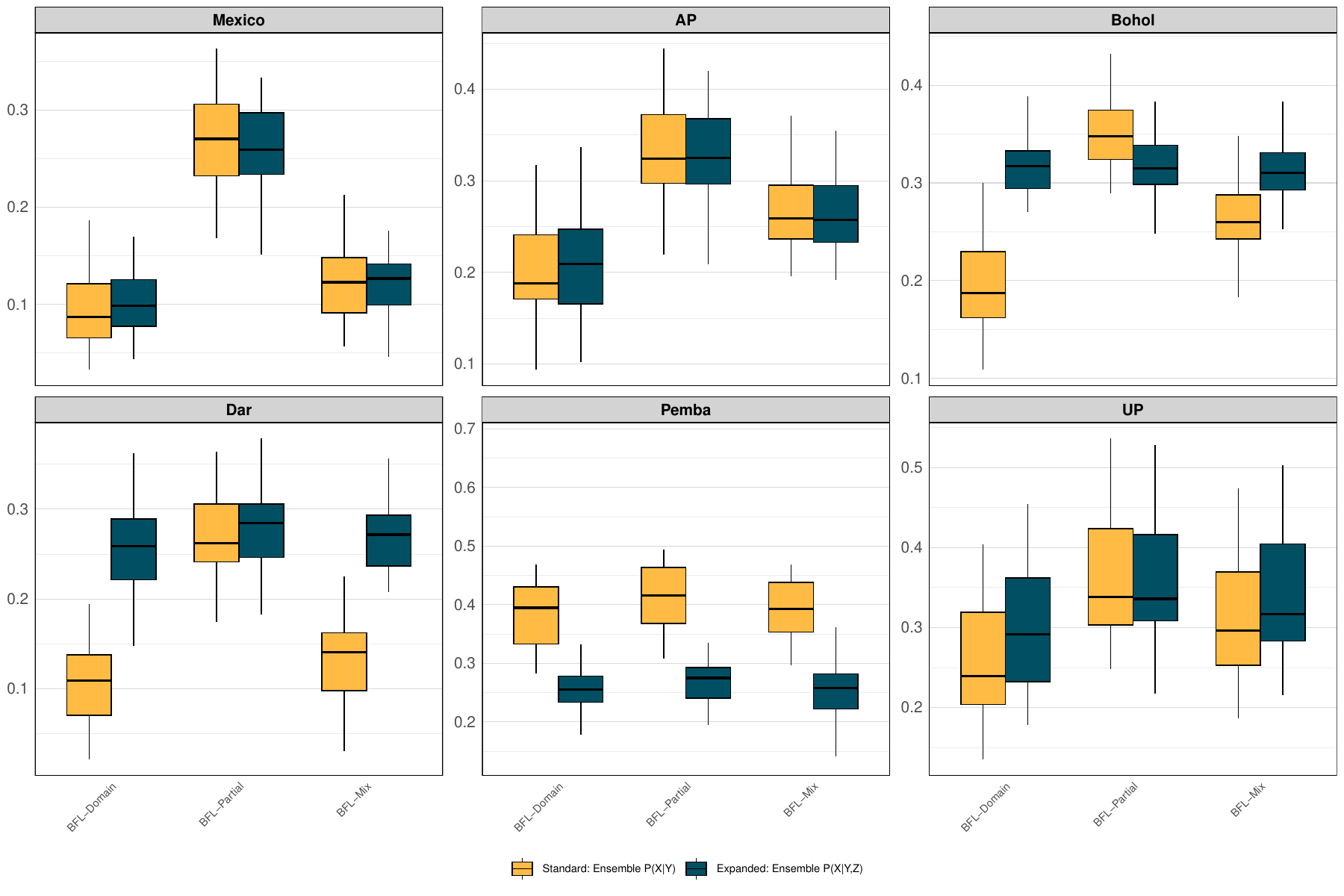}
    \caption{\textbf{Severe within-target label shift. Top cause accuracy}: the two variants are competitive across all sites.}
    \label{fig:yz_sev_acc}
\end{figure}

\begin{figure}[!htb]
    \centering
    \includegraphics[width=\textwidth]{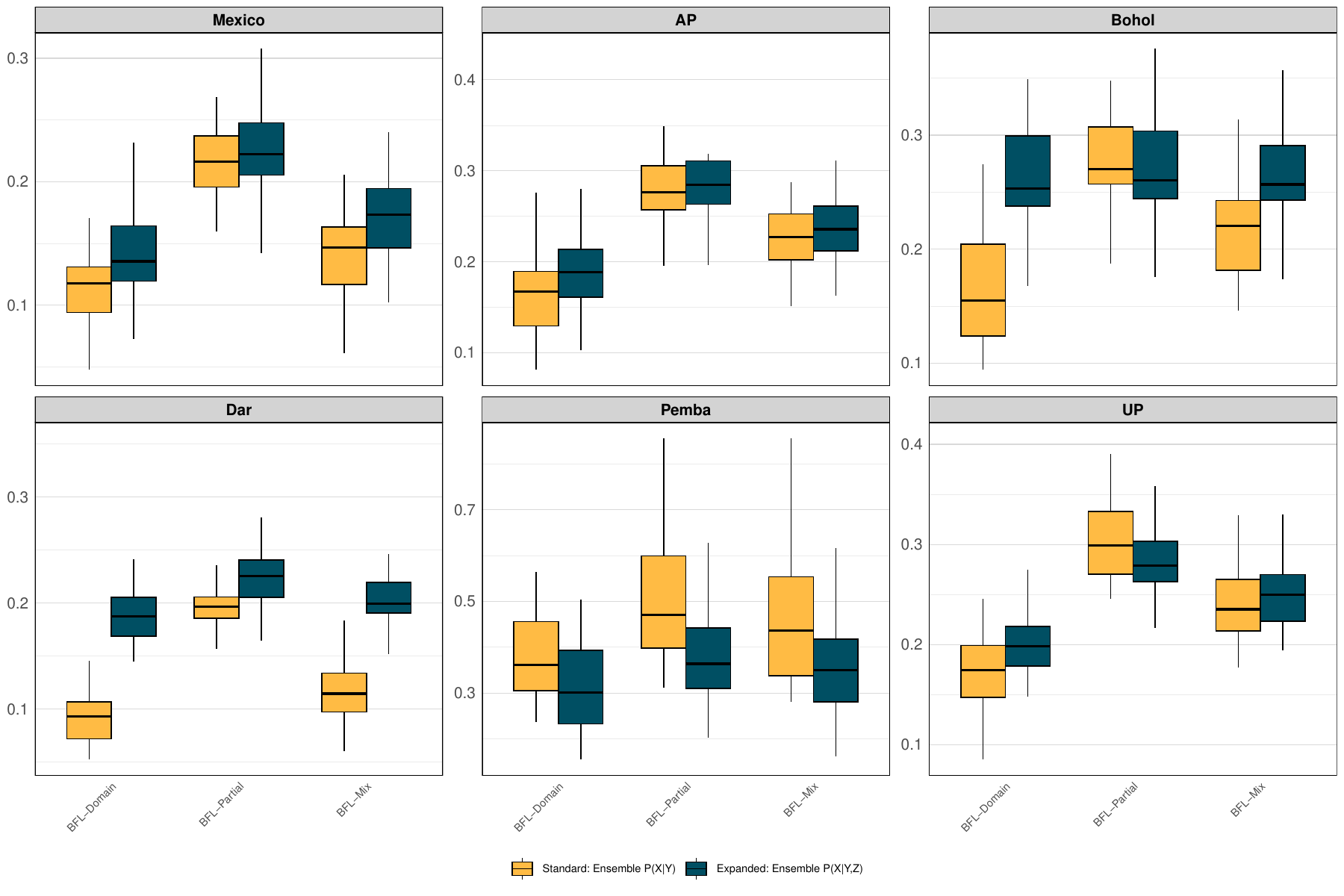}
    \caption{\textbf{Severe within-target label shift. Balanced accuracy}: the two variants are competitive across all sites.}
    \label{fig:yz_sev_bal}
\end{figure}

Overall, incorporating the latent sub-classes leads to similar performance as the BFL model using $P(X\mid Y)$ in the main paper. Across all three label-shift settings the results are largely comparable, and in many cases performs slightly worse. Figures~\ref{fig:yz_no_csmf}--\ref{fig:yz_sev_bal} give the site-level comparison across all metrics and shifts.
\clearpage

\subsection{Additional results using variations of LCVA}
\label{sec:lcva_pool}
\begin{figure}[!htb]
    \centering
    \includegraphics[width=\textwidth]{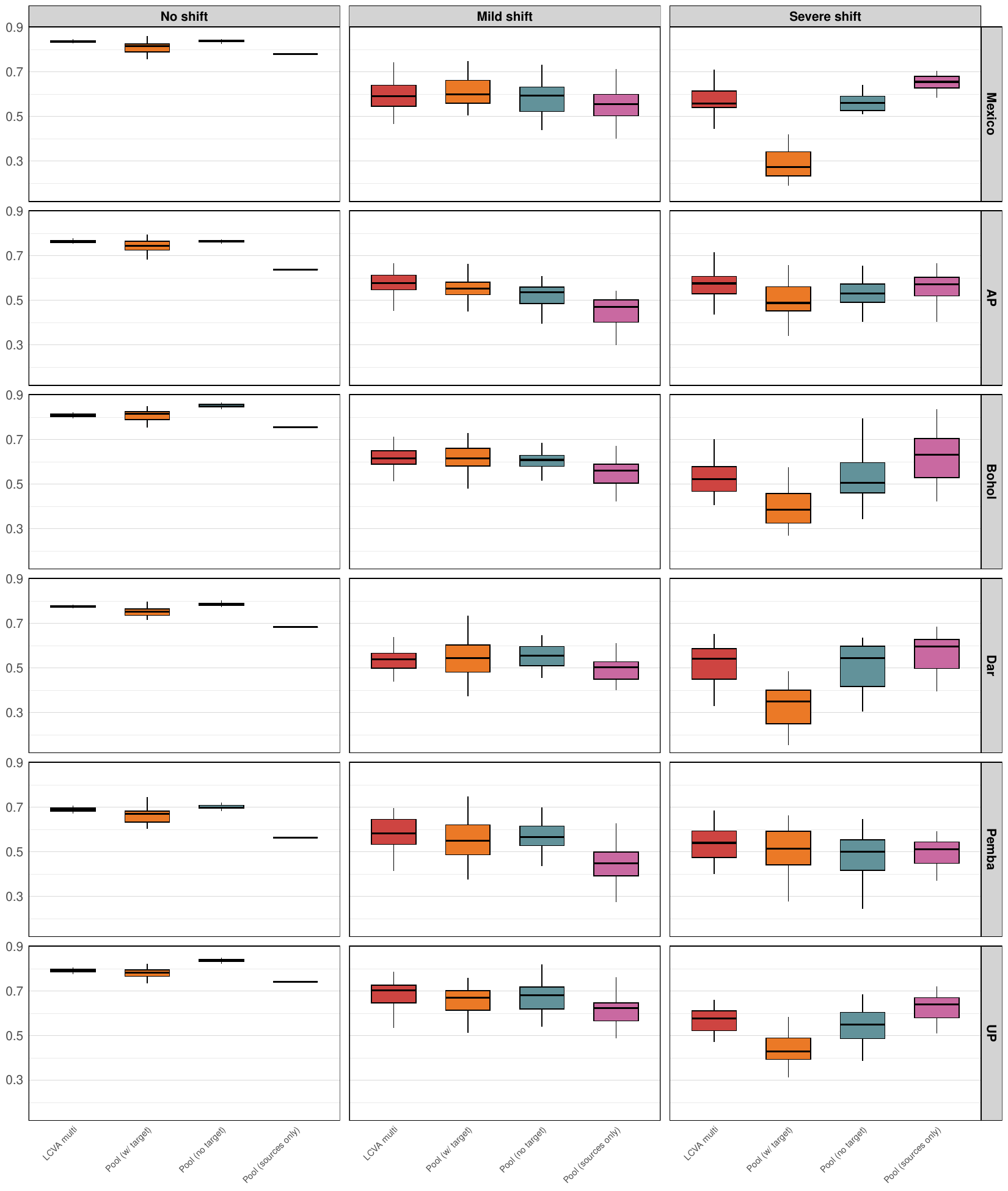}
    \caption{\textbf{CSMF accuracy}: LCVA-multi versus the three pooled LCVA baselines (with target, no target, sources only) across sites and shifts. The pooled and non-pooled variants are nearly identical, with LCVA-multi best on average.}
    \label{fig:lcva_pool_csmf}
\end{figure}

\begin{figure}[!htb]
    \centering
    \includegraphics[width=\textwidth]{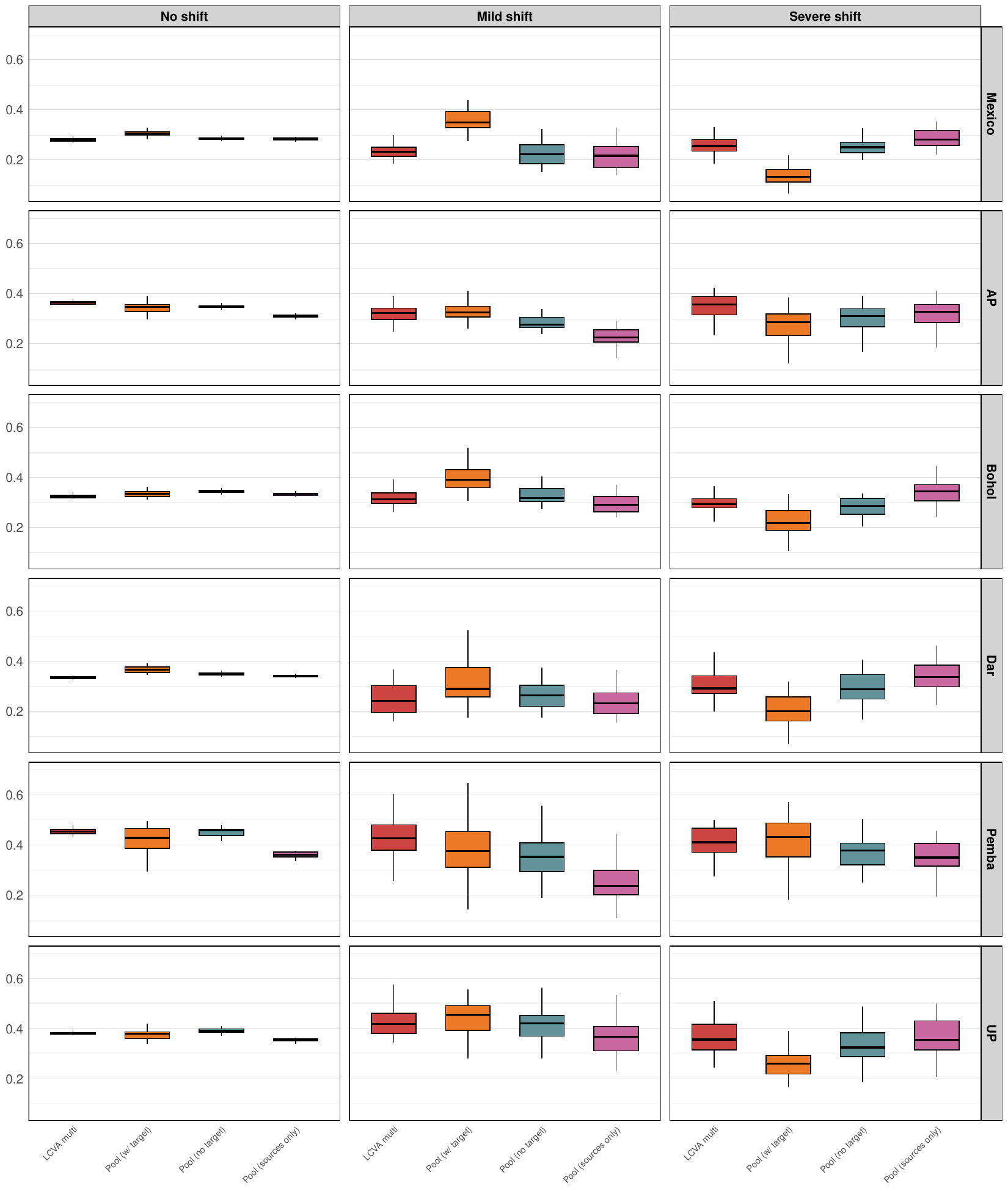}
    \caption{\textbf{Top cause accuracy}: LCVA-multi versus the three pooled LCVA baselines; results are nearly identical across pooling choices.}
    \label{fig:lcva_pool_acc}
\end{figure}

\begin{figure}[!htb]
    \centering
    \includegraphics[width=\textwidth]{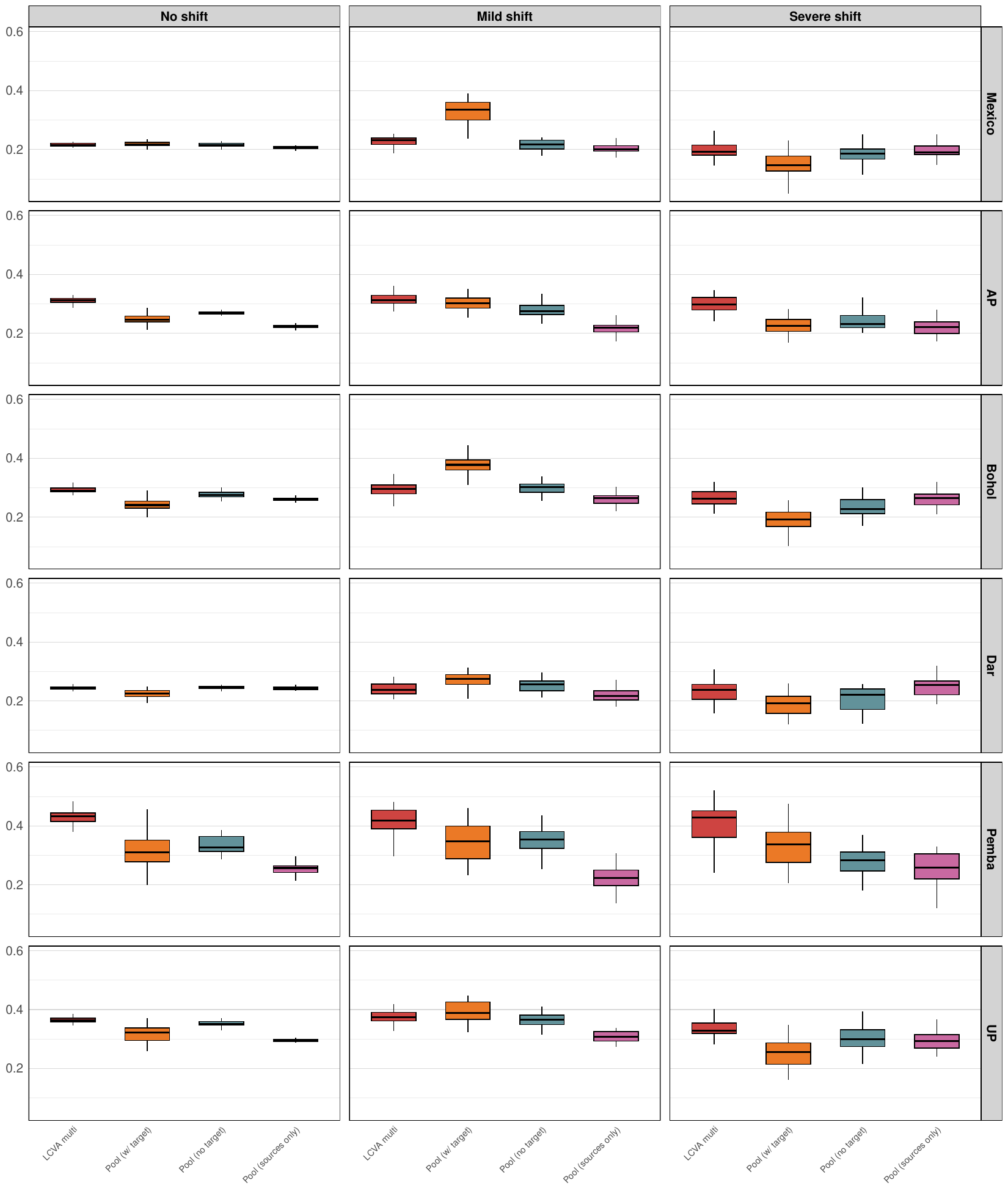}
    \caption{\textbf{Balanced accuracy}: LCVA-multi versus the three pooled LCVA baselines; results are nearly identical across pooling choices.}
    \label{fig:lcva_pool_bal}
\end{figure}

We compare four LCVA baselines that differ in how the source sites are combined and how the target's labels are used:
\begin{itemize}
    \item \textbf{LCVA-multi}: joint multi-domain LCVA that retains the site structure and transfers to the target using the target's partial labels. This is the baseline compared in the main paper
    \item \textbf{LCVA-pool (with target)}: sources pooled into a single domain together with the labeled target rows; the target labels enter through training.
    \item \textbf{LCVA-pool (no target)}: sources pooled into a single domain, with the target's partial labels supplied only at prediction (semi-supervised).
    \item \textbf{LCVA-pool (sources only)}: sources pooled into a single domain with the target data ignored.
\end{itemize}
Pooling versus non-pooling yields mostly similar results. Figures~\ref{fig:lcva_pool_csmf}--\ref{fig:lcva_pool_bal} give the comparison for CSMF accuracy, top cause accuracy, and balanced accuracy.
\clearpage
\subsection{Results with InSilicoVA as base model}
\begin{figure}[!htb]
    \centering
    \includegraphics[width=\textwidth]{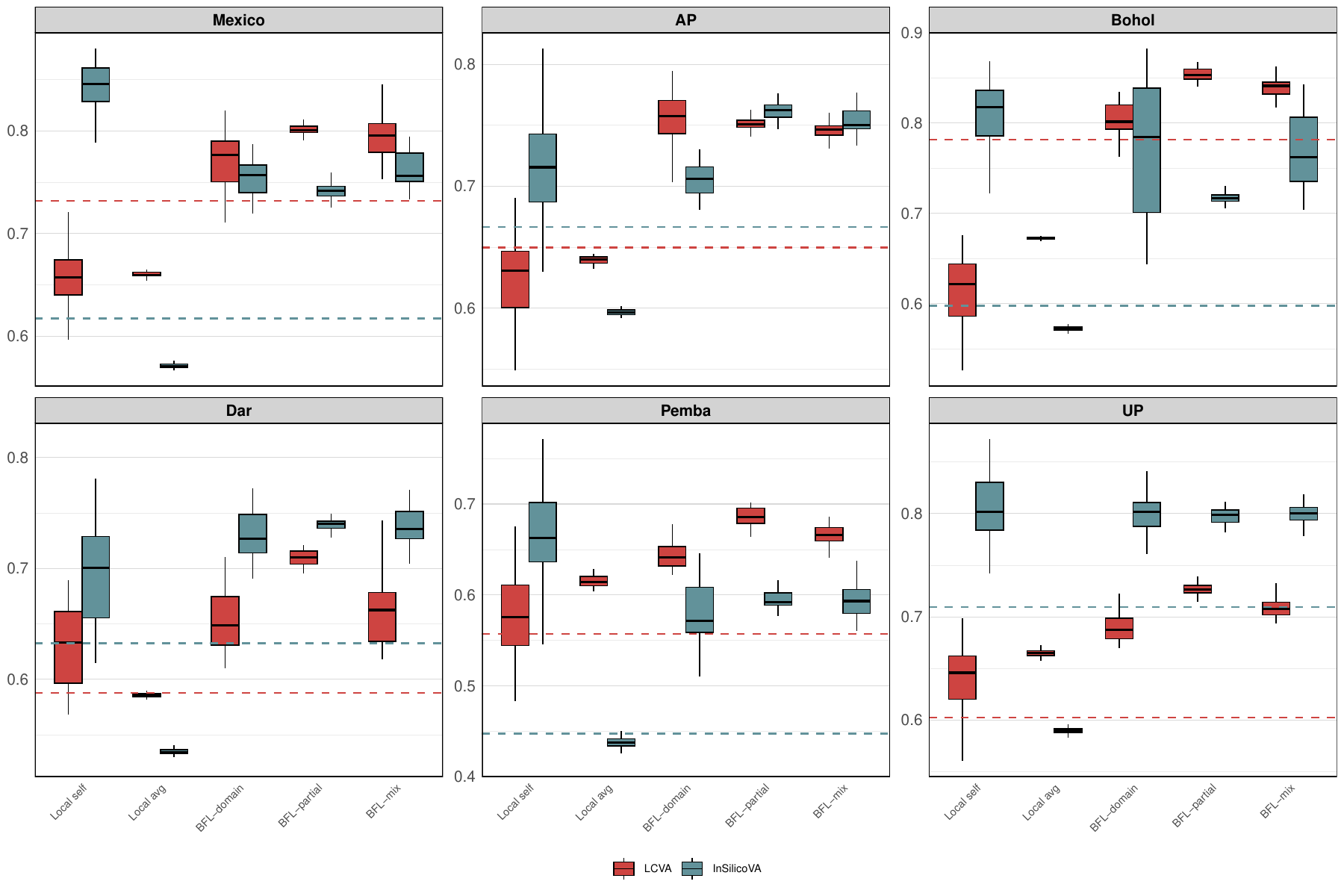}
    \caption{\textbf{CSMF accuracy}: BFL variants with LCVA versus InSilicoVA as the local base model, across methods and sites.}
    \label{fig:insilico_csmf}
\end{figure}

\begin{figure}[!htb]
    \centering
    \includegraphics[width=\textwidth]{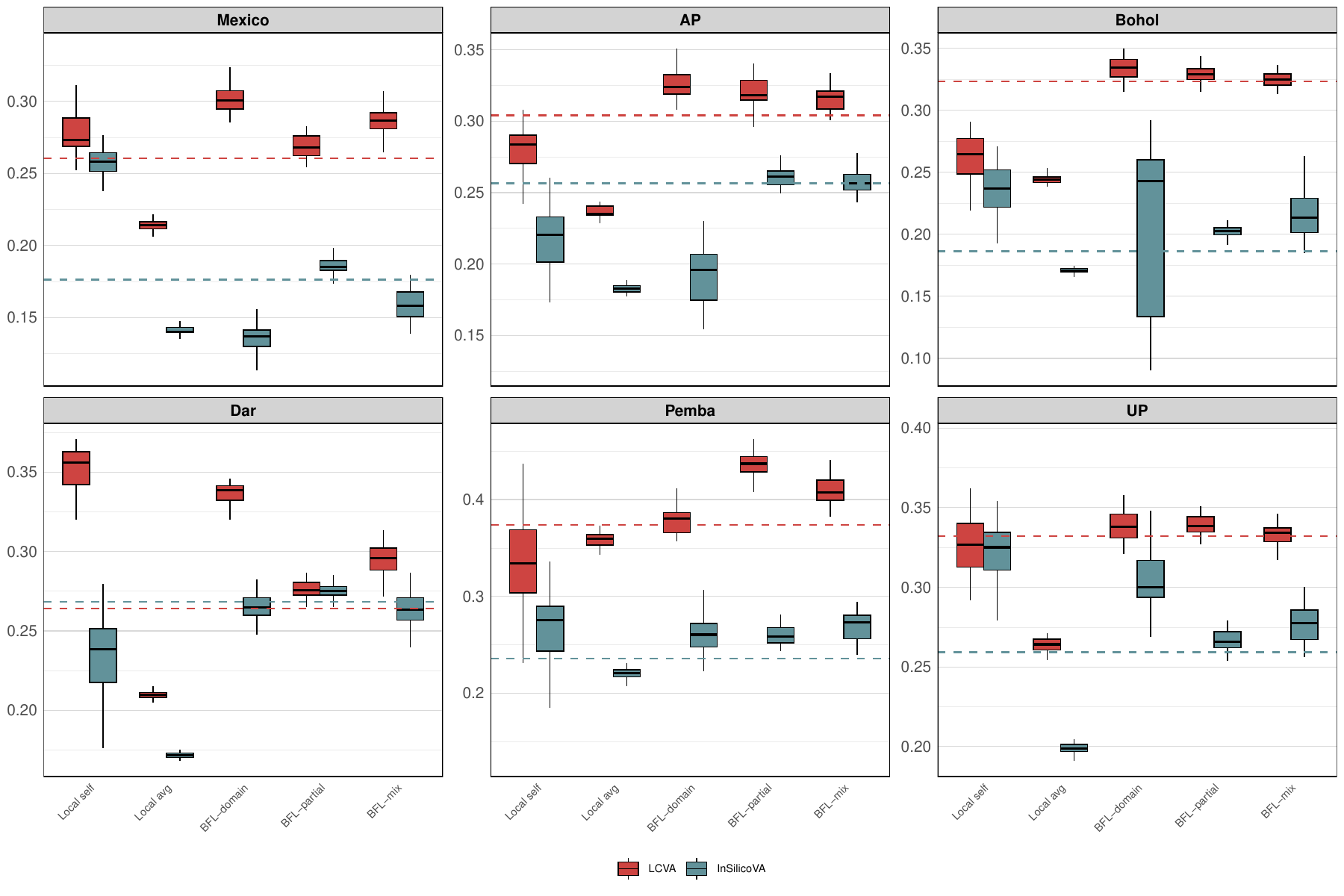}
    \caption{\textbf{Top cause accuracy}: BFL variants with LCVA versus InSilicoVA as the local base model.}
    \label{fig:insilico_acc}
\end{figure}

\begin{figure}[!htb]
    \centering
    \includegraphics[width=\textwidth]{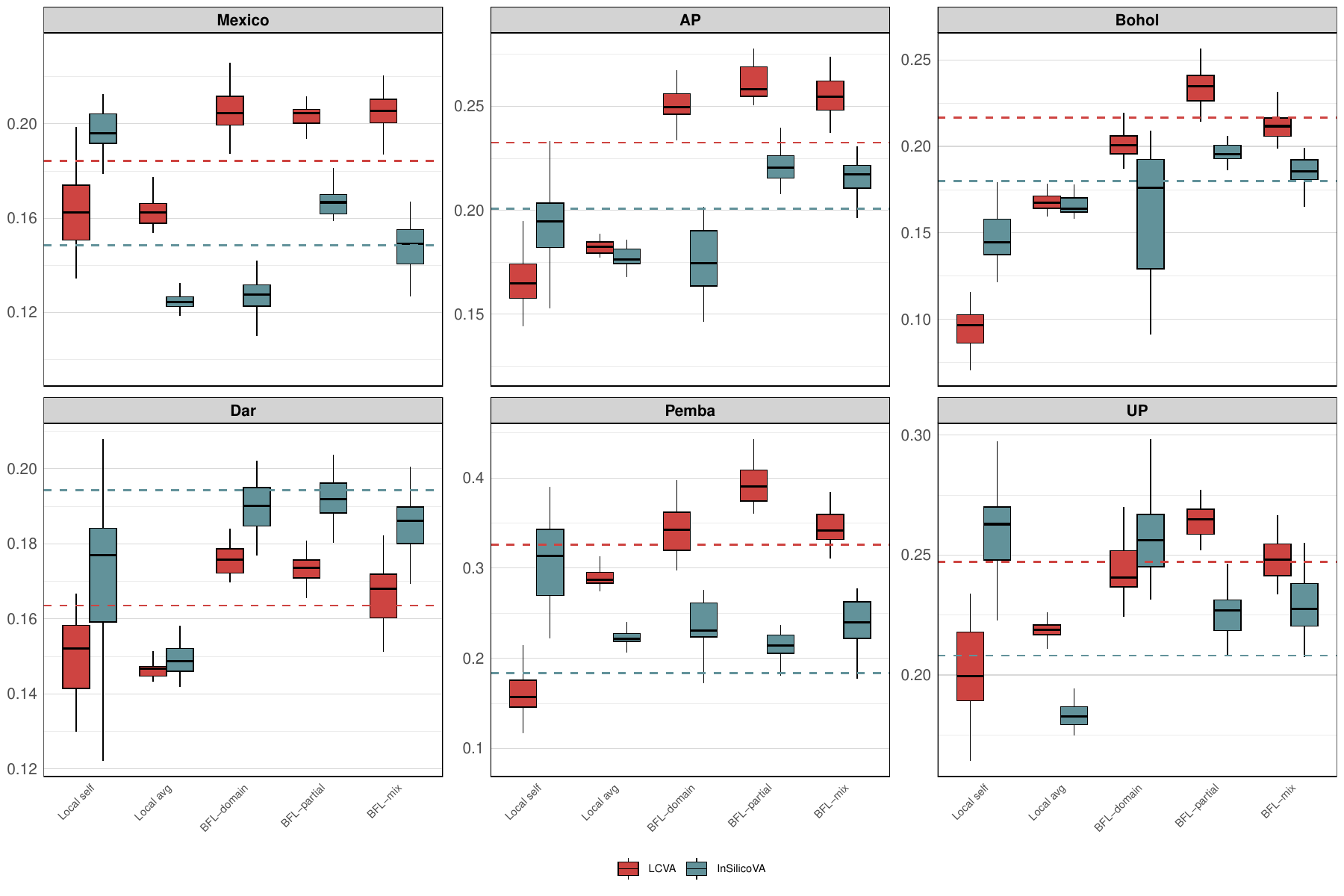}
    \caption{\textbf{Balanced accuracy}: BFL variants with LCVA versus InSilicoVA as the local base model.}
    \label{fig:insilico_bal}
\end{figure}

BFL aggregates site-level model summaries produced by a local base model. To show that BFL is not tied to a particular base model, we repeat the analysis with InSilicoVA in place of LCVA as the local model that generates these summaries, and compare the resulting BFL variants (\emph{BFL-domain}, \emph{BFL-partial}, \emph{BFL-mix}). Figures~\ref{fig:insilico_csmf}--\ref{fig:insilico_bal} report CSMF accuracy, top cause accuracy, and balanced accuracy. BFL with InSilicoVA as base model leads to mostly worse accuracy measures compared to LCVA. In particulary, for Bohol, \emph{BFL-domain} exhibits markedly higher seed-to-seed variance under InSilicoVA (whereas it is stable under LCVA), reflecting the sensitivity of the target self-model to the particular labeled subset when the base model is InSilicoVA.
\clearpage
\subsection{Comparison of samplers}
\label{sec:samplers}
\begin{figure}[!htb]
    \centering
    \includegraphics[width=\textwidth]{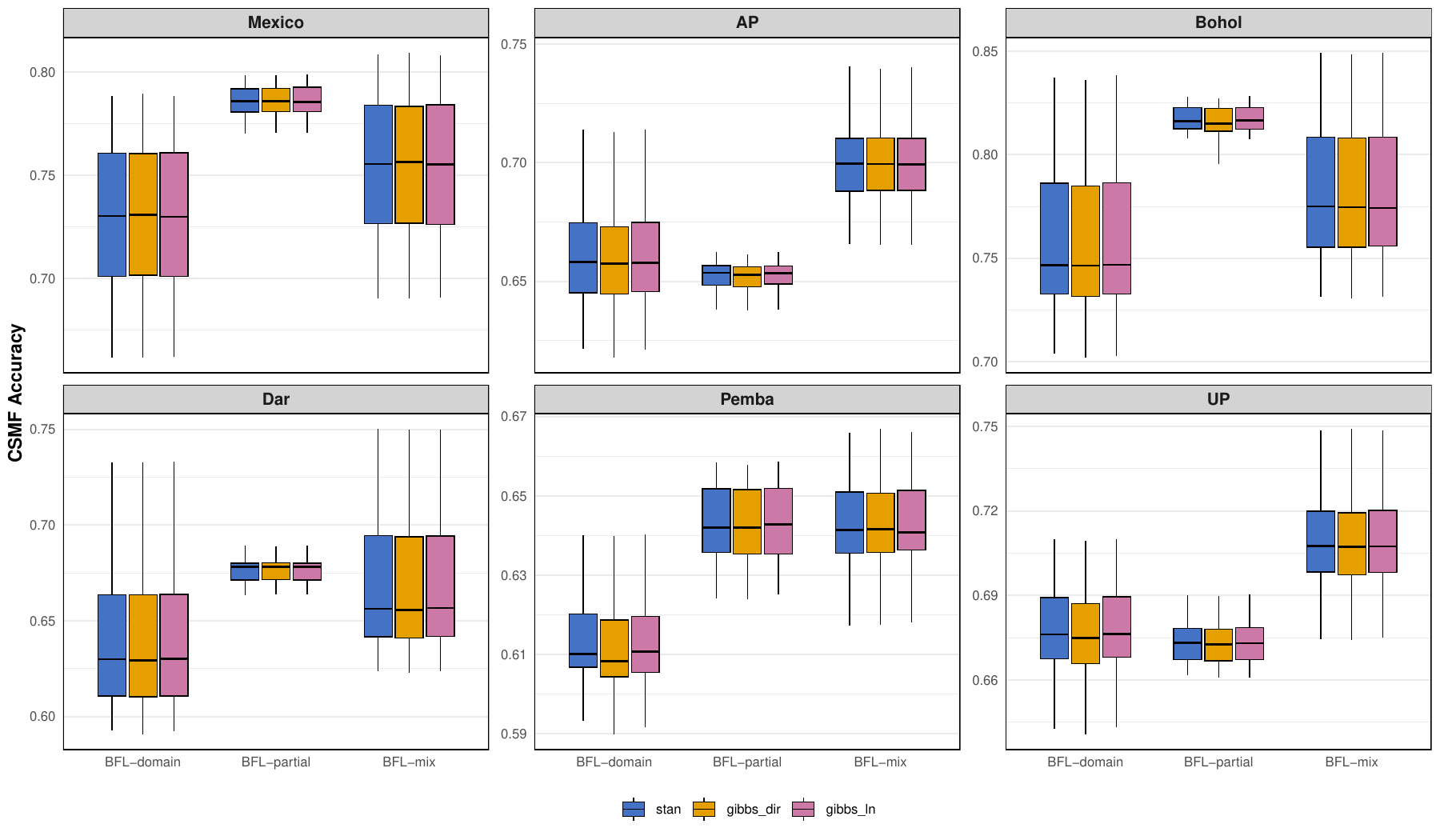}
    \caption{\textbf{No within-target label shift. CSMF accuracy}: Stan versus Gibbs (Dirichlet) versus Gibbs (logistic-normal) for the BFL variants across target sites.}
    \label{fig:sampler_csmf}
\end{figure}

\begin{figure}[!htb]
    \centering
    \includegraphics[width=\textwidth]{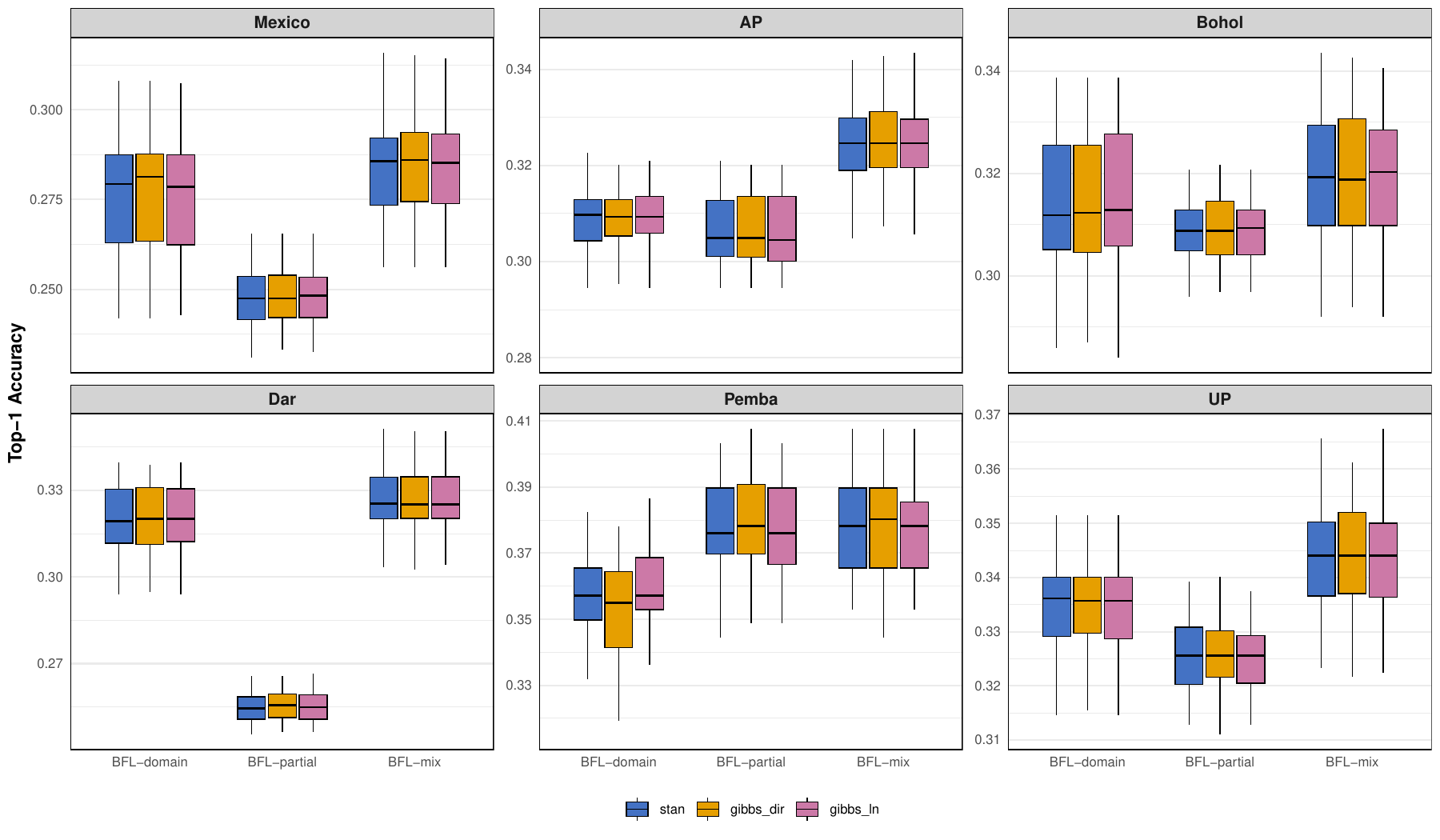}
    \caption{\textbf{No within-target label shift. Top cause accuracy}: the three samplers across the BFL variants and target sites.}
    \label{fig:sampler_top1}
\end{figure}

\begin{figure}[!htb]
    \centering
    \includegraphics[width=\textwidth]{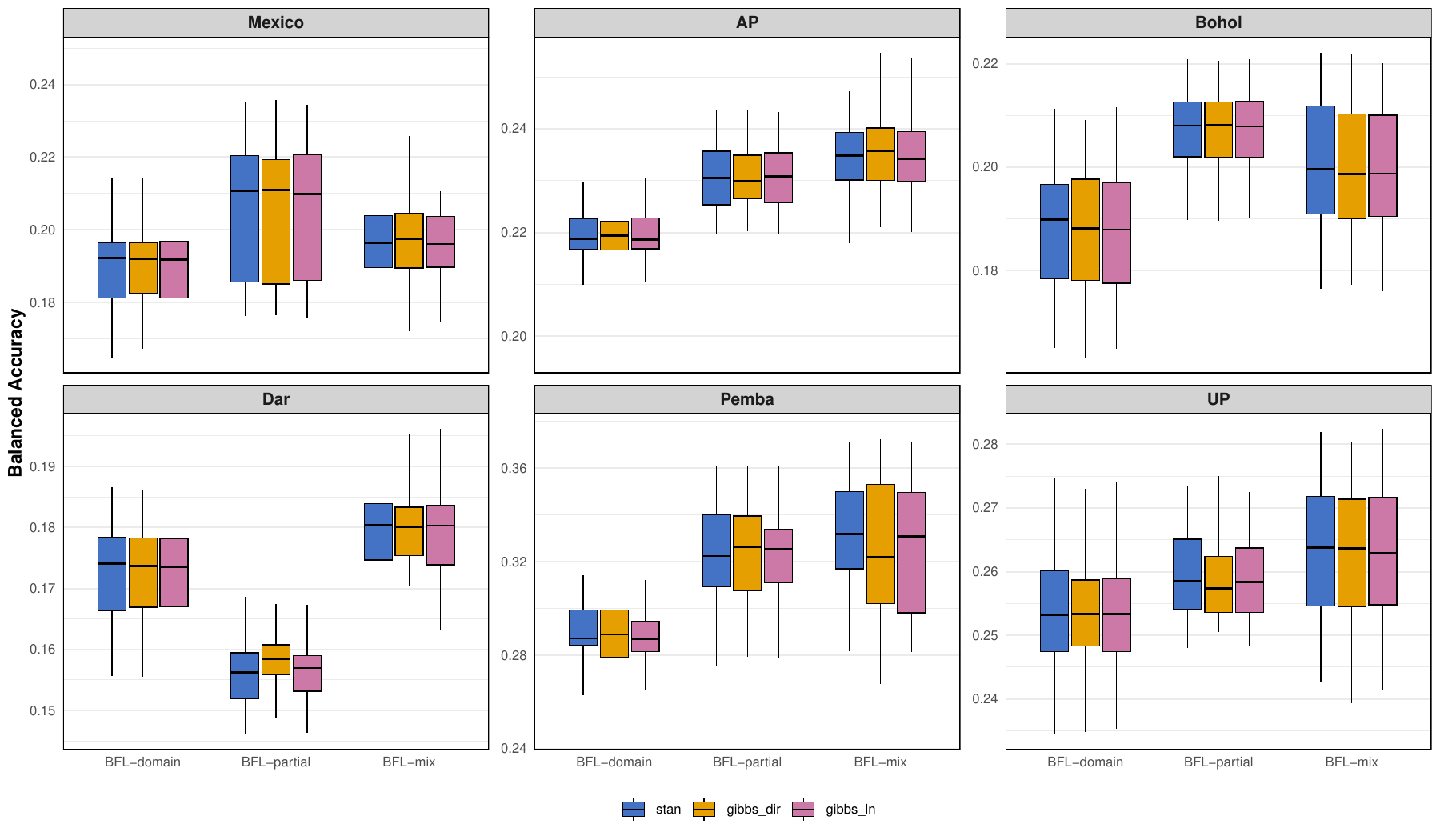}
    \caption{\textbf{No within-target label shift. Balanced accuracy}: the three samplers across the BFL variants and target sites.}
    \label{fig:sampler_balanced}
\end{figure}

The BFL posterior can be drawn with different sampling algorithms. We compare three: Stan (Hamiltonian Monte Carlo), a Gibbs sampler with a Dirichlet prior (\emph{gibbs\_dir}), and a Gibbs sampler with a logistic-normal prior (\emph{gibbs\_ln}). Figures~\ref{fig:sampler_csmf}--\ref{fig:sampler_balanced} compare the three across the BFL variants (\emph{BFL-domain}, \emph{BFL-partial}, \emph{BFL-mix}) at each target site, in the no within-target label-shift setting. The three samplers give near-identical results, so the choice of sampler does not materially affect accuracy.
\clearpage
\subsection{Estimated domain weights under full causes for all six sites}
\begin{figure}[!htb]
    \centering
    \includegraphics[width = \textwidth]{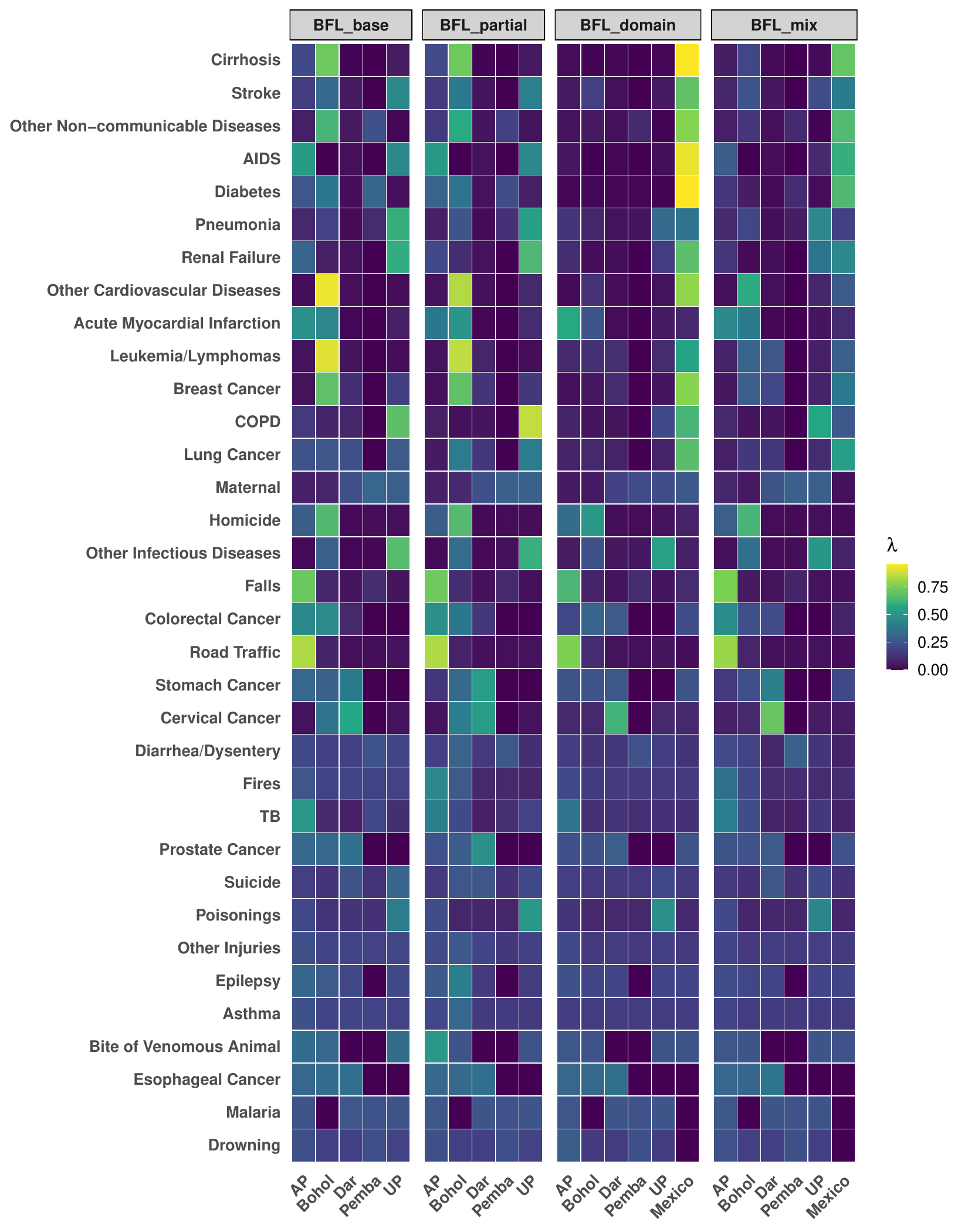}
    \caption{Posterior mean of the estimated domain weights $\blambda$ for predicting Mexico City. The causes in the Mexico City dataset are shown in decreasing order of prevalence from top to bottom. For \emph{BFL-domain} and \emph{BFL-mix}, the additional local model is included in the candidate models (the Mexico column).}
    \label{fig:heatmap_lambda_phmrc_balanced_full_causes_site_Mexico}
\end{figure}

\begin{figure}[!htb]
    \centering
    \includegraphics[width = \textwidth]{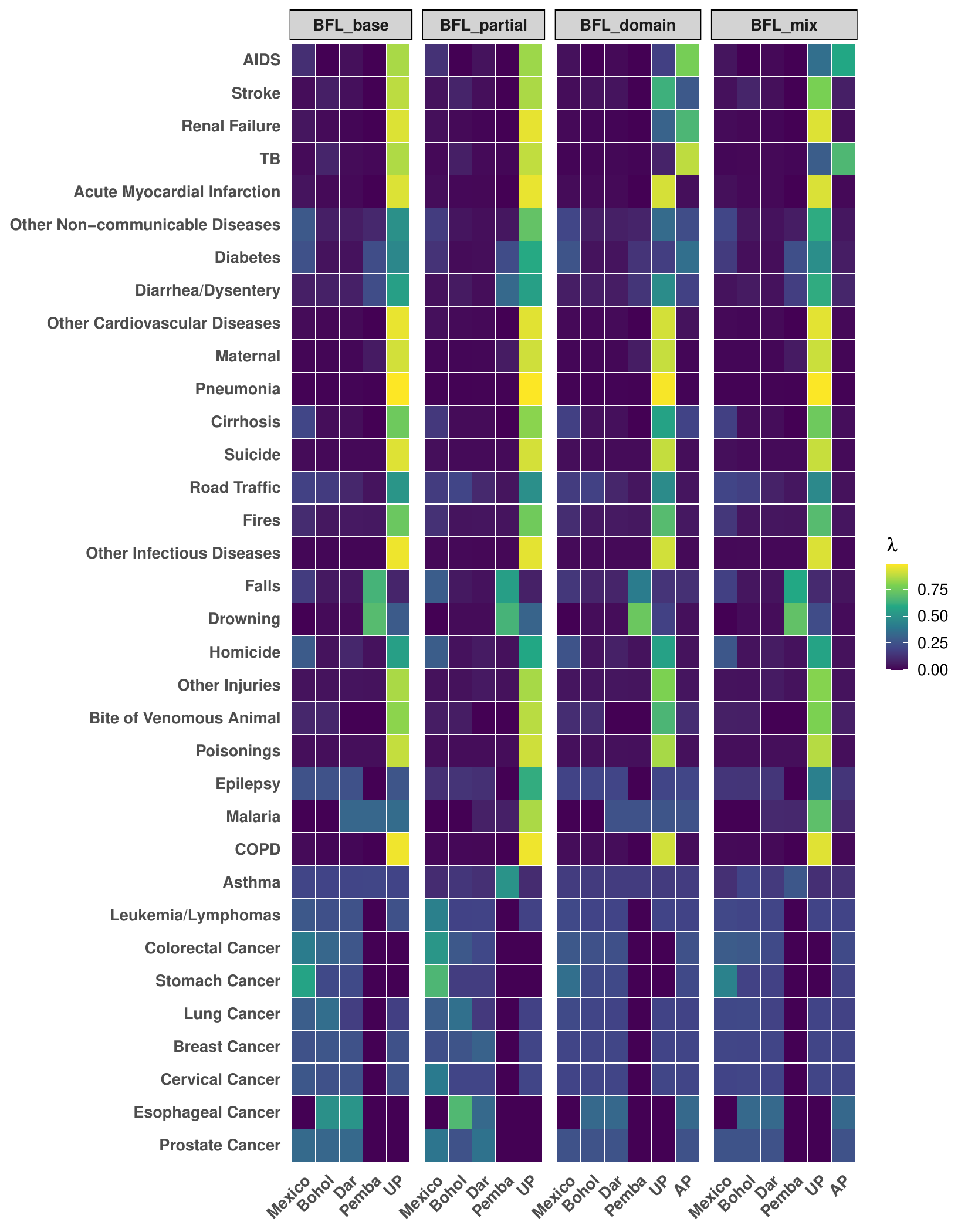}
    \caption{Posterior mean of the estimated domain weights $\blambda$ for predicting Andhra Pradesh. The causes in the Andhra Pradesh dataset are shown in decreasing order of prevalence from top to bottom. For \emph{BFL-domain} and \emph{BFL-mix}, the additional local model is included in the candidate models (the AP column).}
    \label{fig:heatmap_lambda_phmrc_balanced_full_causes_site_AP}
\end{figure}

\begin{figure}[!htb]
    \centering
    \includegraphics[width = \textwidth]{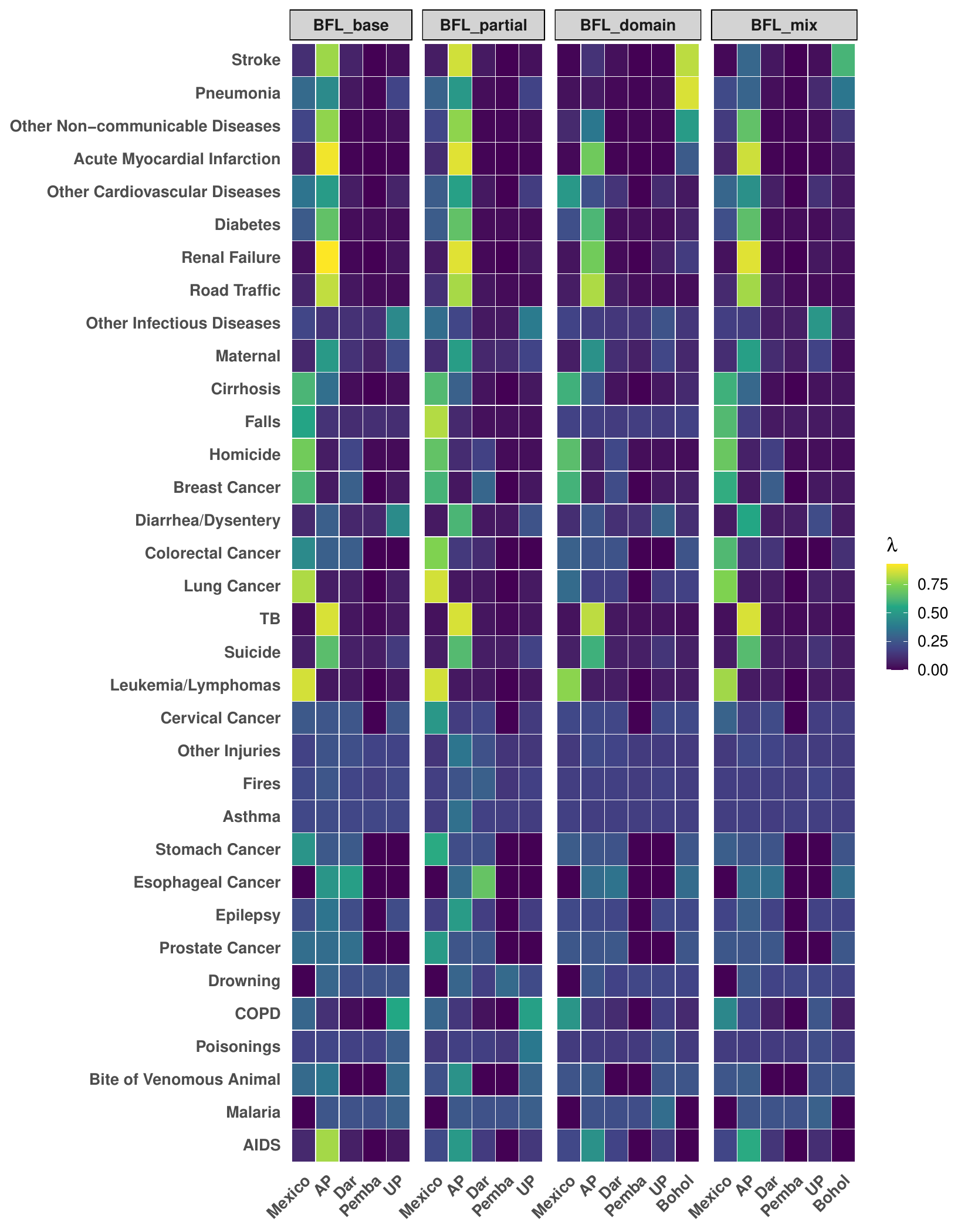}
    \caption{Posterior mean of the estimated domain weights $\blambda$ for predicting Bohol. The causes in the Bohol dataset are shown in decreasing order of prevalence from top to bottom. For \emph{BFL-domain} and \emph{BFL-mix}, the additional local model is included in the candidate models (the Bohol column).}
    \label{fig:heatmap_lambda_phmrc_balanced_full_causes_site_Bohol}
\end{figure}

\begin{figure}[!htb]
    \centering
    \includegraphics[width = \textwidth]{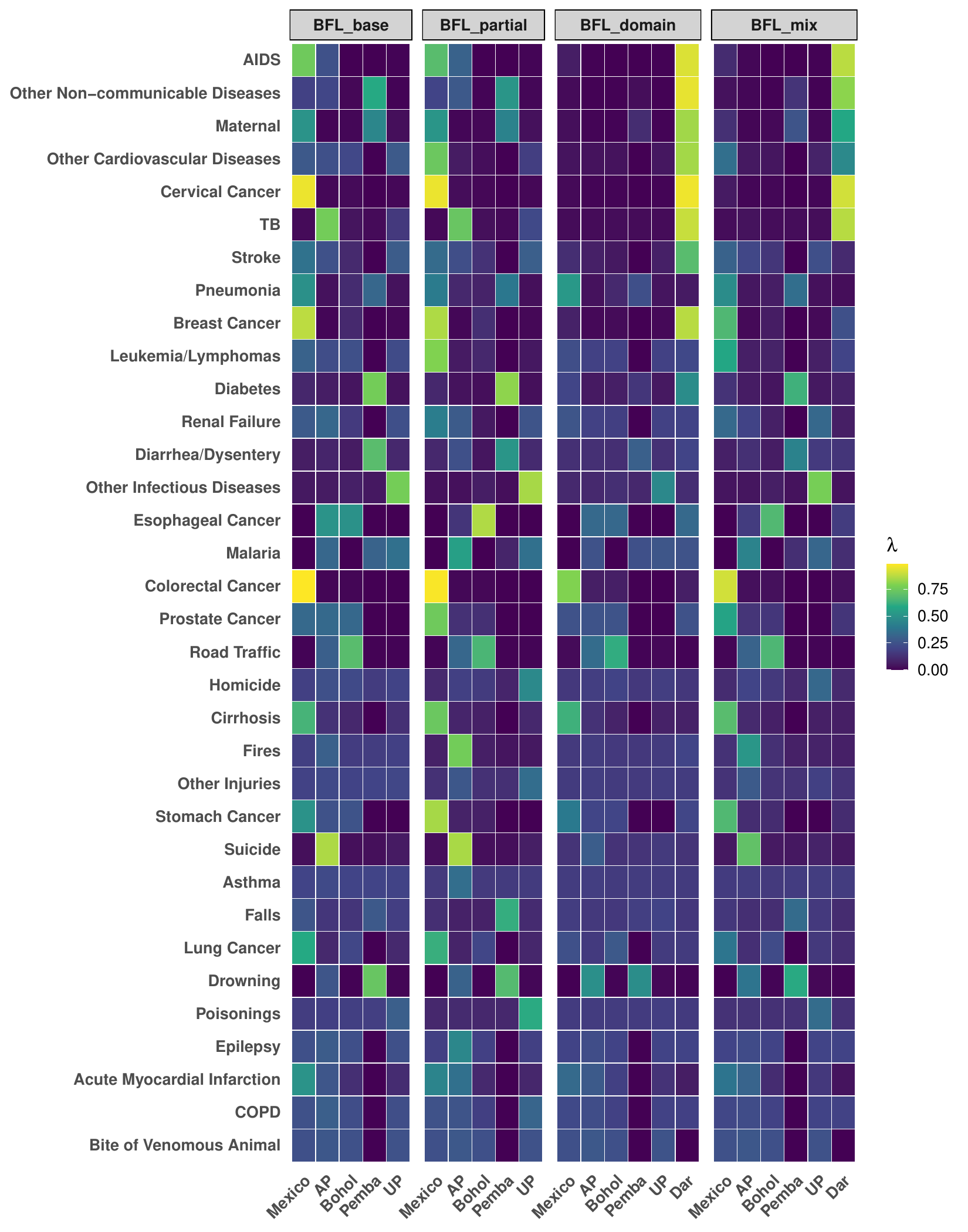}
    \caption{Posterior mean of the estimated domain weights $\blambda$ for predicting Dar es Salaam. The causes in the Dar es Salaam dataset are shown in decreasing order of prevalence from top to bottom. For \emph{BFL-domain} and \emph{BFL-mix}, the additional local model is included in the candidate models (the Dar column).}
    \label{fig:heatmap_lambda_phmrc_balanced_full_causes_site_Dar}
\end{figure}

\begin{figure}[!htb]
    \centering
    \includegraphics[width = \textwidth]{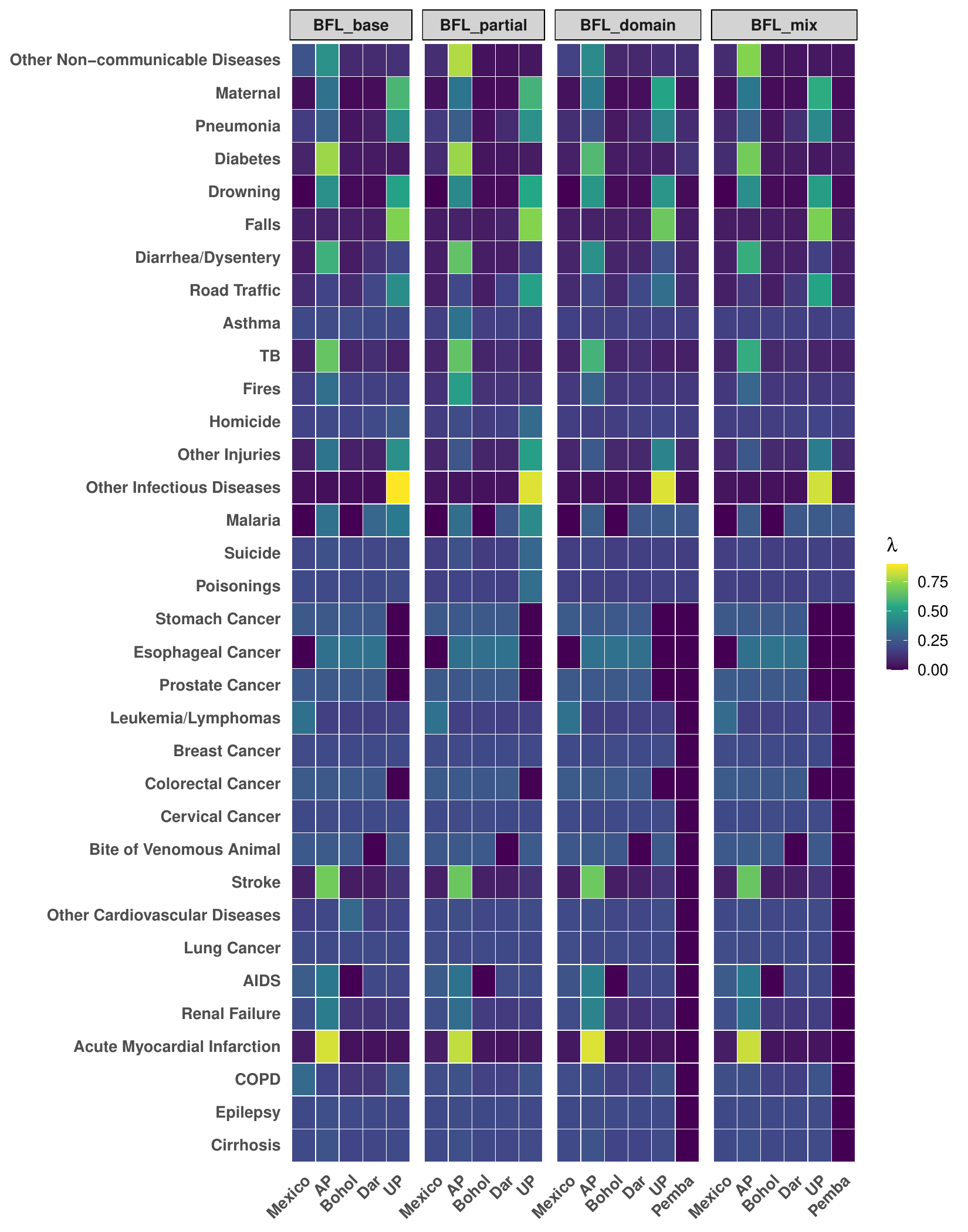}
    \caption{Posterior mean of the estimated domain weights $\blambda$ for predicting Pemba Island. The causes in the Pemba Island dataset are shown in decreasing order of prevalence from top to bottom. For \emph{BFL-domain} and \emph{BFL-mix}, the additional local model is included in the candidate models (the Pemba column).}
    \label{fig:heatmap_lambda_phmrc_balanced_full_causes_site_Pemba}
\end{figure}

\begin{figure}[!htb]
    \centering
    \includegraphics[width = \textwidth]{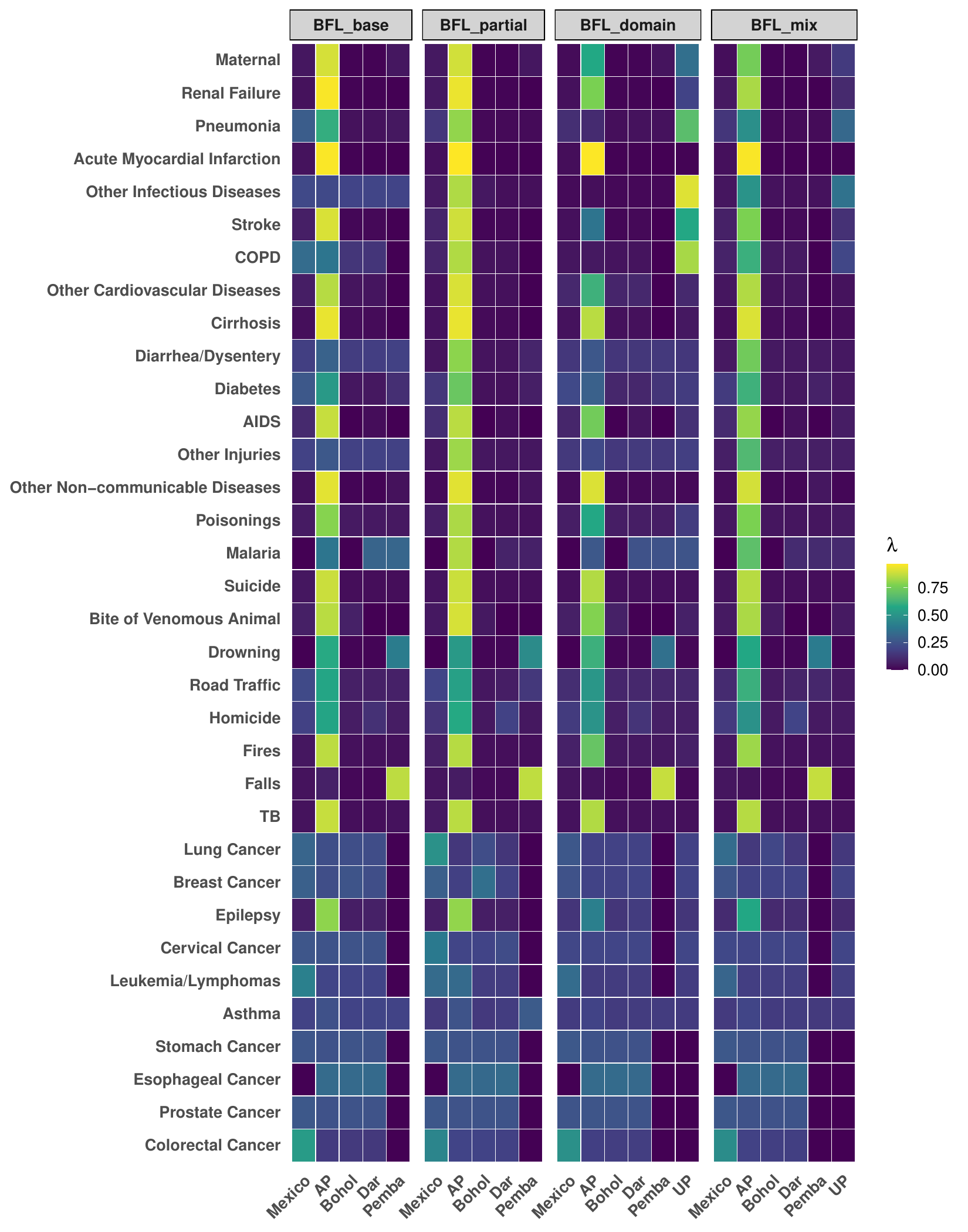}
    \caption{Posterior mean of the estimated domain weights $\blambda$ for predicting Uttar Pradesh. The causes in the Uttar Pradesh dataset are shown in decreasing order of prevalence from top to bottom. For \emph{BFL-domain} and \emph{BFL-mix}, the additional local model is included in the candidate models (the UP column).}
    \label{fig:heatmap_lambda_phmrc_balanced_full_causes_site_UP}
\end{figure}

Figures \ref{fig:heatmap_lambda_phmrc_balanced_full_causes_site_Mexico} to \ref{fig:heatmap_lambda_phmrc_balanced_full_causes_site_UP} show the posterior mean of domain weights under the full cause of death in decreasing order of prevalence among all six different sites.
\clearpage
\section{Additional summaries of the CHAMPS data analysis}
Table \ref{tab:CHAMPS_sample_size} shows the sample size in each of the four domains by causes.
\begin{table}[!htb]
\centering
\begin{tabularx}{1.1\textwidth}{lXXXX|X}
\toprule
 & 
\makecell{Kenya \\(KE)} & 
\makecell{Mozambique \\(MZ)} &  
\makecell{South Africa \\(ZA)} & 
\makecell{Target }  &
\makecell{\textbf{Total}} \\
\midrule
\makecell[l]{
IPRE}  
        & \makecell{167} & \makecell{245} & \makecell{110} & \makecell{284} &  \makecell{806} \\ 
\makecell[l]{prematurity} & \makecell{30} & \makecell{37} & \makecell{154} & \makecell{78}  &    \makecell{299}               \\ 
\makecell[l]{congenital malformation}  & \makecell{4} & \makecell{22} & \makecell{35} & \makecell{46}    &   \makecell{107}       \\ 
\makecell[l]{sepsis} & \makecell{19} & \makecell{29} & \makecell{50} & \makecell{32}    &       \makecell{130}       \\ 
other     & \makecell{13} & \makecell{6} & \makecell{11} & \makecell{26} & \makecell{56}                           \\  
undetermined     & \makecell{8} & \makecell{6} & \makecell{7} & \makecell{17} &     \makecell{38}                      \\  
infection     & \makecell{9} & \makecell{43} & \makecell{30} & \makecell{13}     &       \makecell{95}                \\  
pneumonia     & \makecell{20} & \makecell{9} & \makecell{4} & \makecell{9} &  \makecell{42}                          \\  \hline
\textbf{Total}     & \makecell{270} & \makecell{397} & \makecell{401} & \makecell{505}   &      \makecell{1573}                   \\  
\bottomrule
\end{tabularx}
\caption{Sample size in each of the four domains by causes from the CHAMPS dataset.}
\label{tab:CHAMPS_sample_size}
\end{table}

\bibliographystyle{plainnat}
\bibliography{crms}

\end{document}